\pdfoutput=1
\documentclass[12pt]{elsarticle}
\usepackage{lineno}
\usepackage[numbers]{natbib}
\def\beq{\begin{equation}}
\def\eeq{\end{equation}}

\def\avnu{\langle \nu_{n} \rangle}
\def\avm{\langle M \rangle}

\def\avpt{{\Big\langle} p_t^2{\Big\rangle}}
\def\blangle{{\Big\langle}}
\def\brangle{{\Big\rangle}}
\begin{document}
\begin{frontmatter}
\title{De-Confinement and Clustering of Color Sources in Nuclear Collisions}
\author[address1]{M. A. Braun}
\author[address2]{J. Dias de Deus}
\author[address3]{A. S. Hirsch}
\author[address4]{C. Pajares}    
\author[address3]{R. P. Scharenberg}
\author[address3]{B. K. Srivastava\corref{corrauthor}}
\cortext[corrauthor]{Corresponding author}
\ead{brijesh@purdue.edu}
\address[address1]{Department of High Energy Physics, Saint-Petersburg State University, S. Petersburg, Russia}
\address[address2]{CENTRA, Instituto Superior Tecnico, 1049-001 Lisboa, Portugal}
\address[address3]{Department of Physics and Astronomy, Purdue University, West Lafayette, IN-47907, USA}
\address[address4]{Departamento de Fisica de Particulas, Universidale de Santiago de Compostela and Instituto Galego de Fisica de Atlas Enerxias(IGFAE), 15782 Santiago, de Compostela, Spain}

\date{\today}% It is always \today, today
\begin{abstract}

A brief introduction of the relationship of string percolation to the \\ Quantum Chromo Dynamics (QCD) phase diagram is presented. The behavior of the Polyakov loop close to the critical temperature is studied in terms of the color fields inside the clusters of overlapping strings, which are produced in high energy hadronic collisions. The non-Abelian nature of the color fields implies an enhancement of the transverse
momentum and a suppression of the multiplicities relative to the non overlapping case. 
The prediction of this framework are compared with experimental results from the SPS, RHIC and LHC for $pp$ and AA collisions.
 Rapidity distributions, probability distributions of transverse momentum and
multiplicities, Bose-Einstein correlations, elliptic flow and ridge structures are used to evaluate these comparison. 

The thermodynamical quantities, the temperature, and energy density derived from RHIC and LHC data and Color String Percolation Model (CSPM) are used to obtain the shear viscosity to entropy density ratio ($\eta/s$). It was observed that the inverse of ($\eta/s$) represents the trace anomaly $\Delta =(\varepsilon-3P)/T^{4}$. 
Thus the percolation approach within CSPM can be successfully used to describe the initial stages in high energy heavy ion collisions in the soft region in high energy heavy ion collisions. The thermodynamical quantities, temperature and the equation of state are in agreement with the lattice QCD calculations. Thus the  clustering of color sources has a clear physical basis although it cannot be deduced directly from QCD. 

\end{abstract}
%\begin{keyword}
%\texttt{elsarticle.cls}\sep \LaTeX\sep Elsevier \sep template
%\MSC[2010] 00-01\sep  99-00
%\end{keyword}
\end{frontmatter}
\newpage
\tableofcontents
\newpage
%
%\linenumbers
\section{Introduction}
% Pajares write up starts here
\subsection{Motivation and historical perspective}
 Before the discovery of the quarks there was an interest in the behavior of matter at high density and/or high temperature \cite{par1}. This interest was increased with the formulation of QCD and the possibility of distributing high energy or high nucleon density over a large volume to temporary restore broken symmetries of the physical vacuum creating abnormal dense states of nuclear matter \cite{par2}.
 
Very early after QCD was born, it was pointed out that the asymptotic freedom property of QCD implies the existence of a high density matter formed by deconfined quarks and gluons \cite{par3} and the exponential increasing of the spectrum of Hagedorn was connected to the existence of a different phase, in which quarks and gluons are deconfined \cite{par4}. The thermalized phase of quarks and gluons was called Quark Gluon Plasma (QGP) \cite{par5} and it was realized that the required high density could be achieved in relativistic heavy-ion collisions \cite{par6,bjorken} and several signatures of QGP were proposed. Quarkonium suppression \cite{par8}, anomalous excess of photons and jet quenching \cite{par9,par10} were some of them. At this time, it was pointed out the relevance of percolation in the study of the phase structure of hadronic matter \cite{par11,par12}.

Experimentally there was a large effort to study in laboratory the deconfinement and chiral symmetry restoration phase transitions and to explore the properties of high density matter starting by the AGS and ISR experiments and continued at SPS, RHIC and LHC.
The SPS accelerator experiments already displayed several signatures that hinted at the onset of QGP formation \cite{par13}. The RHIC data show a striking bulk collective elliptic flow, which is generated at relatively early times, since otherwise the spatial anisotropy could not convert into a momentum spatial anisotropy. The flow pattern was consistent with a very low shear viscosity over entropy density ratio $\eta/s$, indicating strongly interacting matter. On the other hand,  jet quenching phenomena were clearly observed indicating that this deconfined
strongly interacting matter was	very opaque \cite{par14,star,phenix,phobos,brahms}. The above mentioned ratio gave rise to an increasing interest on the AdS/CFT correspondence due to the result $\eta/s=1/4\pi$ \cite{kss}.

The recent LHC experiments \cite{par20,par21,par22} have extended the study of the elliptic flow to all the harmonics \cite{par23,par24} confirming that the high quark gluon density matter interacts strongly.
The collective behavior and  the ridge structure observed previously at RHIC in Au-Au and Cu-Cu collisions \cite{par25,par26}, was also observed in pPb \cite{par27,par28,par29} and pp interactions \cite{par30} at high multiplicity at the LHC. The collective flow of pPb and the ridge structure of pPb and pp collisions is a challenge to the usual hydrodynamic descriptions. On the other hand, the data on quarkonium seems each time to confirm more the validity of the combined picture of a sequential melting of the different resonances together with  recombination of  heavy quarks and antiquarks at high energy \cite{par31,par32,par33}. Also a detailed study of the jet quenching for identified particles has been performed \cite{par34} showing interesting features related to the loss of coherence of the gluons emitted in the jet due to the high density medium.
% CGC part 

From the theoretical side in addition to the hydrodynamic studies the Color Glass Condensate(CGC) picture \cite{par35,par35x,par35y,par36,par37} derived directly from QCD, is very appealing and gives a reasonable description of most of the experimental observables. At first sight, due to the non-Abelian nature of QCD, with  gluons carrying color charge, the gluon density 
$ xG(x,Q)$ rises rapidly as a function of the decreasing fractional momentum $ x$, or increasing the resolution $Q_s$. So, the gluon showers generate more gluon showers producing an exponential avalanche toward small $ x$. As the transverse size of the hadron or the nucleus rises slowly at high energies,  the number of gluons and  their density  per unit of area and rapidity increase rapidly as $ x$ decreases. However, there will be fusion of gluons leading to a limited transverse density of gluons at some fixed momentum resolution $Q_{s}$ and that is gluon saturation \cite{par38}. The low $ x$ gluons are closely packed, the distance between them  being very small, hence the
interaction coupling is weak $\alpha_{s} \ll $ 1. Weak couplings systems are adequate to be studied in QCD. In a given collision the multiplicity should be proportional to the number of gluons, which at the  saturation momentum $Q_s$, is \cite{par39,par40}.
\begin{equation}
\frac {dN}{dy}\sim \frac {1}{\alpha_{s}(Q_{s})} Q_{s}^{2} R^{2}.
\label{int1}
\end{equation}
This dense system, called CGC, has a very high occupation number $1/\alpha_{s}(Q_{s})$ and corresponds to a highly coherent state of strong color fields. The high $x$ gluons can be regarded as the sources of the low $x$ gluons. The independence on the cutoff, used to separate the high $x$ gluons from the low $x$ ones, gives rise to a kind of evolution equation. In the infinite momentum frame, these large momentum gluons travel very fast and therefore their time scales are Lorentz dilated. 
%{\it This time dilated is transferred to the low x gluons which evolves very slowly compared to natural times scales, like a glass.}

 Due to the Lorentz contraction the collision of two nuclei can be seen as that of two sheets of colored glass where the color field in each point of the sheets is randomly directed. Taking these fields as initial conditions, one finds that between the  sheets longitudinal color electric and  magnetic fields are formed. The number of these color flux tubes between the two colliding nuclei is $Q_{s}^{2} R^{2}$ forming Glasma \cite{par41}, which has been extensively applied to compare with experimental data.
% Percolation
 
An alternative approach to the CGC is the percolation of strings \cite{ref7,par43,ref8,brapaj2000,par46} subject of this report. The percolation of strings is not directly obtained from QCD and only is QCD inspired. In string percolation, multiparticle production is described in terms of color strings stretched between the partons of the projectile and  target. These strings decay into $q-\bar{q}$ and $qq-\bar{qq}$ pairs and subsequently hadronize producing the observed hadrons. Due to the confinement, the color of these strings is confined to a small area $S_{1}=\pi r_{0}^{2}$, with $r_{0}$=0.2 fm in the transverse space. With increasing energy and/or size and centrality of the colliding objects, the number of strings grows and the strings start to overlap forming clusters, very similar to discs in 2-dimensional percolation theory \cite{isich}. At a given critical density, a macroscopic cluster appears crossing the collision surface, which marks the percolation phase transition. Therefore the nature of this transition is geometrical.

 In string percolation the basic ingredients are the strings, and it is necessary to know their number, rapidity extension, fragmentation and number distribution. All that requires a model and therefore string percolation is model dependent. However most of the QCD inspired models give very similar results for most of the observables in such a way that the predictions are  by a  large measure independent of the model used. 

The string percolation and the Glasma are related to each other \cite{perx}. In the limit of high density there is a correspondence between physical quantities of both approaches. The number of color flux tubes in the Glasma picture, $Q_{s}^{2} R^{2}$, has the same dependence on the energy and on the centrality of the collision that the number of effective clusters of strings in string percolation. In both approaches  for the multiplicity distribution a negative binomial distribution is obtained,  where the parameter ${\it k}$ that controls the width has  the same energy and centrality dependence. The role of the occupation number $1/\alpha_{s}$ in CGC is played by the fraction of the total available surface covered by the strings formed in the collision. The randomness of the color field in the CGC gives rise to a reduction of the expected multiplicity. Similarly the randomness of the direction in color space of the color field of the 
${\it {n}}$ strings of the cluster stipulates that the strength of the resulting field is not ${\it {n}}$ times the strength of the individual color field but $\sqrt{n}$. This reduction also implies  a reduction of the multiplicity of particles produced in the collision. Due to all these similarities in both approaches the predictions for different physical observables are very similar. The string
percolation is able to explore also the region where the limit of high density has not been reached. 

The plan of the review is as follows: After this introduction in the two next chapters  the phase structure of QCD and the main results of lattice QCD are  briefly studied. Percolation studies related to QCD are briefly reviewed and later the color string percolation model  is introduced. In the following sections the applications of the model to the description of transverse momentum distributions, rapidity and  multiplicity distributions,
 azimuthal distributions,  harmonic momenta and ridge structures, Bose-Einstein correlations, $J/\Psi$ suppression, and forward-backward correlations are described. Finally, the thermodynamics in the production of particles in string percolation is studied starting by relating the critical percolation density with the critical temperature. Afterwards we study the equation of state,  the behavior of the trace anomaly and  ratio between the shear viscosity and the entropy density with temperature.
% Phase diagram
\subsection{Phase diagram of nuclear matter}

 What is the behavior of matter when we increase its density? The observed densities of our world expand over many orders of magnitude, from $\sim 10^{-6}$ nucleons/$cm^{3}$ in average in the Universe to $\sim 10^{38}$ nucleons/$cm^{3}$ inside a nucleus and $\sim 10^{39}$ nucleons/$cm^{3}$ in a neutron star. The study of the high density limit, specifically the confinement transition from hadrons to quarks and gluons can be regarded as the place where the high energy collision (two body) probes the short distance limit and therefore meets the thermodynamics (many body) of this short distance dynamics \cite{satzextreme}. 

Let us start by studying the behavior of a high density medium under electromagnetic forces. In a high density medium, the Coulomb potential between two electric charges 
\begin{equation}
V(r) \approx \frac {e^{2}}{r}
\label{int2}
\end{equation}
becomes screened in such a way that the new potential  
\begin{equation}
V(r) \approx \frac {e^{2}}{r}\exp(-r/r_{0}(\rho))
\label{int3}
\end{equation}
disappears for $ r > r_{0}$, where $r_{0}$ is the Debye screening radius, which decreases as the charge density $\rho$ increases.  This potential change occurs because  other charges in the medium are around the two original ones, reducing  the interaction. In this way the interaction range becomes shorter. If we consider a bound electromagnetic state, like a hydrogen atom and increase the density the Debye radius eventually becomes smaller than the binding radius of the atom $r_{0} < r$ and the force between the proton and  electron is reduced so that the two particles can no longer be bound. An insulator material consisting of bound electrical charges at high density becomes a conductor, it undergoes a phase transition, the Mott transition \cite{par49}. The charge screening dissolves the binding of the constituents giving rise to a new state of the matter, the plasma. In this transition, the electrons change their mass from the usual physical mass in the insulator phase to an effective medium electron mass. In fact, the electrons in the conducting phase move in the periodic field of the ions, which is  very different from the vacuum. 
In the case of QCD, we have confinement, which means that the lines of the color field between two color charges are no longer expanded into all the space like in the case of Coulomb potential between electrical charges but
 are confined to a rather small space region, forming hadrons, which are colorless bound states of colored quarks.
\begin{figure}[thbp]
\centering        
\vspace*{-0.2cm}
\includegraphics[width=0.65\textwidth,height=3.0in]{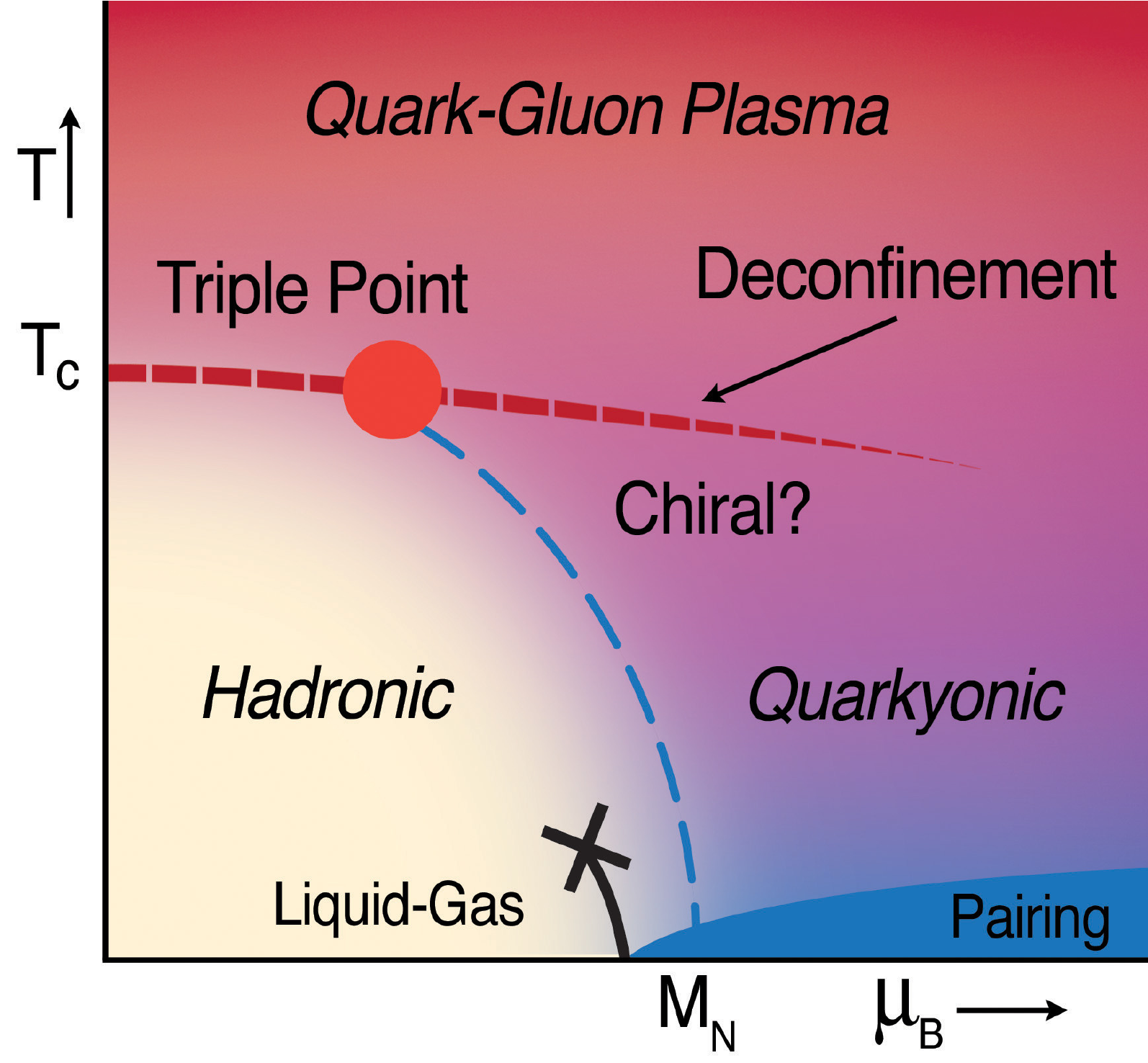}
%\vspace*{-1.0cm}
\caption{Phase diagram of the nuclear matter. Temperature vs baryonic chemical potential \cite{phase}.}
\label{phasediagram}
\end{figure}
However, at high color charge density we expect that color screening takes place resulting in an exponential damping of the potential and removing all the long range effects of the confining potential. In this way, we expect that color screening transforms a color insulator into a color conductor, leading to  deconfinement of quarks from hadrons \cite{par50}. This transition, like the insulator to conductor in the electromagnetic case, is a collective effect, in which  many charges participate, and we expect a behavior like a phase transition. 

In a similar way to the electron mass, the quark masses are expected to change between the two phases. In the confined phase the quarks are dressed with gluons, acquiring an effective constituent mass of around 300 MeV (1/3 of the proton mass). For light quarks this constituent mass is much larger than their mass appearing in the QCD Lagrangian where it is close to zero. In another words, the light quark mass in the confined phase is generated by the confinement interaction, and when deconfinement occurs  this additional mass disappears.

 The role of gluons would be to generate the effective quark mass, maintaining spontaneous chiral symmetry breaking. There will be need of a higher temperature $T$ or baryonic potential $\mu$ to evaporate or melt the gluonic density of quarks leading to deconfined quarks and gluons with restored chiral symmetry. According to this picture, from the hadronic phase, first it is the dressed quark phase \cite{par51} and after the deconfined and the restoration of chiral symmetry. In this picture, there are two scales: First, the hadronic scale $R_{h}=$ 1 fm when the density $\rho$ is below
\begin{equation}
\rho_{h}\sim \frac {1}{(4/3) \pi R_{h}^{3}}.
\label{int4}
\end{equation}
and the medium is of hadronic nature. For densities $\rho$ 
\begin{equation}
\rho_{h} <  \rho < \rho_{q} \sim \frac {1}{(4/3) \pi R_{q}^{3}}
\label{int5}
\end{equation}
with $R_{q} \sim 0.3$ fm the medium consists of deconfined massive quarks. Finally, for densities $\rho > \rho_{q}$ the medium becomes quarks and gluons phase with restoration of chiral symmetry.

This picture could be true if the scale $R_{q}$ were independent of $T$ and $\mu$. However, the lattice studies have shown, that near $\mu=0$, color deconfinement and chiral symmetry restoration coincide. Hence, in a medium of low baryon density, the mass of the constituent quark vanishes at the deconfinement point $T_{c}$ and  also the screening radius of the gluon cloud size vanishes. At low $ T$ and high $\mu$, there is no reason to expect a similar behavior and probably there will be an intermediate region of massive dressed quarks between the hadronic phase and the
deconfined and chiral symmetry restoration phase. At low $ T$ and high $\mu$, other possibilities could exist as quarkyonic matter and color 
superconductivity \cite{par52,par53,par54}. A possible phase diagram is shown in Fig.~\ref{phasediagram} \cite{phase}.

\subsection{Lattice QCD}
In finite temperature lattice QCD, the deconfinement order parameter is provided by the vacuum expectation value of the Polyakov loop $L(\vec{x})$ defined in
 Euclidean space
\begin{equation}
L (\vec{x}) = {\rm Tr}\prod_{t=1}^{N_{t}}A_{4}(\vec{x},t)
\label{lat1}
\end{equation}
$L(\vec{x})$ is the ordered product of the SU(3) temporal gauge variables $A_{4}(\vec{x},t)$ at a fixed spatial position, where $N_{t}$, is the number of lattice points in time direction and Tr denotes the trace over color indices. The Polyakov loop corresponds to a static quark source and its vacuum expectation value is related to the free energy
 $F_{q}$ of a single quark
\begin{equation}
L (\vec{x}) \sim \exp\Big(- \frac {F_{q}}{T}\Big).
\label{lat2}
\end{equation}
Below the critical temperature $T_{c}$ quarks are confined and $F_{q}$ is infinite implying  $\langle L(\vec{x}) \rangle$=0. In a deconfined medium color screening among the gluons makes $F_{q}$ finite, hence for $T > T_{c}$ $\langle L(\vec{x}) \rangle \neq 0$.  
The breaking of chiral symmetry is controlled by the chiral condensate $X(T) \equiv \langle \bar{\Psi}\Psi\rangle \sim M_{q}$ which measures the constituent quark masses obtained from a Lagrangian with massless quarks. At high temperature this mass melts, therefore 
\begin{equation}
X(T)_{  = 0\, T > T_{x}}^{ \neq 0\, T < T_x}. 
\label{lat3}
\end{equation}
$T_{x}$ defines another critical temperature.
The corresponding derivatives with respect to $T$ of $\langle L \rangle $ and $X$, the susceptibilities, have been studied in finite temperature lattice QCD at vanishing baryon number, showing a sharp peak that defines respectively $T_{c}$ and $T_{x}$. The two temperatures, within errors, coincide. Also seen is a sharp transition, i.e. a crossover. The quoted value of $T_{c}$ is 155 $\pm 9$ MeV \cite{par55,par56,peter}.

\begin{figure}[thbp]
\centering        
\vspace*{-0.2cm}
\includegraphics[width=0.999\textwidth,height=2.6in]{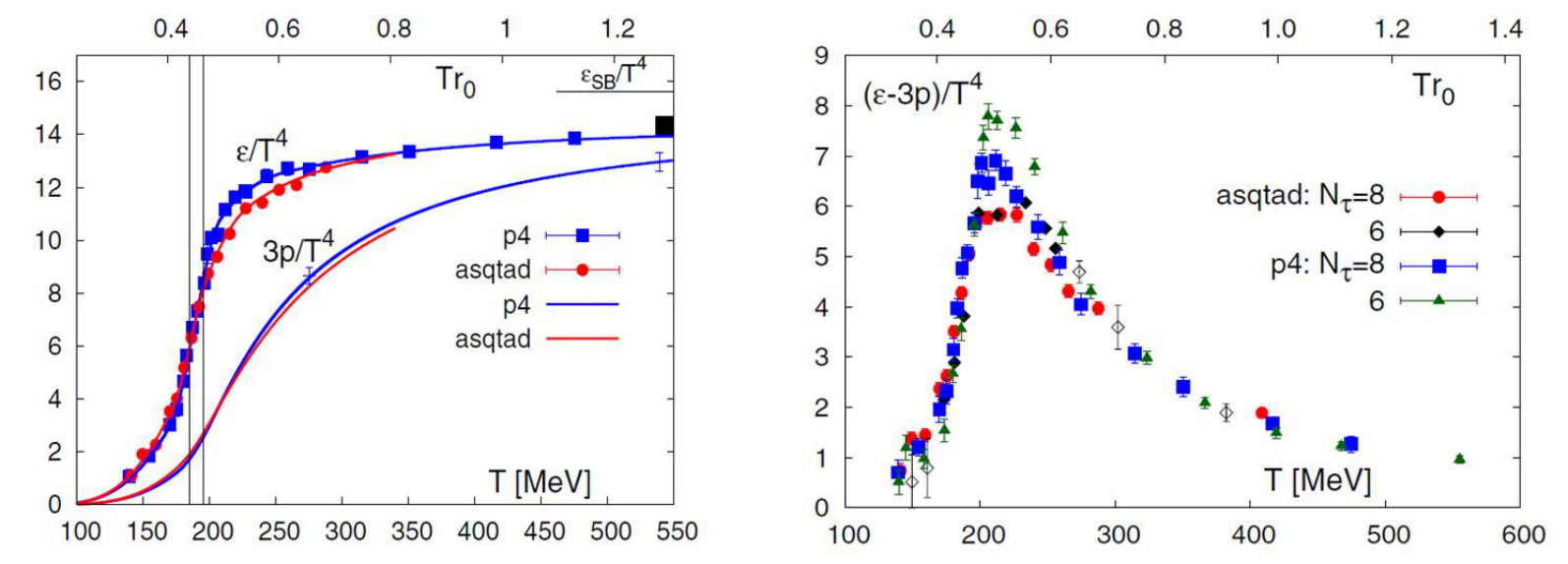}
\caption{The energy density and the pressure as function of temperature (left). The energy density shows a sharp rise in the temperature region 170-200 MeV. The interaction measure calculated using different staggered fermion actions (right)  \cite{peter}.}
\label{hotqcd}
\end{figure}

Energy densities resulting from lattice QCD are shown in Fig.~\ref{hotqcd}(left) indicating that even for $T > 3 T_{c}$ its values are far from the energy density of a free gas of quarks and gluons, namely
\begin{equation}
\varepsilon = \frac{\pi^{2}}{30}[g_g+\frac{7}{8}(q_q+q_{\bar{q}})]T^{4}.
\label{lat4}
\end{equation}
where $g_g$, $g_q$ and $g_{\bar{q}}$ are the degeneracy numbers of the gluons, quarks and antiquarks. 
This fact indicates that in a rather broad range of
temperatures the deconfined phase is a strongly interacting medium. This is better seen looking at the interaction measure 
\begin{equation}
\Delta = (\varepsilon -3 P)/T^{4}.
\label{lat5}
\end{equation}
The energy momentum tensor trace 
\begin{equation}
T_{\mu}^{\mu} = \frac {B(g_{s})}{2g_{s}}G_{\mu \nu}^{a}G^{\mu \nu a}+[1+r(g_{s})]m_{q}\bar\Psi_{q} \psi_{q}
\label{lat6}
\end{equation}
is ($\varepsilon - 3 P$) and even for massless quarks $T_{\mu}^{\mu} \ne 0$ as a consequence of the introduction of the scale in the re-normalization 
process breaking the conformal symmetry (trace anomaly). In Fig.~\ref{hotqcd} (right) the results of lattice QCD  are shown. $\Delta$ decreases with ${\it T}$ very slowly, even less than $1/T^{2}$. The deconfined medium interacts strongly for a broad range of temperatures above ${\it T}$.

\subsection{Percolation}
\begin{figure}[thbp]
\centering        
\vspace*{-0.2cm}
\includegraphics[width=0.99\textwidth,height=2.1in]{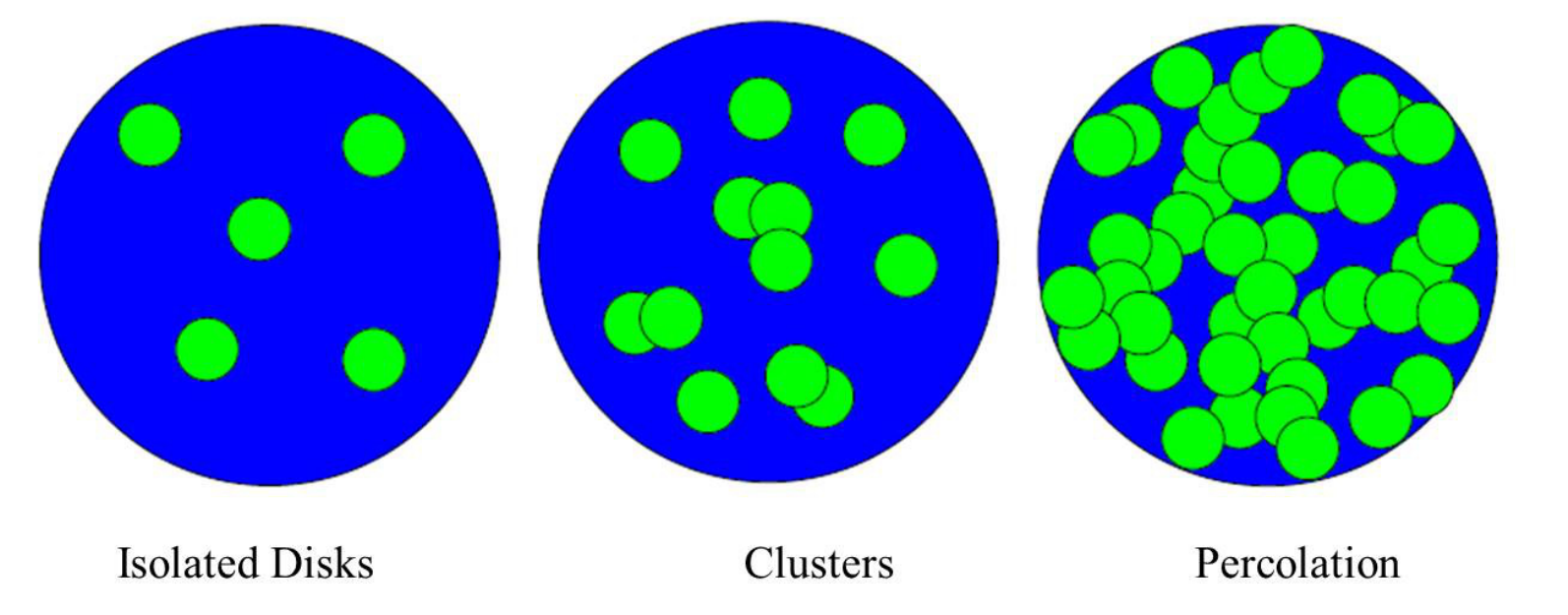}
%\vspace*{-2.0cm}
\caption{Left panel: Disconnected discs, Middle: Cluster formation, Right panel: Overlapping discs forming a cluster of communication \cite{satzextreme}.}
\label{cluster}
\end{figure}

 A simple example of percolation  is the 2-dimensional continuous percolation, which will be used extensively studying string percolation \cite{isich,stauf}.
Let us distribute small discs of area $\pi r_{0}^{2}$ randomly on a large surface, allowing overlap between them. As the number of discs increases clusters of overlapping discs start to form. If we regard the discs as small drops of water, how many drops are needed to form a puddle crossing the considered surface? Given $N$ discs, the disc density is $\rho = N/S$ where $S$ is the surface area. The average cluster size increases with $\rho$, and at a certain critical value $\rho_{c}$ the cluster 
spans the whole surface as shown in  Fig.~\ref{cluster} \cite{satzextreme} .

 The critical density for the onset of continuum percolation has been determined by numerical and Monte-Carlo simulations for different systems, which in 2- dimensional case gives
\begin{equation}
\rho_{c}=\frac {1.13}{\pi r_{0}^{2}}.
\label{per1}
\end{equation}
In the thermodynamical limit corresponding to  $N \rightarrow \infty $, keeping $\rho$
fixed, the distribution of  overlaps of the discs is Poissonian with a mean value 
 $\xi = \rho \pi r_{0}^2$, $\xi$ being a dimensionless quantity
\begin{equation}
P_{n}= \frac{\xi^{n}}{n!}\exp (-\xi).
\label{per2}
\end{equation}
Hence the fraction of the total area covered by discs is $1-\exp(-\xi)$  \cite{brapaj2000}.
 For the critical value of 1.13 around 2/3 of the area is covered by discs. This number 1.13 is obtained for the case of an homogeneous surface \cite{stauf,isich}. In cases of non homogeneous surface profiles, this factor changes. For example, in the cases of circular surfaces with Gaussian or Wood-Saxon profiles the critical percolation is reached at
\begin{equation}
\rho_{c}=\frac {1.5}{\pi r_{0}^{2}}.
\label{per3}
\end{equation} 
 Also the fraction of the area covered by strings is no longer given by $1-\exp(-\xi)$, but  by the function \cite{par59}
\begin{equation}
\frac {1}{1+a \exp(-(\xi-\xi_{c})/b)}
\label{per4}
\end{equation}
where $\xi_{c}$ =1.5. The parameters $ a$ and $ b$ depend on the profile function. In particular $ b$  controls the ratio between the width of the border of the profile (2$\pi R$) and the total area ($\pi R^2$), therefore is proportional to $1/R$.

Three-dimensional percolation has been applied to study the phase boundaries of high density matter. As mentioned earlier, even before the discovery of the quarks, Pomeranchuk \cite{par1} realized that above a certain high density hadrons lost their identity. In fact, when the density of a gas of hadrons is increased by raising either the
temperature or the baryon density, a quark of a given hadron will be closer to some quark or antiquark of other hadrons than to the original partners. The identity of hadronic matter is lost and now there is deconfined quark and antiquark matter. At $\mu =0$, the percolating density for mesons and low density baryons is
\begin{equation}
\rho_{c} \simeq \frac {1.24}{(4/3)\pi R_{h}^{3}}\simeq 0.6 fm^{-3},
\label{per6}
\end{equation}
where $R_{h}$= 0.8 fm is the hadron radius. This is the density of the percolating cluster at the onset of perco1ation. 
If we place overlapping spheres in a large volume, the critical density for the percolating spheres is \cite{isich,par59,par60}
\begin{equation}
\rho_{c} \simeq \frac {0.34}{(4/3)\pi R_{h}^{3}}\simeq 0.16 fm^{-3}.
\label{per7}
\end{equation}
However at this point only 29$\%$ of the space is covered by the overlapping spheres and 71$\%$ remain empty, very different from that in 2-dimensional percolation. Here, both spheres and empty space form infinite connected networks or clusters. The density given by Eq. (\ref{per7})
gives  the normal nuclear matter density. The more accurate critical percolation density for the onset of the deconfinement transition is given by Eq. (\ref{per6}), leading to a density 4 times larger than nuclear matter. The existence of two percolation thresholds, one for the formation of a spanning cluster of spheres and another for the disappearance of a spanning vacuum cluster is a general feature of percolation in three or more dimensions.

 Assuming that for the density of 0.6-0.8 $fm^{-3}$ the cluster is formed by an ideal gas of all hadrons and resonances, the temperature of such a gas at this density is 170-190 MeV \cite{par61} which is not far from the critical temperature obtained in lattice QCD of $\sim$ 155 MeV.
The critical density Eq. (\ref{per7}) implies that the average distance between quark and antiquark at the deconfined point is $d_{q}\simeq 1/\rho_{c}^{1/3}\simeq $ 1.2 fm. 

At high $\mu$ and $T$ = 0 one should consider  percolation of nucleons having an impenetrable spherical hard core, of radius R around $R_{h}/2$. Each sphere defines a volume 
$V = \frac {4/3}{\pi R_{h}^{3}}$, which is not accessible to the center of any other sphere.
The spheres can only overlap partially and the distance between their centers must be larger than $ R =2R$. Then we have again two percolation thresholds. Numerical studies show that in the case of the spheres forming a spanning cluster there is no variation in the value of the critical density, 
however for the case of the vacuum percolation threshold, now we have \cite{par61}
\begin{equation}
\rho = \frac {2}{(4/3)\pi R_{h}^{3}}\simeq 0.93 fm^{-3}.
\label{per8}
\end{equation}
The disappearance of the vacuum cluster  for hard spheres requires a higher density than needed  for permeable  spheres.
At this point, it is interesting to explore the percolation of constituents of mass $M_{q} \sim$ 300 MeV and radius $R_{q}\sim$ 0.3 fm. In this case the critical density will be
\begin{equation}
\rho = \frac {1.2}{(4/3)\pi R_{q}^{3}}\simeq 3.5 fm^{-3}.
\label{per9}
\end{equation}
This value is 4 times larger than the critical value for nucleon percolation (0.93 $fm^{-3}$) and 22 times the normal nuclear matter density (0.16 $fm^{-3}$). According to this, massive deconfined quarks exist between the hadronic matter and the deconfined quarks and gluons state. In this intermediate state the quarks are deconfined, but the gluons are bound into the constituents quarks. In the lower baryon density limit, the quarks bind into nucleons. In the higher baryon density limit a connected medium of deconfined quarks and gluons   is obtained. Also at fixed baryon density and increasing 
$ T$ the deconfined quarks and gluons and the restoration of the chiral symmetry is found.

\begin{figure}[thbp]
\centering        
\vspace*{-0.2cm}
\includegraphics[width=0.90\textwidth,height=3.0in]{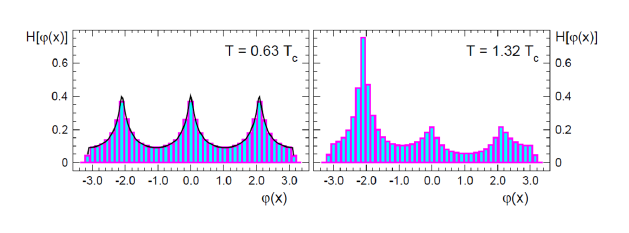}
\vspace*{-1.0cm}
\caption{Histograms for the distributions of the phase $\phi (x)$ of the local loops  $\langle L(\vec{x}) \rangle$. The left plot shows the distribution below $T_{c}$ while distribution in deconfined phase is shown in the right plot \cite{par62}.}
\label{poly}
\end{figure}

Let us  mention  that  in the  SU(3) lattice gauge theory, spatial clusters can be identified  as those where the local Polyakov loops
 $\langle L (\vec{x}) \rangle$ have values close to some element of the center. The elements of the center group $Z_{3}$, are a set of three phases $[0,2\pi/3, -2\pi/3]$ \cite{par62}.  
Below  $T_{c}$, ($\langle L (\vec{x}) \rangle = 0 $) the values of $\langle L(\vec{x}) \rangle$ grouped around these three phases, show three pronounced peaks located at the center phases.
 Above $T_{c}$, $\langle L(\vec{x}) \rangle \neq 0$ the distribution changes. One of the peaks grows and the other two shrink. A spontaneous symmetry breaking occurs, which leads to a non vanishing $\langle L(\vec{x}) \rangle$. Spatial clusters can be defined grouping the sites with  a very similar value of $\langle L(\vec{x}) \rangle$. The weight of the largest cluster increases sharply at $ T=T_{c}$ as seen in Fig.~\ref{poly}. Also the diameter of the largest cluster, that remains constant below $ T_{c}$, starts to rise quickly at $ T=T_{c}$ being 40 times larger at $ T= 1.2T_{c}$ indicating that the cluster percolates. These results do not change using ensembles with different lattice spacing. Therefore in the pure gauge SU(3) theory the deconfinement transition is a percolation phase transition (of the second order).

In the case of full QCD \cite{par63}, the fermions break the center symmetry explicitly and act as an external magnetic field in the Ising model, which favors the phase 0 of the Polyakov loop, although the other two phases ($\pm 2\pi /3)$ 
remain populated. As in the pure gauge case at $ T=T_{c}$ there  is a pronounced increase of the dominant phase. The explicit breaking of the center symmetry leads to a crossover type of transition.
This suggests that the chiral symmetry restoration phase transition also can be related to a generalized percolation phase transition \cite{par64}.
These properties of the Polyakov loop, giving rise to domain like structure or clusters of deconfined matter could explain its large opacity as well as its near ideal fluid properties, being the origin of the elliptic flow \cite{par65}.
In  experimental collisions at high energy, we expect that color strings are formed between the projectile and target partons. These color field configurations must have a small transverse size due to confinement. In this way, the strings are seen as small discs in the total available surface of the collisions. As the number of strings grows with energy and centrality degree of the collision, these strings would overlap forming clusters which eventually percolate. This 2-dimensional percolation and its phenomenological consequences in relation to SPS, RHIC and LHC pp, pA and AA data is the main subject of this review. In this case the critical  percolation density is given by Eq. (\ref{per1}) or Eq. (\ref{per3}) in case of realistic profiles.
\subsection{String models}
The phenomenology of string percolation takes its main ingredient, the strings, from models, even though most of the predictions are not dependent on the details of the models. Majority of the models roughly coincide in basic postulates such as the number of strings and its dependence on  energy and centrality, which is taken from the Glauber-Gribov model.
Strings models can be divided in models with color exchange between projectile and target as the Dual Parton Model (DPM) \cite{ref1x,ref1,ref4}, Quark Gluon  String Model (QGSM) \cite{ref2}, VENUS \cite{ref4}, EPOS \cite{par70}, DPMJET \cite{par71} and models without color exchange where the interaction excites the projectile and target producing  strings between the partons of both. Examples of this kind of models are HIJING \cite{par72}, PYTHIA \cite{par73}, AMPT \cite{par74}, HSP \cite{par76}, and URQMD \cite{par76}. We will concentrate in models with color exchange, essentially the DPM, which are based on the $1/N_{c}$ QCD expansion and  are in correspondence via unitarity with the Gribov Reggeon calculus. The DPM or QGSM have been extensively compared to the experimental data of ISR, SPS and FermiLab obtaining an overall agreement \cite{ref1x}. 

\begin{figure}[thbp]
\centering        
\vspace*{-0.2cm}
\includegraphics[width=0.65\textwidth,height=2.0in]{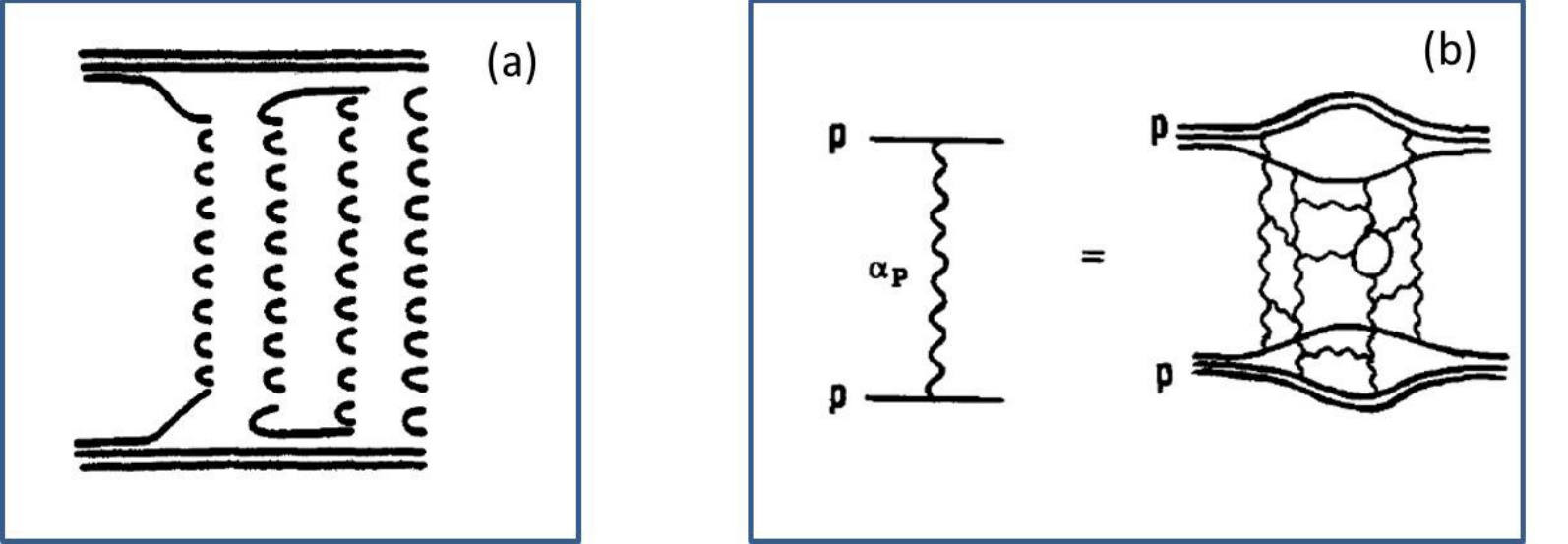}
\caption{(a):Two cut Pomeron diagram (four chain) for proton-proton collisions.
(b):Single Pomeron exchange and its underlying cylindrical topology. This is a dominant contribution to proton-proton elastic scattering at high energies \cite{ref1x}.}
\label{pom1}
\end{figure}

In DPM or QGSM, the multiplicity distribution dN/dy of pp collisions is described  by the formation and fragmentation of ${\it 2k}$ strings \cite{ref1x} as shown in Fig.~\ref{pom1}(a).
\begin{equation}
\frac{dN^{pp}}{dy}=\frac{1}{\sigma}\sum{\sigma_{k}[N_{k}^{qq-q}(s,y)+N_{k}^{q-qq}(s,y)+(2k-2)N_{k}^{q-\bar{q}}(s,y)]},
\label{stri1}
\end{equation}
where $N_{k}^{q-qq}$ and $N_{k}^{q-qq}$ are the inclusive spectra of hadrons produced in the strings stretched between a valence diquark of the projectile (target) and a quark of the target (projectile) and $N_{k}^{q-\bar{q}}$ are the inclusive spectra of the strings stretched between sea quarks and antiquarks.
The single particle distribution of each string can be obtained by folding the momentum distribution  of the partons at the end of the string with the fragmentation function  of the string
\begin{equation}
N_{1}^{qq-q}(s,y)= \int_{0}^{1}\int_{0}^{1} dx_{1}dx_{2}\rho_{k}(x_{1})\rho_{k}(x_{2})\frac{dN^{qq-q}}{dy}(y-\bar{\Delta},s_{s}),
\label{stri2}
\end{equation}
Here $\sqrt {s_{s}}$ is the invariant mass of the string, $s_{s} = s x_{1}x_{2}$, where $x_1$ and $x_2$ are  the light-cone momentum fractions of the constituents at the ends of the string.  $\bar\Delta$  is the rapidity shift necessary to go from the overall pp center of mass (CM) frame to the CM of one string,
\begin{equation}
\bar \Delta = \frac{1}{2}\log(\frac{x_{1}}{x_{2}})
\label{stri3}
\end{equation}
The momentum distributions used for the valence quarks, sea quarks or antiquarks and valence diquarks are $x^{-1/2}$, $x^{-1}$, and $x^{3/2}$ respectively. In general, the distribution  of  2$k$ partons in the proton is
\begin{equation}
\rho_{k}(x_{1},x_{2k};x_{2},x_{3},....x_{2k-1})=C_{k}^{\rho}x_{1}^{-1/2}x_{2}^{-1}...x_{2k-1}^{-1}x_{2k}^{3/2}\delta(1-\sum_{1}^{2k}x_{i}),
\label{stri4}
\end{equation}
where $C_{k}^{\rho}$ can be found by normalizing $\rho_{k}$ to unity.

With these momentum distributions  the	$N^{qq-q}$ and $N^{q-qq}$ strings are long  (due to $x^{3/2}$ and $x^{-1/2}$ behavior of their extremes) centered at a rapidity point shifted with respect to the CM. The $N^{q-q}$ strings are short,  centered at the CM (due to the $x^{-1}$ of their constituents ends). Concerning the fragmentation functions different methods have been used even within the same model. In string percolation, mostly the Schwinger mechanism or the Lund fragmentation is used. In  Eq. (\ref{stri1}) $\sigma_{k}$ is the cross section for producing 2$k$ strings resulting from cutting $k$ Pomerons. 

 As the Pomeron has the topology of a cylinder its
cutting gives rise to two strings (Fig.~\ref{pom1}(b)).
Using the AGK cutting rules \cite{par77}, the dependence of $\sigma_{k}$ with the energy is given by
\begin{equation}
\sigma_{k}=\frac{8\pi g \exp(\Delta y)}{kz}[1-e^{-z}\sum_{l=0}^{k-1}\frac{z^{l}}{l!}],\ \     k > 0,
\label{stri5}
\end{equation}
where 
\[
z=\frac {2g C \exp(\Delta y)}{R^{2}+ \alpha^{'} y}
\]
and $g$ is the coupling of the Pomeron to the proton, 1+$\Delta$ the intercept of the trajectory of the Pomeron, $\alpha^{'}$ its slope and $\mathcal C $ a parameter describing the inelastic diffractive states ($\mathcal C= $1  means only elastic scattering without diffractive states). The total cross section is obtained summing over $k$
\begin{equation}
\sigma_{tot}= \exp(\Delta y) \sum_{k=0}^{\infty}\sum_{l=k, l > 0}^{\infty}(\frac{-z}{2})^{l-1}\frac{8\pi g}{l!}[\delta_{k0}+(-1)^{1-k}2^{l-1}(^{l}_{k})].
\label{stri6}
\end{equation}

The rise of dN/dy is mainly due to  short strings, whose number grows with energy. On the other hand, outside the region of central rapidity there is no contribution of these short strings and the rise with energy is much slower. 
Assuming a Poisson distribution for cutting $k$ Pomerons
\begin{equation}
P_{k}(n) = \frac {(kN)^n}{n!}\exp(-kN),
\label{stri7}
\end{equation}
where $N$ is the mean multiplicity produced when cutting one Pomeron, the multiplicity distribution is
\begin{equation}
P_{k}(n) = \sum_{k} w_{k} P_{k}(n),\  \ w_{k}= \frac{\sigma_{k}}{\sigma}.
\label{stri8}
\end{equation}
Very often $<n>P(n)$ is plotted as a function of  $n/<n>$. When the result is independent of energy one has the well-known Koba-Nielsen-Olesen (KNO) scaling. KNO scaling is roughly obeyed up to the highest ISR energy ($\sqrt{s}$ = 63 GeV) but  it is clearly violated at SPS, Fermilab, RHIC and LHC. The origin of KNO violation can be understood in DPM easily. The contributions of multistrings diagrams become increasingly important when ${\it s}$ increases, and since they contribute mostly to high multiplicities they push upwards the high multiplicity tail. The increase with $ s$ of the multistrings contributions is due both to the increase  of the invariant mass of short strings and to the ${\it s}$-dependence of the weights. Hence, we expect a larger KNO violation at central rapidity region where the short strings contribute than in the whole rapidity range or close to the
ends of the phase space where the short strings do not contribute.

The width of the KNO multiplicity distributions is related to the fluctuations of the
number of strings, which also control the forward-backward correlations. These correlations can be described by the approximate linear expression
\begin{equation}
<n_{b}> = a + b n_{f},
\label{stri9}
\end{equation}
where $n_{f(b)}$ is the number of particles observed in the forward (backward) rapidity window  and $a$ and $b$ are given by
\[
a=\frac{<n_{b}><n_{f}>^{2}-<n_{f}n_{b}><n_{f}>}{<n_{f}^{2}>-<n_{f}>^{2}},\]\begin{equation} b=\frac{<n_{f}n_{b}>-<n_{f}><n_{b}>}{<n_{f}^{2}>-<n_{f}>^{2}}.
\label{stri10}
\end{equation}
Usually the forward and backward rapidity intervals are taken separated by a central rapidity window $|y| < y_{c}$ in such a way that the short range correlations are eliminated 
 ($y_c >$ 0.5). Consider symmetric forward and backward intervals, $\delta y_{F}=\delta y_{B} = \delta y$. In any multiple scattering model the
origin of long range correlations is the fluctuations in the number of multiple scatterings \cite{par78,par79,par80}.
Let $N$ strings decay each  into $\mu$ particles on the average. The slope $b$  can be split into short range correlations (SR) and long range correlations (LR) \cite{par81}
\begin{equation}
b= b_{SR}+b_{LR} = \frac {\beta}{1+C}+\frac{C}{1+C},
\label{stri11}
\end{equation}
where $\beta$ is the ratio between the forward-backward variance and the forward-forward variance of the distribution of particles produced in a single string:
\begin{equation}
\beta = \frac{D_{FB}}{D_{FF}} = \frac{<\mu_{f}\mu_{b}>-<\mu_{f}><\mu_{b}>}{<\mu_{f}^{2}>-<\mu_{f}>^{2}}
\label{stri12}
\end{equation}
and
\begin{equation}
C= \frac {D_{N}\mu_{F}^{2}}{<N>D_{FF}},\ \  D_{N}= <N^{2}>-<N>^{2}.
\label{stri13}
\end{equation}
For a large rapidity window between the forward and backward intervals, there are no long range correlations in a single string, $D_{FB}$ = 0 and we have
\begin{equation}
b = \frac{1}{1+\frac{<N>D_{FF}}{D_{N}\mu_{F}^{2}}}.
\label{stri14}
\end{equation}
Usually it is assumed that the multiplicity distribution of a single string is Poissonian, $D_{FF}= \mu_{F}$ and $b$ becomes
\begin{equation}
b = \frac{1}{1+\frac{<N>}{D_{N}\mu_{F}}}.
\label{stri15}
\end{equation}
According to Eq. (\ref{stri15}), at low energy there are no fluctuations in the number of strings, $D_{N}$ = 0 and ${\it b}$ = 0. As the energy increases, $D_{N}$ increases as well as $b$. On the other hand, if we fix the multiplicity we eliminate many of the string fluctuations and therefore $b$ is smaller.

The generalization of DPM to nucleus-nucleus collisions is obtained as follows \cite{par82}. Consider a collision of a nucleus A with nucleus B in a configuration with $N_{A}$ participating nucleons of A, $N_{B}$ participating nucleons of B  (assume that $N_A\leq N_B$) and a total number $N_c$ of inelastic collisions. In this configuration hadrons are produced in 2$N_{c}$ strings (2 strings for each inelastic collision). Of these 2$N_{A}$ are stretched between valence quarks and valence diquarks ($q_{v}^{A}-qq_{v}^{B}$ and $qq_{v}^{A}-q_{v}^{B} )$. The remaining $N_{B}-N_{A}$ valence quarks and diquarks of B have no valence partner of A and have to form $2N_{B}-2N_{A}$ strings with sea quarks and antiquarks of A ($q_{s}^{A}-qq_{v}^{B}$ and $\bar {q}_{s}^{A}-q_{v}^{B} $).  The remaining $2N_{c}-2N_{B}$ strings are formed between sea quarks and antiquarks of A and B ($q_{s}-\bar q_{s}$)
%\begin{eqnarray*}
\[
\frac{dN^{AB}}{dy}=\frac{1}{\sigma_{AB}}\sum_{N_{A},N_{B},N_{c}} [\sigma_{N_{A},N_{B},N_{c}}^{AB}\theta(N_{B}-N_{A})N_{A}(N^{qq_{v}^{A}-q_{v}^{B}}(y)+N^{q_{v}^{A}-qq_{v}^{B}}(y)) + \]
\[ 
 (N_{B}-N_{A})(N^{\bar q_{s}^{A}-q_{s}^{B}}(y)+N^{q_{s}^{A}-qq_{v}^{B}}(y)) + \]
\begin{equation}
 (N_{c}-N_{B})(N^{q_{s}^{A}-\bar q_{s}^{B}}(y)+ N^{\bar {q_{s}}^{A}-q_{s}^{B}}(y))+ sym(N_{A}\leftrightarrow N_{B})],
\label{stri16}
\end{equation}
%\end{eqnarray}
where $\sigma_{N_{A},N_{B},N_{C}}^{AB}$ is the cross section for $N_{C}$ inelastic nucleon-nucleon collisions involving $N_{A}$ nucleons of A and $N_{B}$ nucleons of B. This cross section have been studied extensively \cite{par83,par84,par85} together with its different approximations needed for its evaluation. The inclusive spectra, as in the pp case, are given by a convolution of momentum distribution and fragmentation functions.
In the case of A=B we have approximately
%\begin{eqnarray*}
\[
\frac {dN^{AA}}{dy}\approx <N_{A}>(2N^{qq-q_{v}}(y)+(2<k>-2)N^{q_{s}-\bar q_{s}}(y)) +\]
\begin{equation}
(<N_{c}>-<N_{A}>)2<k>N^{q_{s}-\bar q_{s}}(y),
\label{stri17}
\end{equation}
%\end{eqnarray}
where we have introduced the possibility of having $k$ multiple scattering in the individual nucleon-nucleon interactions, which was neglected in the Eq. (\ref{stri16}). Notice that in the term proportional to the number of collisions it is not specified if the collision is soft or hard. In fact there are many inelastic soft collisions included in this term. Sometimes it is wrongly assumed  that the term proportional to the number of collisions contains only hard collisions. We observe that in the central rapidity region we have $ 2Nk$ strings which for high energy and heavy nuclei is a very large number (more than 1500). Due to that we expect interaction among them, and therefore they will not fragment in an independent way. We study later such interactions. 

In hadron-nucleus interactions  in DPM, QGSM or VENUS the  multiplicity distribution $P(n)$ can be approximated by a negative binomial distribution.

%
% %%%%%%%%%%%%%%%%%%%%%%%%%%%%%%%%%%%%%%%%%%%%% 
%
%  Braun's contribution
%
%%%%%%%%%%%%%%%%%%%%%%%%%%%%%%%%%%%%%%%%%%%%%%%%%%%%%
\section{Color strings with fusion and percolation}
\subsection{Introduction}
As indicated in Section 1.5, multiparticle production
at high energies can be  described in terms of
color strings stretched between the projectile and target
~\cite{ref1,ref4,ref2,ref3,ref5,ref6}.
Hadronization of these strings produces the observed hadrons.
The basic characteristic feature of the color string model with
fusion and percolation, which is the subject of this review, is that
strings are provided with a finite area in the transverse space.
In terms of gluon color field they can be considered as the color flux tubes
stretched between the colliding partons,  which in the
transverse space are restricted to a finite disc of a given radius, dictated
by the confinement. The mechanism of particle creation is then similar to the
one in the well-known Schwinger mechanism of pair creation in a constant
electric field covering all the space, except that now the space is finite
in the transverse plane. Note that pair creation actually splits the space
filled with the chromoelectric field into two parts, each of them attached
to one of the initial and one of the created partons. In this way the dynamics
of the string evolution consists of successive breaking into more strings.

So creation of particles
goes via emission of $q\bar q$ pairs in the field of the string.
From the start it is relevant to mention one of the important properties of
this mechanism. The transverse dimension of the string is a characteristic
independent of the form of the distribution of created partons in the
transverse momentum, in contrast to what one may think considering the
string itself as a distribution of partons (say gluons). In the latter case
the average transverse momentum of emitted particles is obviously inverse to
the transverse dimension of the source. However it is apparently not so for
the Schwinger mechanism. With the field constant in the transverse plane
the spectrum of constituents (photons in the QED case) is just the
$\delta$-function. However the emitted electrons have non-zero transverse
momenta whose average is determined by the strength of the field
(although the total transverse momentum of the pair is indeed zero).
Likewise in our color string picture emitted partons have average transverse
momenta determined by the strength of the chromoelectric field
and do not depend on the transverse dimension of the string.

At low energies for collision of hadrons and nuclei with relatively small
atomic numbers the fact that strings have finite dimension has no influence
on the results. In the transverse plane strings are projected as discs
at large distances from one another and particle creation does not feel
their interaction. However with growing energy and/or
atomic number of colliding particles, the number of strings grows.
Once strings have a certain nonzero dimension in the transverse space they
start to overlap  forming clusters, very much like discs in the
2-dimensional percolation theory. The geometrical behavior of strings in
the transverse plane then follows that of percolating discs. In particular
at a certain critical string density a macroscopic cluster appears
(infinite in the thermodynamic limit), which marks the percolation phase
transition ~\cite{ref7,par43,ref8}.

The percolation theory governs the geometrical pattern of the string
clustering.
Its observable implications however  require introduction of some
dynamics to describe string interaction,
that is, the behavior of a cluster formed by several overlapping strings.

One can study different possibilities.

A most naive attitude is to assume that nothing happens as strings overlap,
in other words,
they continue to emit particles independently without being affected by their
overlapping neighbors. This is a scenario of non-interacting strings,
which closely
corresponds to original calculations in the color strings approach, oriented
at comparatively small energies (and numbers of strings).
This scenario  however contradicts the idea that strings are areas of the
transverse
space filled with color field and thus with energy, since in the overlapping
areas the energy should have grown.

In another limiting case one may assume that a cluster of several overlapping
strings
behaves as a single string with an appropriately higher color field (a string
of higher color, or a ``color rope'' ~\cite{ref10}). This fusion scenario was proposed
by the authors
and later realized as a Monte-Carlo algorithm nearly  decades ago
~\cite{ref11,ref12,ref12x}. It predicts
lowering of total multiplicities and forward-backward correlations (FBC)
and also strange baryon enhancement, in a reasonable agreement with the  known
experimental trends.

\begin{figure}[thbp]
%\hspace{1.5cm}
%\epsfig{file=prmf1.pdf,width=15 cm}
\centering        
\vspace*{-0.2cm}
\includegraphics[width=0.60\textwidth,height=2.0in]{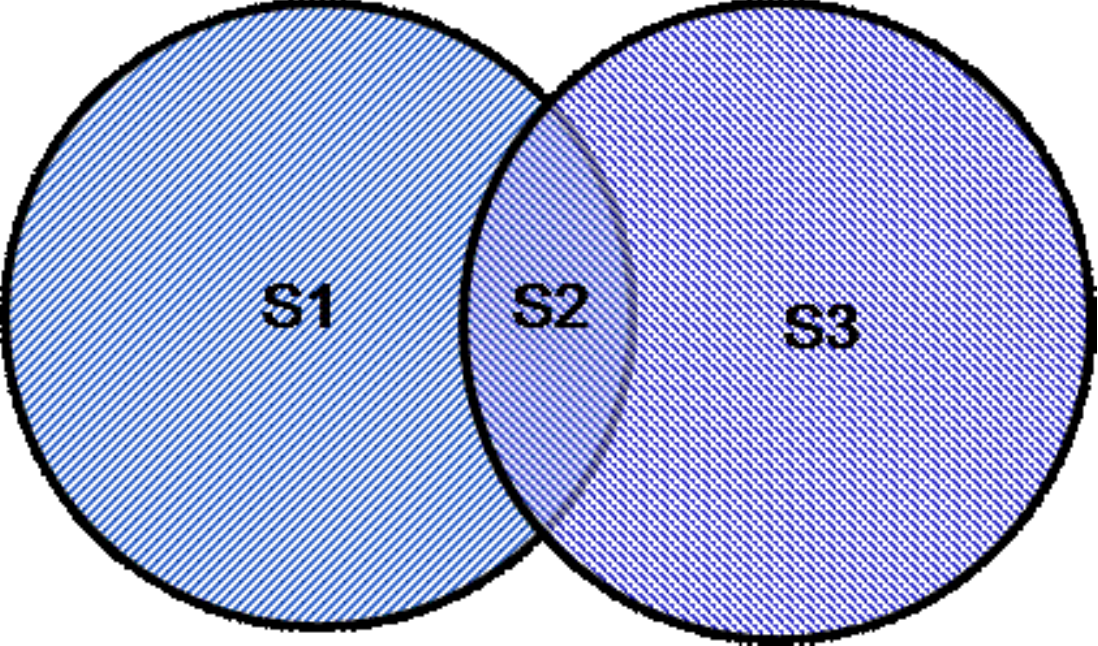}
\vspace*{-0.3cm}
\caption{ Projections of two overlapping strings onto the transverse plane \cite{brapaj2000}.}
\label{fig1}
\end{figure}

However both discussed scenarios are obviously of a limiting sort. In a typical
situation
strings only partially overlap and there is no reason to expect them to fuse
into a single
stringy object, especially if the overlap is small. The
transverse space occupied by a cluster of overlapping strings splits into
a number of areas
in which different number of strings overlap, including areas where no
overlapping takes place.
In each such area color fields coming from the overlapping strings will add
together.
As a result the total cluster area is split in domains with different color
field strength.
As a first approximation, neglecting the interaction at the domain frontiers,
one may assume
that emission of $q\bar q$ pairs in the domains proceeds independently,
governed by the field strength (``the string tension'') in a given domain.
This picture implies that  clustering  of strings actually leads to
their proliferation, rather than fusion,
since each particular overlap may be considered as a separate string.
Evidently newly formed strings differ not only in their
colors but also in their transverse areas.
As a simple example consider a cluster of two partially overlapping
strings (Fig. \ref{fig1}) \cite{brapaj2000}.

One distinguishes three different regions: regions 1 and 3 where no
overlapping takes place and the color field remains the same as in a single string, and the overlap region 2 with color fields of both strings summed. In our picture particle production will proceed independently from these three areas, that is, from three different
``strings'' corresponding to areas 1, 2 and 3. In this sense string interaction
has split two
strings into three of different color, area and form in the
transverse space.

We stress that these dynamical assumptions are  rather independent of the
geometrical
picture of clusterization. In particular, in each of the scenarios discussed
above, at a
certain string density there occurs the percolation phase
transition. However its experimental signatures crucially depend on the
dynamical contents of
string interaction. With no interaction, clustering does not change physical
observables,
so that the geometric percolation will not be felt at all. With the
interaction between strings
turned on, clustering (and percolation) lead to well observable implications.

In this chapter we  shall review these implications for the most
immediate
and important observables, multiplicities and transverse momenta spectra
of produced particles.

\subsection{Multiplicity and transverse momentum for overlapping strings}
As stated in the Introduction, the central dynamical problem is to find
how the observables change when several strings form a cluster partially
overlapping. Let us consider a ``simple'' string
stretched between a quark and antiquark with a transverse area $S_{1}$.
It emits partons with the transverse momentum distribution
\begin{equation}
I_0(y,p)\equiv \frac{4\pi d\sigma}{dy d^2p}
=Ce^ {-\frac{m_\perp^2(p)}{t_1}}
\label{eq1a}
\end{equation}
where $t_1$ is the tension of the simple string,
$m_\perp^2=m^2+p^2$ and $m$ and $p$ are the mass and
transverse momentum of the emitted parton. In the following we mostly
consider emitted pions when we take $m=0$.
In accordance with the
Schwinger picture of particle emission we assume that tension $t_1$
is proportional to the field responsible for emission
and thus to the color charge at the string ends ~\cite{ref10,ref13}.
For the simple string stretched between the quark and antiquark it is
proportional to the quark color charge squared $Q_0^{2}$.
According to Eq. (\ref{eq1a}) the average transverse momentum squared
is given by tension $t_1$: $\avpt_{1}=t_{1}$ and so is proportional to $Q_{0}$.
We denote the multiplicity of produced particles per unit rapidity
as $\mu_{1}$. It is also proportional to the color charge ~\cite{ref10,ref13}.

Now consider two simple
strings, of areas $S_1$ in the transverse space each,
partially overlap in the area $S^{(2)}$ (region 2
in Fig. \ref{fig1}), so that
$S^{(1)}=S_1-S^{(2)}$ is the area in each string not
overlapping with the other.
A natural assumption seems to be that
the average color density $q$ of
the simple string
in the transverse plane is a constant $q=Q_0/S_1$.
For partially overlapping strings
the color in
each of the two non-overlapping areas will then be
\begin{equation}
 Q_1=q S^{(1)}=Q_0(S^{(1)}/S_1).
\end{equation}
In the overlap area each string will have color
\begin{equation}
 \bar{Q}_2=q S^{(2)}=Q_0(S^{(2)}/S_1).
\end{equation}
The total color in the overlap area $Q_2$ will be a vector sum of the two
overlapping colors
$q S_2$. In this summation the total color squared should be conserved
~\cite{ref10}.
Thus $Q_2^2=({\bf Q}_{ov}+{\bf Q}'_{ov})^2$ where ${\bf Q}_{ov}$ and
${\bf Q}'_{ov}$ are the
two vector colors in the overlap area. Since the colors in the two strings
may generally
be oriented in an arbitrary manner respective to one another, the average  of
${\bf Q}_{ov}{\bf Q}'_{ov}$ is zero. Then $Q_2^2=Q^2_{ov}+{Q'_{ov}}^2$, which
leads to
\beq
Q_2=\sqrt{2}\,q S^{(2)}=\sqrt{2}\,Q_0(S^{(2)}/S_1).
\eeq
One observes that, due to its vector nature, the color in the overlap
is less than the
sum of the two overlapping colors. This phenomenon was first
mentioned in ~\cite{ref10} for the so-called color ropes.

As mentioned, the simplest observables, the multiplicity $\mu$ and the average
transverse momentum squared $\avpt$, are directly related to the field strength in the string and thus to its generating color. They are both proportional to the color \cite{ref10,ref13}.
Thus, assuming independent emission from the three regions 1, 2, and 3 in Fig. \ref{fig1} we get for the multiplicity
\begin{equation}
\mu/\mu_1=2(S^{(1)}/S_1)+\sqrt{2}\,(S^{(2)}/S_1),
\end{equation}
where $\mu_1$ is a multiplicity for a single string.
To find $\avpt$ one has to divide the total transverse momentum squared of
all observed particles by the total multiplicity. In this way for our cluster 
 of two strings we obtain

\[
\avpt/\avpt_1= \frac{2(S^{(1)}/S_1)+\sqrt{2}\sqrt{2}\,(S^{(2)}/S_1)}
{2(S^{(1)}/S_1)+ \sqrt{2}\,(S^{(2)}/S_1)}\] \beq =\frac{2}{2(S^{(1)}/S_1)+\sqrt{2}\
(S^{(2) }/S_1)},
\eeq
where $\avpt_1$ is the average transverse momentum squared for  a simple
string and we have
used the evident property $ 2 S^{(1)}+2S^{(2)} = 2 S_{1} $ in the second equality.

Generalizing to any number $N$ of overlapping strings we find the total
multiplicity as
\beq
\mu/\mu_1=\sum_i\sqrt{n_i}\,(S^{(i)}/S_1),
\label{eq6}
\eeq
where the sum goes over all individual overlaps $i$ of $n_i$ strings having
areas $S^{(i)}$.
Similarly for the $\avpt$ we obtain
\begin{equation}
\avpt/\avpt_1=\frac{\sum_i n_i\,(S^{(i)}/S_1)}{\sum_i\sqrt{n_i}\,
(S^{(i)}/S_1)}=
\frac{N}{\sum_i\sqrt{n_i}\,(S^{(i)}/S_1)}.
\label{eq7}
\end{equation}
In the second equality we again used an evident identity
$\sum_i n_i\,S^{(i)}=NS_1$.
Note that Eqs. (\ref{eq6}) and (\ref{eq7}) imply a simple relation between the multiplicity and transverse momentum
\begin{equation}
\frac{\mu}{\mu_1}\frac{\avpt}{\avpt_1}=N,
\end{equation}
which evidently has a meaning of conservation of the total transverse momentum
produced.

Equations (\ref{eq6}) and (\ref{eq7}) do not look  easy to apply. To calculate the
sums over $i$ one
seems to have to identify all individual overlaps of any number of strings
with their
areas. For a large number of strings the latter may have very complicated
forms and
their analysis presents great calculational difficulties. However one
immediately recognizes that such individual tracking of overlaps is not at all necessary. One can combine  all terms with a given number of overlapping strings $n_i=n$ into a single term, which sums all such overlaps into a total area of exactly $n$ overlapping strings $S^{tot}_n$. Then one finds instead of Eq. (\ref{eq6})
and (\ref{eq7})
\begin{equation}
\mu/\mu_1=\sum_{n=1}^N\sqrt{n}\,(S^{tot}_n/S_1)
\label{eq9}
\end{equation}
and
\begin{equation}
\avpt/\avpt_1=
\frac{N}{\sum_{n=1}^N\sqrt{n}\,(S_n^{tot}/S_1)}.
\label{eq10}
\end{equation}

In contrast to individual overlap areas $S^{(i)}$ the total ones $S_n^{tot}$ can be easily calculated (see next subsection). Let the projections of the strings onto the transverse space be distributed uniformly in the total interaction
area $S$ with a density $\rho$. Introduce a dimensionless parameter
(``percolation parameter'')
\beq \xi=\rho S_1=N S_1/S.
\label{eq11}
\eeq
%The ``thermodynamic limit'' corresponds to taking the number of the strings
%$N\rightarrow\infty $ keeping $\xi$ fixed. In this limit one readily finds
%that the distribution of the overlaps of $n$ strings is Poissonian with a
%mean value
$\xi$:
% Equation is same as \per2
%\beq
%{\cal P}_n=\frac{S_n^{tot}}{S}=\frac{\xi^n}{n!}e^{-\xi}.
%\label{eq12}
%\eeq
From Eq. (\ref{eq9}) we then find that the multiplicity is damped due to overlapping by a factor
\begin{equation}
F(\xi)=\frac{\mu}{N\mu_1}=\frac{\blangle\sqrt{n}\brangle}{\xi},
\label{eq13}
\end{equation}
where the average is taken over the Poissonian distribution Eq. (\ref{per2}).

\begin{figure}[thbp]
\centering        
\vspace*{-0.2cm}
\includegraphics[width=0.85\textwidth,height=3.0in]{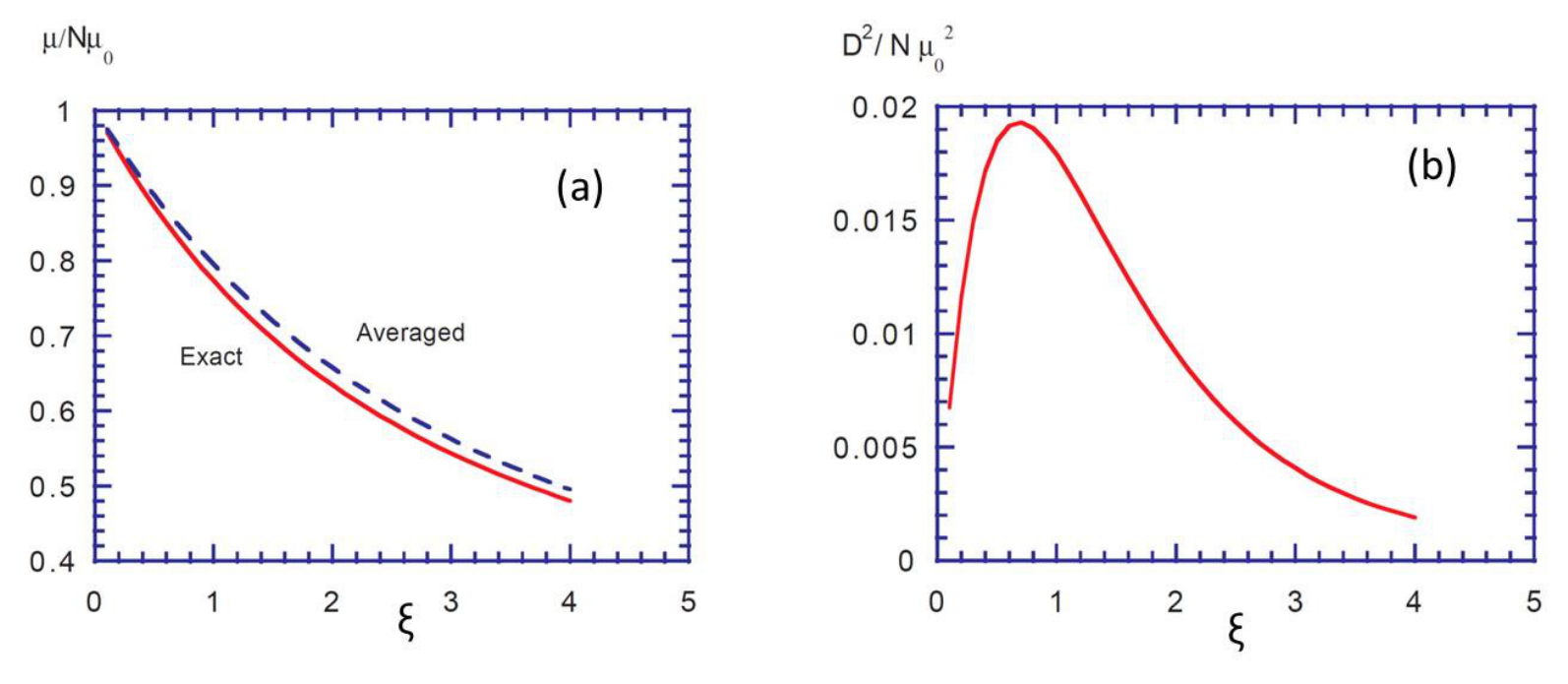}
\caption{(a). Damping of the multiplicity as a function of $\xi$.
 (b): Percolation dispersion squared of the multiplicity per string in units $\mu_1^2$ as a function of $\xi$ \cite{brapaj2000}.}
\label{fig2}
\end{figure}

The behavior of $F(\xi)$ is shown in Fig. \ref{fig2}(a). It smoothly goes down from unity at $\xi=0$ to values around 0.5 at $\xi=4$ falling as $1/\sqrt{\xi}$ for larger $\xi$'s. According to Eq. (\ref{eq10}) the inverse of $F$ shows the
rise of the $\avpt$.

Note that a crude estimate of $F(\xi)$ can be done from the overall
compression of the string area due to overlapping. The fraction of the total 
area occupied by the strings according to Eq. (\ref{per2}) (see also ~\cite{ref14}) is given by
\beq
\sum_{n=1}{\cal P}_n=1-e^{-\xi}.
\label{eq14}
\eeq
The compression is given by Eq. (\ref{eq14}) divided by $\xi$. According to our picture the multiplicity is damped by the square root of the compression factor, so that the damping factor is
\begin{equation}
F(\xi)=\sqrt{\frac{1-e^{-\xi}}{\xi}}.
\label{eq15}
\end{equation}
For all the seeming crudeness of this estimate, Eq. (\ref{eq15}) is very close to 
the exact result, as shown in Fig. \ref{fig2}(a) by a dashed curve.

\subsection{Multiplicities and their dispersion}
In our picture in the transverse space simple strings are represented
by discs of radius $r_{0}$
and area $S_{1}=\pi r_{0}^{2}$  homogeneously distributed in the
total area $ S$. We normalize $ S$ assuming that centers of the discs
are inside the unit circle
of area $S_0=\pi$ so that $S=\pi (1+a)^2$, where $a$ is the normalized disc radius. The disc density is
$\rho=N/S$ and the percolation parameter is $\xi=\rho S_{1}= N S_{1}/S$. In the thermodynamic limit $N\rightarrow\infty$, so that at fixed
$\xi$ the radius of the discs goes to zero.
For fixed $\xi$,  $a^2=\xi/(\sqrt {N}- \sqrt {\xi})^{2}$,
so that with our normalization at large $N$ $a\sim 1/\sqrt{N}$ and
$S_1\sim1/N$.
Since the discs are distributed homogeneously, the probability that their
centers are at points $r_i$, $i=1,...N$ inside the unit circle is
independent of $r_i$ and  is  given by
\beq
P(r_i)=1/S_0^{N}.
\label{eq47}
\eeq
With the disc centers at points
$r_i$.
the overlap area of exactly $n$ discs is given by the integral
\beq
S_n^{tot}(r_1,...r_N)=\int_Sd^2r\sum_{\{i_1,..i_n\}\subset\{i_1,...i_N\}}
\prod_{k=1}^{n}
\theta(a-|{\bf r}-{\bf r}_{i_k}|)
\prod_{k=n+1}^N\theta(|{\bf r}-{\bf r}_{i_k}|-a).
\label{eq48}
\eeq
The average of $S_n^{tot}$ will be given by a multiple integral over $r_i$ with the probability Eq. (\ref{eq47}):
\beq
\blangle S_n^{tot}\brangle=\frac{1}{S_0^N}\int_{S_0}\prod_{i=1}^N d^2r_iS_n^{tot}(r_1,...r_N)=
C_N^n\int_Sd^2rF^n(r)(1-F(r))^{N-n},
\label{eq49}
\eeq
where
\beq
F(r)=(1/S_0)\int_{S_0}d^2r_1\theta(a-|{\bf r}-{\bf r}_1|).
\label{eq50}
\eeq

The function $S_0
F(r)$ gives an area occupied by a circle $C$ of radius $a$ with a center
at
$r$ which is inside the unit circle $S_0$. If $r<1-a$ then $C$
is always inside $S_0$ so that
\beq
F(r)=\sigma_0/S_0,\ \ \ 0<r<1-a.
\label{eq51}
\eeq
However for $r>1-a$ a part of $C$ turns out to be outside the unit circle,
and
\beq
F(r)=\sigma(r)/S_0,\ \ \ 1-a<r<1+a,
\label{eq52}
\eeq
where $\sigma(r)\leq S_{1}$ is the overlap of the two discs $C$ and $S_0$.

Generally, the overlap of two circles of radii $r_1$ and $r_2$ with a
distance $r$ between their centers is given by
\beq
\sigma(r_1,r_2,r)=
(1/2)r_1^2(\alpha_1-\sin \alpha_1)+(1/2)r_2^2(\alpha_2-\sin \alpha_2),
\label{eq53}
\eeq
where
\[
\cos (\alpha_1/2)=\frac{1}{2r_1}\left(r+\frac{r_1^2-r_2^2}{r}\right),
\]
\beq
\cos (\alpha_2/2)=\frac{1}{2r_2}\left(r-\frac{r_1^2-r_2^2}{r}\right).
\label{eq54}
\eeq
The function $\sigma(r)$ in Eq. (\ref{eq52}) is just $\sigma(1,a,r)$.

Equations (\ref{eq49}) -(\ref{eq54}) allow to calculate numerically the average
$\blangle S_n^{tot}\brangle$ for any finite value of $N$ without much difficulty.

In the thermodynamic limit, $N\rightarrow\infty$ with $\xi $ being fixed,
 the calculation of $\langle S_n\rangle$ becomes trivial.
Indeed then one can neglect the part of integration in $r$ with $r>1-a$
altogether, with an error $\sim 1/a\sim1/\sqrt{N}$. With the same precision
one then finds
\beq
\langle S_n^{tot}\rangle=S C_N^n(\sigma_0/S)^n(1-\sigma_0/S)^{N-n},
\eeq
where we have put $S_0\simeq S$. 
The physically relevant values of $n$ remain finite as
$N\rightarrow\infty$. So we can approximately take
\beq
C_N^n=N^n/n!,\ \ \ (1-\sigma_0/S)^{N-n}=\exp(-N\sigma_0/S).
\eeq
We then find that in the thermodynamic limit the distribution of overlaps
in $n$ is Poissonian with the mean value given by $\xi$ (Eq. (\ref{per2})).

%%%%%%%%%%%%%%%%%%%%%%%%%%%%%%%%%%%%%%%5
Calculation of the multiplicity dispersion requires knowledge of the
average of its
square. With the centers of the discs at $r_1,...r_N$ it has the form
\beq
\mu^2(r_1,...r_N)=(1/\sigma_0^2)(\sum_n\sqrt{n}S_n^{tot}(r_1,...r_N))^2,
\eeq
where $S_n^{tot}(r_1,...r_N)$ is given by Eq. (\ref{eq48}). Taking the average over the
discs centers
positions we now come to a double integral in $r$ and $r'$
\[
\langle\mu^2\rangle=\frac{1}{\sigma_0^2}\sum_{m,n}\sqrt{mn}
\int_S d^2rd^2r'\frac{1}{S_0^N}\int_{S_0}\prod_{i=1}^N d^2r_i\]\[
\sum_{\{i_1,..i_n\}\subset\{1,...N\}}\prod_{k=1}^{n}
\theta(a-|{\bf r}-{\bf r}_{i_k}|)
\prod_{k=n+1}^N\theta(|{\bf r}-{\bf r}_{i_k}|-a)\]\beq
\sum_{\{j_1,..j_m\}\subset\{1,...i\}}\prod_{l=1}^{m}
\theta(a-|{\bf r}-{\bf r}_{j_l}|)
\prod_{l=m+1}^N\theta(|{\bf r}-{\bf r}_{j_l}|-a).
\label{eq58}
\eeq

This complicated expression, however, continues to be factorized in all
$r_i$ and can be substantially simplified. Leaving the details to the
original derivation in ~\cite{brapaj2000} we present here the final
expression in the form of the sum
\[
\blangle\mu^2\brangle=\frac{1}{\sigma_0^2}\sum_{n,m,p}\sqrt{(n+p)(n+p)}\]\beq
C_N^{n,m,p}\int_Sd^2rd^2r'\phi^p(r,r')\chi^n(r,r')\chi^{m}(r',r)
\zeta^{N-n-m-p}(r,r').
\label{eq62}
\eeq
where
\beq
C_N^{n,m,p}=\frac{N!}{n!m!p!(N-n-m-p)!}
\label{eq59}
\eeq
\[
\chi(r,r')=F(r)-\phi(r,r'),
\]
\[
\zeta(r,r')=1-F(r)-F(r')+\phi(r,r'),
\]
\[
\phi(r,r')=(1/S_0)\int_{S_0}\theta(a-|{\bf r}-{\bf r}_1|)
\theta(a-|{\bf r'}-{\bf r}_1|),
\]
with $F(r)$  defined before by Eq. (\ref{eq50}).

This expression is exact and may serve as a basis for the calculation of the
average square of the multiplicity at finite $N$. However the new function
$\phi$  becomes very complicated when both variables $r$ and $r'$
are greater than $1-a$.
For this reason rather than analyze the general expression Eq. (\ref{eq62})  for
finite $N$ we shall
immediately take the thermodynamic limit $N\rightarrow \infty$. We are in fact
interested in the dispersion, not in the average square of multiplicity. It is
important, since the leading terms in $N$ cancel in the dispersion.
So we shall study
the difference
\beq
D^2=\blangle\mu^2\brangle-\blangle\mu\brangle^2
\label{eq63}
\eeq
in the limit $N\rightarrow\infty$, $\xi$ finite. As we shall see,
although both terms
in the right-hand side of of Eq. (\ref{eq63}) behave as $N^2$ separately, their
difference grows
only as $N$.

Separating from Eq. (\ref{eq62}) the term with $p=0$ and combining it with the second term on
the right-hand side of Eq. (\ref{eq63}) we present the total dispersion squared as a
sum of two
terms
\beq
D^2=D_1^2+D_2^2,
\label{eq64}
\eeq
where
\[
D_1^2=
\frac{1}{\sigma_0^2}\sum_{n,m}\sqrt{nm}
\int_Sd^2rd^2r'[C_N^{n,m}
\chi^n(r,r')\chi^{m}(r',r)\zeta^{N-n-m}(r,r')-\]\beq
C_N^nC_N^mF^n(r)F^M(r')(1-F(r))^{N-n}(1-F(r'))^{N-m}]
\label{eq65}
\eeq
and $D_2^2$ is given by Eq. (\ref{eq62}) with a restriction $p\geq 1$.

Leaving again the details to the original calculations
in ~\cite{brapaj2000} we present the final results
in the thermodynamic limit.

The first part of the dispersion squared is
\[
D_{1}^2/N=
-\frac{(\xi\langle\sqrt{n}\rangle-\langle n\sqrt{n}\rangle)^2}{\xi^2}\]\beq
+\frac{2}{\xi}e^{-2\xi}\sum_{n,m}\frac{\sqrt{nm}}{n!m!}\xi^{n+m}
\int_0^2RdR[(1-\lambda(R))^{n+m}e^{\xi\lambda(R)}-1].
\label{eq71}
\eeq
In the first term the averages are to be taken over the Poissonian
distribution.
\beq
\lambda(R)=\frac{1}{\pi}(\alpha-\sin \alpha),\ \
\alpha=2{\rm arccos}\frac{R}{2}.
\label{eq70}
\eeq
The second one can be easily evaluated numerically.
The second part is
\[
D_{2}^2/N=\frac{2}{\xi}e^{-2\xi}\sum_{n,m}\sum_{p=1}
\frac{\sqrt{(n+p)(m+p)}}{n!m!p!}\eta^{n+m+p}\]\beq\times
\int_0^2RdR\lambda^p(R)(1-\lambda(R))^{n+m}e^{\xi\lambda(R)}.
\label{eq72}
\eeq
This part is evidently positive. Its numerical evaluation shows that it
nearly cancels the large negative contributions from $D_1^2$.
So the numerical calculation of the dispersion requires some care.

As mentioned in Sec.1.4 percolation is a purely classical mechanism.
Overlapping strings form clusters.
At some critical value of the parameter $\xi$ a phase transition of
the 2nd order occurs: a cluster appears which
extends over the whole surface (an infinite cluster
in the thermodynamic limit).
The critical value of $\xi$  is found to be
$\xi_c\simeq 1.12-1.20$ ~\cite{isich}.
Below the phase transition point, for $\xi<\xi_c$, there is no infinite
cluster. Above the transition point, at $\xi>\xi_c$ an infinite cluster appears
with a probability
\beq
P_{\infty}=\theta (\xi-\xi_c) (\xi-\xi_c)^{\beta}.
\label{eq16}
\eeq
The critical exponent $\beta$ can be calculated from Monte-Carlo
simulations.
However the universality of critical behavior, that is, its independence of
the percolating substrate, allows to borrow its value from lattice
percolation, where $\beta=5/36$.

Cluster configuration can be characterized by the
occupation numbers $\avnu$, that is average numbers of clusters made of $n$
strings.
 Their behavior at all values of $\xi$ and $n$
is not known. From scaling considerations in the vicinity of the phase
transition  it has been found ~\cite{ref14}
\beq
\avnu= n^{-\tau}F(n^{\sigma} (\xi-\xi_c)),\ \ |\xi-\xi_c|<<1,\ n>>1,
\label{eq17}
\eeq
where $\tau=187/91$ and $\sigma=36/91$ and the function $F(z)$ is finite
at $z=0$ and falls off exponentially for $|z|\rightarrow\infty$.
Eq. (\ref{eq17}) is of limited value, since near $\xi=\xi_c$ the bulk of the
contribution is still due to low values of $n$, for which Eq. (\ref{eq14})
 is not
valid. However from Eq. (\ref{eq17}) one can find non-analytic parts of other
quantities of interest
at the transition point. In particular, one finds a singular part of
the total number of clusters $M=\sum\nu_n$ as
$\Delta\avm=c|\xi-\xi_c|^{8/3}$.
This singularity is quite weak: not only $\avm$ itself but also its two
first derivatives in $\xi$ stay continuous at $\xi=\xi_c$ and only the
third blows up as $|\xi-\xi_c|^{-1/3}$. So one should not expect that
the percolation phase transition will be clearly reflected in some
peculiar behavior of standard observables.

Indeed we observed that neither the total multiplicity nor $\avpt$ show any
irregularity in the vicinity of the phase transition, that is, at $\xi$
around unity.
This is not surprising since both quantities reflect the overlap structure
rather
than the cluster one. The connectedness property implied in the latter has no
 effect on these global observables.

It is remarkable, however, that the fluctuations of these observables carry some
information about the phase transition. As discussed above, the dispersion of the multiplicity due to overlapping and clustering
can easily be calculated in the thermodynamic limit. The result is
shown in Fig. \ref{fig2}(b). The dispersion shows a clear maximum around $\xi=1$ (in fact at $\xi\simeq 0.7$). This value is somewhat lower than the critical one $\xi \sim 1.2$ but still conveys certain information about the percolation phase transition around this point 
%So some information of the percolation phenomenon is passed to the total multiplicity,
 in spite of the fact that it basically does not feel the connectedness properties of the formed clusters. Of course, due to relation Eq. (\ref{eq10}), the dispersion of $\avpt$ has a similar behavior.

We have to warn against a simplistic interpretation of this result.
 The dispersion shown in Fig. \ref{fig2}(b) is only part of the total one, which besides includes contributions
from the fluctuations inside the strings and also in their number. Below we
 shall discuss  the relevance and magnitude of these extra contributions.

An intriguing question is a relation between the percolation and formation of
 the quark-gluon plasma. Formally these phenomena are different. Percolation is
 related to the connectedness property of the strings. The (cold) quark-gluon plasma formation is
related to the density of the produced particles (or, equivalently, the
density of their transverse energy). However in practice percolation and plasma
formation go together. In fact, the transverse energy density inside a single
string seems to be sufficient for the plasma formation. Percolation makes the total area occupied by strings comparable to the total interaction area, thus, creating, a sizable area with energy densities above the plasma formation 
threshold. 

\begin{figure}[htbp]
\centering        
\vspace*{-0.2cm}
\includegraphics[width=0.65\textwidth,height=3.0in]{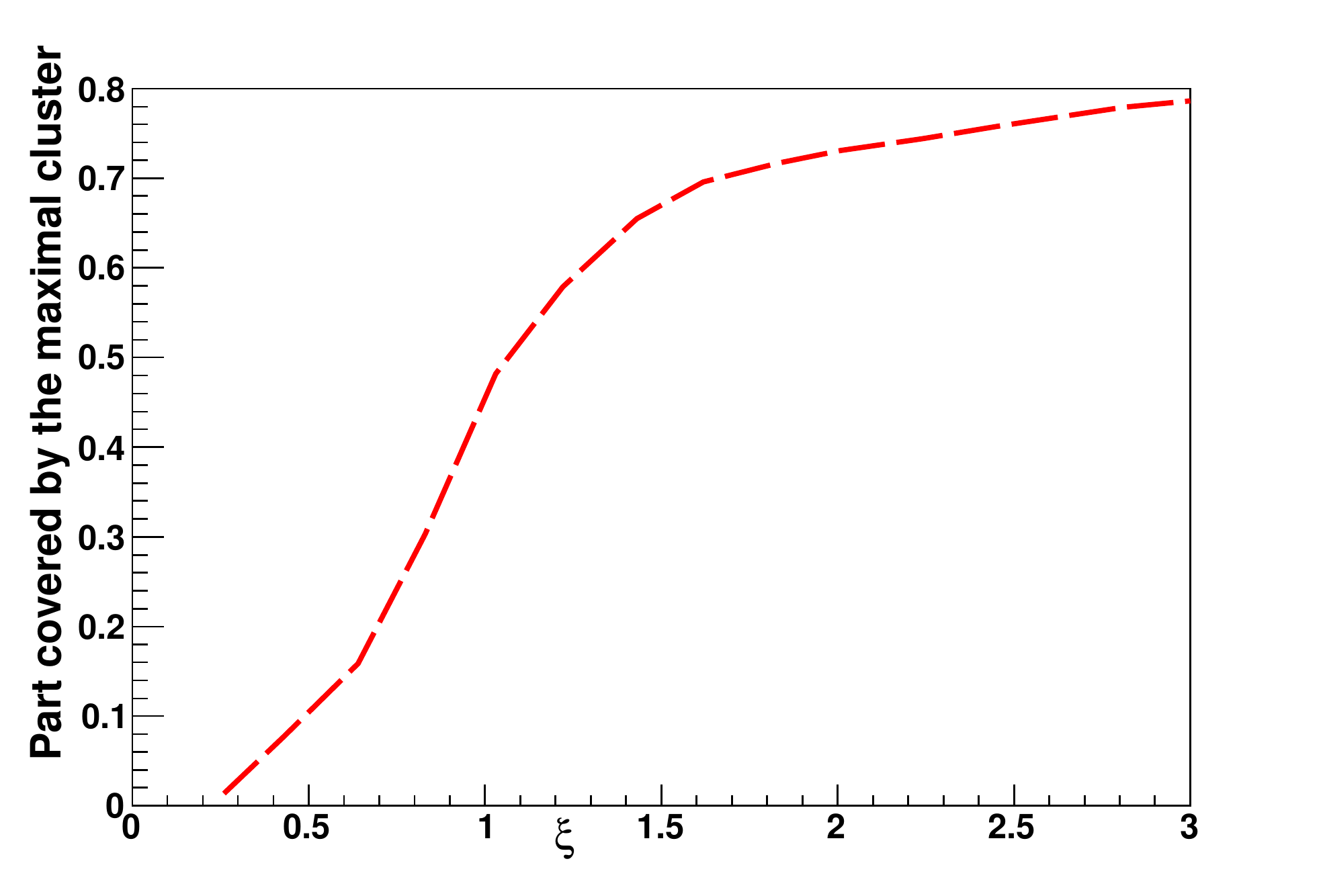}
\caption{Fraction of the total interaction area covered by the maximal
cluster as a function of $\xi$ \cite{brapaj2000}.}
\label{fig4}
\end{figure}

Let us make some crude
estimates. Comparison with the observed multiplicity densities in
$pp(\bar{p})$ collisions
at present energies fix the number of produced (charged) particles
per string per unit
rapidity at approximately unity. Taking the average energy of each
particle as 0.4 GeV
(which is certainly a lower bound), formation length  in the Bjorken
formula ~\cite{bjorken}
as 1 fm and the string transverse radius as 0.2 fm ~\cite{ref7} we get the
transverse energy density inside the string as $\sim 3$ $GeV/{\rm fm}^3$.
The plasma
threshold is currently estimated to be at $1$ $GeV/{\rm fm}^3$. So
it is tempting to say
that the plasma already exists inside strings. This however has little
physical sense because a very small area is occupied by a string. One can speak of a plasma only when
the total area occupied by a cluster of strings reaches a sizable fraction
of the total
interaction area. In Fig. \ref{fig4} we show this fraction for a maximal cluster as a function of
$\xi $ calculated by Monte -Carlo simulations in a system of 50 strings.
It grows with $\xi$ and the fastest growth occurs precisely in the region of the
percolation phase transition: as $\xi$ grows from 0.8 to 1.2  the fraction
grows from 0.3 to 0.6. With a string cluster occupying more than half of the interacting area,
one can safely speak of a plasma formed in that area.

\subsection{Distribution in the transverse momentum and quenching}
Clusters of strings may take quite  complicated forms in the transverse
plane varying from a disc corresponding to the simple string to
long chains of such discs. The question arises what the distribution
in transverse momenta of emitted partons will be from a cluster.
As we have seen in Subsection 2.2 the average transverse momentum from
a cluster can be determined in a comparatively simple manner, especially in the
thermodynamic limit:
\beq
\avpt=\avpt_1/F(\xi).
\label{eq2a}
\eeq
However this does not fix the distribution in $p_t$ uniquely.

Let us return to our simple picture of the fusion of two simple strings
Fig. \ref{fig1}. We assume that partons are emitted independently
from the three areas, corresponding to the overlap and two remaining
areas. Each of these areas can be visualized as a set of strings of
elementary transverse area $d^2r$. The total transverse momentum
distribution from
these three areas will be given by the sum of the three integrals
\[
\frac{dp^{tot}}{dyd^2p}=
C\Big\{\int_{S^{(1)}\oplus S^{(3)}}d^2re^{-\frac{m_\perp^2}{t_1}}
+\sqrt{2}\int_{S^{(2)}}d^2re^{-\frac{m_\perp^2}{t_1\sqrt{2}}}\Big\}\]\beq
=C\Big\{2S^{(1)}e^{-\frac{m_\perp^2}{t_1}}+\sqrt{2}S^{(2)}
e^{-\frac{m_\perp^2}{t_1\sqrt{2}}}\Big\}
\label{eq3a}
\eeq
and the distribution of emitted particles from the cluster will be
\beq
I(y,p)=\frac{1}{\mu}\,\frac{dp^{tot}}{dyd^2p}=
C\frac{2S^{(1)}e^{-\frac{m_\perp^2}{t_1}}
+\sqrt{2}S^{(2)}e^{-\frac{m_\perp^2}{t_1\sqrt{2}}}}
{2S^{(1)}+\sqrt{2}S^{(2)}}.
\label{eq4a}
\eeq

Generalizing to many clusters made of different number of
simple strings
we find the transverse momentum distribution in the general case.
\beq
\avpt=C\frac{\sum_{n=1}\sqrt{n}(S_n^{tot}/S_1)
e^{-m_\perp^2/(t_1\sqrt{n})}}
{\sum_{n=1}\sqrt{n}(S_n^{(tot}/S_1)}.
\label{eq5a}
\eeq
We recall that $S_n^{tot}$ is the total area of overlaps of $n$ simple
strings.

We stress that the distribution from the clusters remains isotropic
in the transverse space in spite of the fact that clusters themselves
have different forms and their distribution may not be isotropic at all.

Equation (\ref{eq5a}) may be used in the Monte-Carlo simulation.
For many practical problems it can be calculated in the thermodynamic limit
\beq
\avpt=C\frac{\blangle\sqrt{n}e^{-m_\perp^2/(t_1\sqrt{n})}\brangle}
{\blangle\sqrt{n}\brangle},
\label{eq6a}
\eeq
where averaging is done with the Poissonian distribution Eq. (\ref{per2}).
With reasonable accuracy it may be further simplified to
\beq
\avpt=Ce^{-\frac{m_\perp^2}{t_1\blangle\sqrt{n}\brangle}},
\label{eq7a}
\eeq
which implies that the distribution has the same Gaussian form
as for a simple string with appropriately enhanced tension.

Experimental data indicate however that the transverse momentum spectra
of emitted particles are not isotropic. Their dependence on
azimuthal angle together with the anisotropy of the string distribution
leads to the well-known azimuthal flows (Section 3.3). In view of this
fact our string picture requires certain refinement. As a source of
azimuthal anisotropy one may introduce quenching of produced partons
in the external chromoelectric field created by strings.

\begin{figure}[htbp]
\centering        
\vspace*{-0.2cm}
\includegraphics[width=0.65\textwidth,height=3.0in]{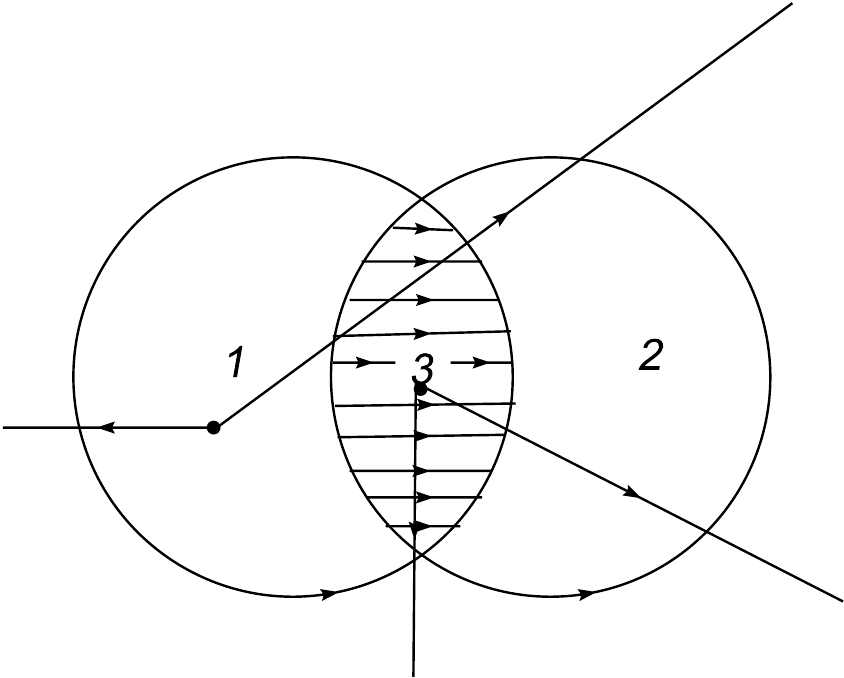}
\caption{Emission of particles in different azimuthal directions
from two overlapping strings.}
\label{fig1a}
\end{figure}

Turn again to fusion of two strings (Fig. \ref{fig1}). Let the observed
parton be emitted in different azimuthal directions either from the
overlap  or from the remaining part of one of the
strings (Fig. \ref{fig1a}). It is clear that the emitted partons have
to travel paths of different longitude before they go out and are
observed. Besides, partons going through the overlap meet stronger field
than those going only through the field of simple strings.
So one concludes that if partons loose their energy passing through the
field their observed distribution will depend on their azimuthal angle although
initially they were emitted isotropically.
This implies that the distribution in the transverse momentum has the
form Eq. (\ref{eq1a}) with $p\to p_0$ where $p_0({\bf p})$ is the parton
momentum at the instant of its creation and ${\bf p}$ is the observed momentum
\beq
P({\bf p})=Ce^{-\frac{p_0({\bf p})^2}{t}}.
\label{prob}
\eeq
The dependence of $p_0({\bf p})$ on the observed momentum will depend
both on the longitude of the path traveled by the parton and
on the strength
of the field it meets along this path. The concrete form of this
dependence is determined by the mechanism of quenching.

Radiative energy loss has been extensively studied for a parton passing
through the nucleus or quark-gluon plasma as a result of multiple collisions
with the medium scattering centers ~\cite{baier,baierx}. In our case the situation is somewhat different:
 the created parton moves in the external gluon field
inside the string. In the crude approximation this field can be taken as
being constant and  orthogonal to the direction of the parton propagation.
In the same spirit as taken for the mechanism of pair creation,
one may assume that the reaction force due to radiation is similar to the
one in the QED when a charged particle is moving in the external
electromagnetic field. This force causes a loss of energy, which for an
ultra-relativistic particle is proportional to
[its momentum $\times$field]$^{2/3}$ ~\cite{niri}:
\beq
\frac{dp(x)}{dx}=-0.12e^2 \Big(eEp(x)\Big)^{2/3},
\label{vbb}
\eeq
where $E$ is the external electric field.
Eq. (\ref{vbb}) leads to the quenching formula
\beq
p_0(p,l)=p\Big(1+\kappa p^{-1/3}t^{2/3}l\Big)^3,
\label{quench1}
\eeq
where we identified $eE/\pi=t$ as the string tension and $l$ is
the longitude of the path traveled by the parton in the field with
tension $t$.
The quenching coefficient $\kappa$ has to be adjusted to
the experimental data. In our practical calculations
it was chosen to give the experimental value for coefficient $v_2$
in mid-central Au-Au collisions at 200 GeV, integrated over the
transverse momenta.

Of course the possibility to use electrodynamics formulas for
the chromodynamic case may raise certain doubts. However in
~\cite{mikhailov} it was found that at least in the $N=4$ SUSY
Yang-Mills case the loss of energy of a colored charge moving
in the external chromodynamic field was given by essentially the same
expression as in the QED.

Note that from the moment of particle creation to the moment of its
passage through other strings a certain time elapses depending on the
distance and particle velocity. During this time strings decay and
the traveling particle will meet another string partially
decayed, with a smaller color $Q$ than at the moment of its formation.
So one has to consider a non-static string distribution
with string colors evolving in time and gradually diminishing until strings
disappear altogether.
To study the time evolution of strings we  again turn to the Schwinger
mechanism.  For it one has the probability
of pair creation in unit time and unit volume as ~\cite{nikishov}
\beq
\Gamma_{VT}=\frac{1}{4\pi}t^2e^{-\frac{p_0^2}{t}},
\eeq
where again $t$ stands for $eE/\pi$ in QED. For a realistic string the volume
$V=SL_z$ where $S$ is the string transverse area and $L_z$ is
the longitudinal dimension of the string.
For the single string of color $Q$  we have  $t=Qt_1$ where
$t_1$ is the string tension of the ordinary string with $Q=1$. The average
transverse momentum squared of the emitted quark-antiquark pair
$<p_0^2>$ is just $t$. To estimate $L_z$ we assume that the string emits a
pair when its energy
is equal to $2<p_0>=\sqrt{t}$, which gives $L_z=1/\sqrt{t}$, so that
we get the average probability in unit time
\beq
\Gamma_T=\frac{1}{4\pi}t^{3/2}S.
\eeq
The string color diminishes by unity with each pair production.
So we find an
equation which describes the time evolution of the string color $Q(T)$
\beq
\frac{dQ(T)}{dT}=-\alpha Q^{3/2}(T)
\eeq
with the solution
\beq
Q(T)=\frac{Q_0}{(1+\frac{1}{2}\alpha T \sqrt{Q_0})^2},
\label{qtime}
\eeq
where $Q_0$ is the initial color at the moment of the string creation.
Coefficient $\alpha=t_0^{3/2}S/(2\pi)$ depends on the string transverse
area $S$. In practical calculations we use the picture in which
the fused string is in fact modeled by a set of ``ministrings'' formed at
intersections of simple strings with the same area as the simple string, but
greater color. This gives $\alpha= 0.03$ $fm^{-1}$.
The average color of ministrings is of the order 2 - 3. So it changes
only by  30 -50 \% even when the emitted parton travels 5 $fm$ of distance.
So the time scale of string
evolution is estimated to be  considerably greater  than time intervals characteristic
for partons traveling inside the string matter. However the effect of
string decay with time is noticeable and we take it into account in our
calculations.
In fact it
practically does not change the results but changes the value
of the quenching coefficient $\kappa$, which in any case is to be adjusted,
as explained above. In this sense our results are practically independent
of the concrete choice of $\alpha$ in the reasonable interval of values.
%%%%%%%%%%%%%%%%%%%%%%%%%%%%%%%%%%

Finally we note that the Schwinger formula Eq. (\ref{eq1a})
describes well the spectra only at very soft $p_0$. To extend
its validity to
higher momenta one may use the idea that the string tension fluctuates, which
transforms the Gaussian distribution into the thermal one ~\cite{bialas,pajares3}:
\beq
P({\bf p})=Ce^{-\frac{p_0({\bf p})}{\sqrt{t/2}}}.
\label{probb}
\eeq

\subsection{Rapidity dependence}
Color strings are stretched between partons into which  colliding
hadrons or nuclei pass before collisions. Let
CM energy squared of the collision be
$s=2p\bar{p}\simeq 2p_+\bar{p}_-$ where $p$ and $\bar{p}$ are
the 4-momenta of
colliding nucleons and we neglect here the nucleon mass $m$ at high energies.
A string carries a fraction of $s$ depending on the  momenta
of the partons
$k$ with $k_+=x_+p_1$ and $\bar{k}$ with$\bar(k)=x_-\bar{p}$, so that its CM energy squared is $ s_s=sx_{+}x_-$.
Defining upper and lower rapidities as
\beq
y=\frac{Y}{2}+\ln x_+,\ \ \bar{y}=-\frac{Y}{2}-
\ln x_-
\eeq
where $Y=\ln (s/m^2)$ is the overall rapidity
one can say that the string is stretched in the final interval of
rapidity between $\bar{y}$ and $y$. If this interval is large then
the probability of particle emission at rapidity $y_p$ from the string
will be practically independent of rapidity while $\bar{y}<y_p<y$.
This does not imply that the observed particle spectrum will be rapidity
independent. In fact, as explained in Section 1.5, partons forming the string are distributed in the
colliding hadrons with probabilities $\rho(y)$ and $\rho (\bar{y})$, which are obtained from 
Eq. (\ref{stri4})
upon integration over spectator variables.
The final distribution in rapidity $y_p$ will then be governed by the
factor
\beq
{\cal P}(y_p)=\int_{y_p}^Ydy\int_{\bar{y}}^{y_p}d{\bar y}
\rho(y){\rho}(\bar{y}).
\label{py}
\eeq
and is thus totally determined by the distribution of partons in the
colliding hadrons.

Now let us consider string fusion. Let two strings be stretched between
partons from the projectile with momenta $k_1$ and $k_2$ and target
momenta $\bar{k}_1$ and $\bar{k}_2$ with
$k_{i+}=x_{i+}p$ and $\bar{k}_{i-}=x_{i-}$, $i=1,2$.
Conservation of momentum dictates that if the two strings completely
fuse then the ends of the fused string have light-cone momenta
$(x_{1+}+x_{2+})k_+$ and $(x_{1-}+x_{2-})\bar{k}_-$.
In terms of rapidities the new ends will be
\[
y_{fused}=\ln \Big(e^{y_1}+e^{y_2}\Big),\ \
\bar{y}_{fused}=-\ln \Big(e^{-\bar{y}_1}+e^{-\bar{y}_2}\Big)
\]
where $y_i$ and $\bar{y}_i$, $i=1,2$ are ends of the fusing strings.
This result trivially generalizes to fusion of arbitrary
number of strings. If end rapidities of the strings are $y_i$ and
$\bar{y}_i$ then if they completely fuse the ends of the fused string
will be
\[
y_{fused}=\ln \Big(\sum_ie^{y_i}\Big),\ \
\bar{y}_{fused}=-\ln \Big(\sum_ie^{-\bar{y}_i}\Big)
\]
\label{yfus}
For illustration, if all the fusing strings are the same then
\[
y_{fused}=y+\ln n,\ \ \bar{y}_{fused}=\bar{y}-\ln n\]

The situation obviously complicates if strings are not fused completely,
as in Fig. \ref{fig1}. Then one has to consider the three parts with
areas $S^{(1)}, S^{(2)}$, and $S^{(3)}$ as independent strings and for each
of them determine ends in rapidity separately taking into account the
sum of the parton momenta in each of the three.
\section{Model Results and comparison with experiments}
\subsection{Multiplicity distributions}
The multiplicity distributions in the DPM or QGSM in pp and AA collisions are given by Eqs. (\ref{stri1}-\ref{stri2}) and Eqs. (\ref{stri16}-\ref{stri17}) respectively. However, as the energy or the centrality of the collision increases one expects interaction among the strings stretched between the projectile and target partons. As discussed earlier, due to the randomness of the color field in color space, the strength of the resulting color field in a cluster of $n$ strings, is only $\sqrt {n}$ times the strength of the color field of a single string, giving rise to a suppression of the multiplicity of  particles produced in the decay of the cluster. The same reason lies at the origin of the enhancement of the mean transverse momentum  with string density. The corresponding equations for both quantities, were obtained in the previous section, 
Eq. (\ref{eq13}) and Eq. (\ref{eq2a}) respectively. 
\begin{figure}[htbp]
\centering        
\vspace*{-0.2cm}
\includegraphics[width=0.65\textwidth,height=3.0in]{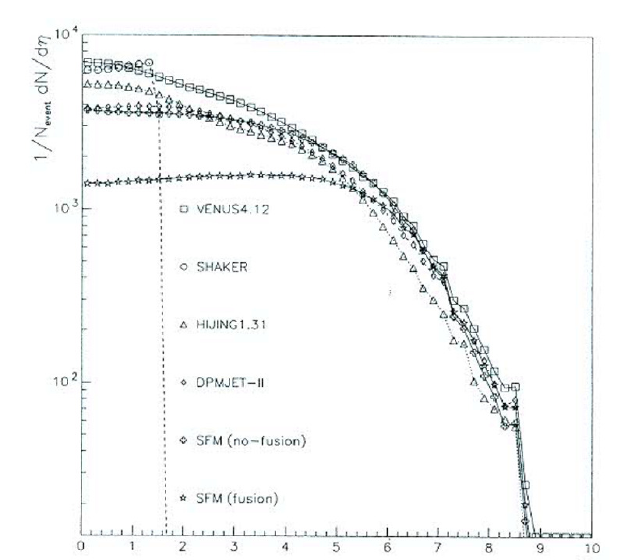}
\caption{Pseudorapidity distribution of charged particles for central Pb-Pb events at a beam energy of 3 TeV per nucleon \cite{ref12A}.}
\label{figmult1}
\end{figure}
At present, most of the  models have incorporated the suppression of multiplicities compared with the superposition of independent scatterings 
result by means of different mechanisms, but this was not so twenty years ago, when the predictions of all the models for central Pb-Pb collisions at LHC energy were a factor two higher than the prediction of the string fusion model (SFM) \cite{ref12,ref12A}, which was a previous version of the percolation string model, as it is seen in Fig. \ref{figmult1}. The SFM prediction is close to the LHC data for Pb-Pb collisions at 2.76 TeV.
According to  Eq. (\ref{eq13}) the multiplicity distribution in pp collisions in the central rapidity region is given by
\begin{equation}
\frac {dN^{pp}}{dy}= F(\xi_{p}) N_{p}^{S}\mu_{1},\ \   \xi_{p} = N_{p}^{s}\frac{S_{1}}{S},
\label{multe1}
\end{equation}
where  factor $F$ is given by  Eq. (\ref{eq15}), $N_{P}^{S}$ is the number of strings in the central rapidity region, 
 $S_{1}=\pi r_{0}^{2}$ with $r_0=0.2$ fm and $S$ is the transverse area of the proton. In the DPM in nucleus-nucleus collisions the number of strings stretched between the sea quark and antiquarks in the central rapidity region is proportional to $N_{A}^{4/3}-N_{A}$ and is given by  Eq. (\ref{stri17}). Here $N_{A}^{4/3}$ is the total number of nucleon-nucleon collisions. Hence, the total number of strings in a central heavy ion collisions can be very large, actually more than $\sim 10^{3}$ strings are produced in Au-Au at RHIC energies. However, each string must have a minimum of energy to be produced and decay subsequently into particles (at least two pions). On the other hand, the total energy available in a collision grows as $A$, whereas the number of strings $\sim A^{4/3}$ in central collisions. Hence, at not very high energy (for instance RHIC energy) the energy is not  sufficient to produce such a large number of strings. In order to take into account this energy momentum conservation effect one may  reduce the number of sea quark and antiquark strings, changing \cite{mult4}
\beq
N_{A}^{4/3} \rightarrow N_{A}^{1+\alpha (\sqrt s)}
\label{multe2}
\eeq
with
\beq
\alpha (\sqrt s) = \frac  {1}{3} \Big(1- \frac {1}{1+\ln(\sqrt(s/s_{0}) +1)}\big).
\label{multe3}
\eeq
One can thus write
\beq
\frac {dN^{AA}}{dy} \sim N_{A} (N_{A}^{\alpha (\sqrt s)-1}) \frac {dN^{pp}}{dy}. 
\label{multe4}
\eeq
Parameter  $s_{0}$ marks the energy squared below which  energy-momentum conservation effects become small and  $\alpha \rightarrow $1/3. Up to here we do not take into account the interaction among the strings. If we do take into account the interaction of strings then  we can write a closed formula for the multiplicity distribution in $ AA$ collisions in terms of the multiplicity distribution in pp collisions, namely \cite{mult4}
\beq
\frac{1}{N_{A}}\frac{dN}{dy}{\Big |}_{y=0} =\frac {dN^{pp}}{dy}{\Big |}_{y=0}
\Big(1+\frac {F(\xi_{N_A})}{F(\xi_p)}
N_{A}^{\alpha (\sqrt {s})-1}\Big)
\label{multe5}
\eeq

where 
\beq
\xi_{N_{A}} = \xi_{p} N_{A}^{\alpha (\sqrt s)+1} \frac {S_1}{S_{N_A}}
\label{multe6}
\eeq
and $S_{N_{A}}$ is the transverse area of the collision region formed when there are $N_{A}$ wounded nucleons of the projectile and $N_{A}$  nucleons of the target. $S_{N_{A}}$ depends on $N_{A}$ and A. 
The dependence of the multiplicity on the center of mass collision energy 
 $\sqrt s$ is fully specified once the average number of strings in a pp collision $N_{p}^{s}$ is known. At low energy $N_{p}^{s}$ is approximately equal to 2, growing with energy as $(\sqrt s/m_{p})^{2\lambda}$  so that
\beq
N_{p}^{s} = 2+ 4 (\frac {r_{0}}{R_{p}})^{2}(\frac {\sqrt{s}}{m_{p}})^{2\lambda}.
\label{multe7}
\eeq
Here a single  parameter $\lambda$ describes the rise of the multiplicity with energy for both pp and AA multiplicity distributions, even though
in   AA central collisions the multiplicity increases  faster than in pp collisions due to the energy dependent factor $\alpha$, 
arising from energy momentum conservation.

A fit to pp collisions data in the range  53 $\leq \sqrt s \leq$ 7000 GeV and to AA collisions
(Au-Au, Cu-Cu, and Pb-Pb) at different centralities for 19.6 $\leq \sqrt s \leq$ 2760 GeV has been done. The values obtained from the fit for the two parameters are $\sqrt s_{0}$ = 245 GeV and $\lambda$ =0.201.

\begin{figure}[htbp]
\vspace*{-0.5cm}
\centering        
\includegraphics[width=0.65\textwidth,height=3.0in]{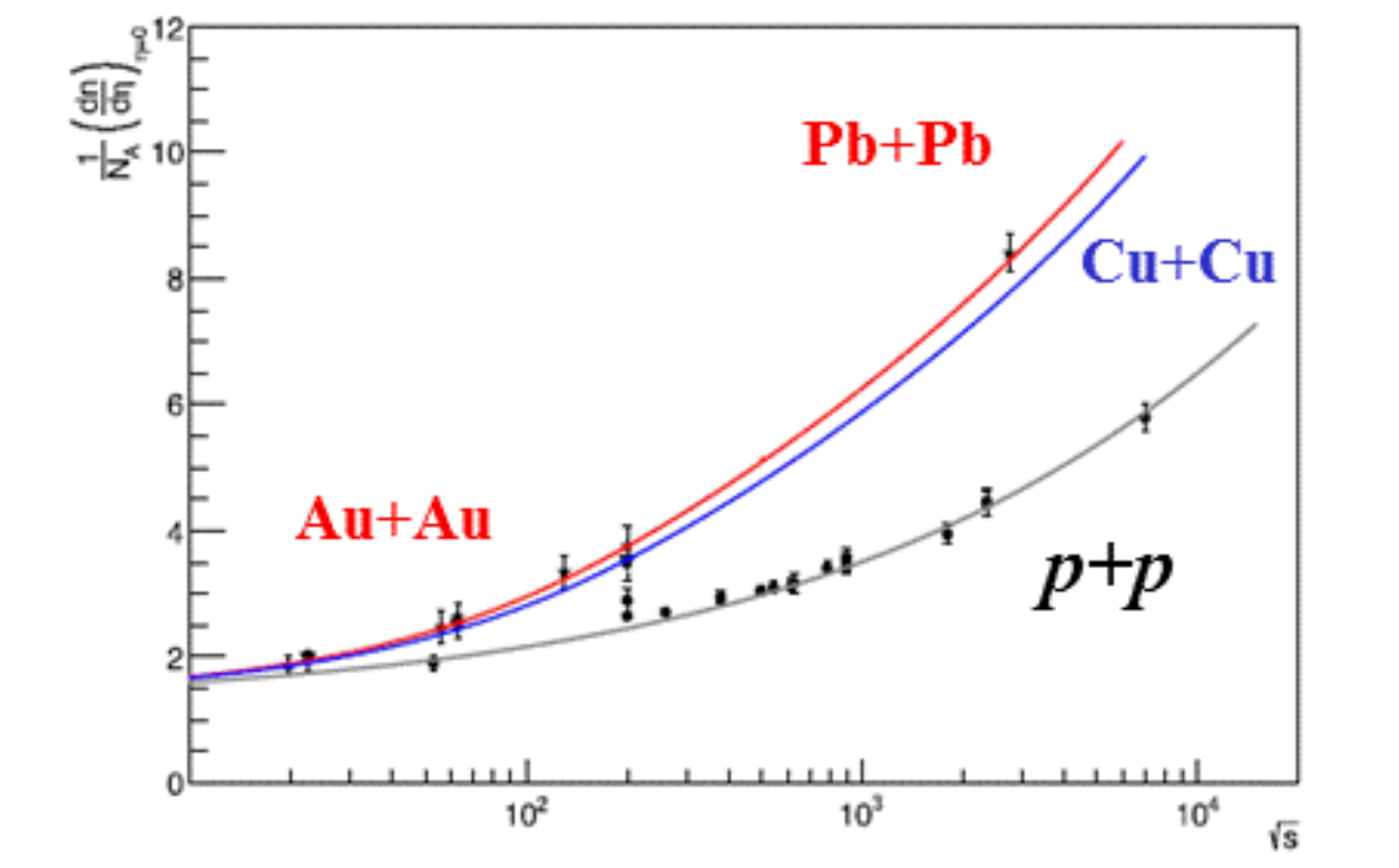}
\caption{Comparison of the evolution of the mid-rapidity multiplicity with energy from the CSPM and data for pp and A-A collisions. Lines are from the model for
 pp (gray), Cu-Cu (blue) and red lines for Au-Au/Pb-Pb \cite{mult4}.}
\label{figmult2}
\end{figure}

\begin{figure}[htbp]
\centering        
\vspace*{-0.2cm}
\includegraphics[width=0.80\textwidth,height=3.0in]{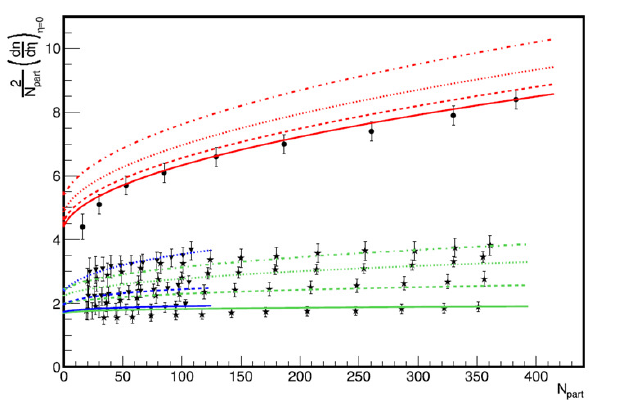}
\caption{Multiplicity dependence on centrality ($N_{part}$). Cu-Cu (triangles), Au-Au (stars) and Pb-Pb (circles).Curves represent the model calculations. Blue line for Cu-Cu, green line for Au-Au and red for Pb-Pb \cite{mult4}.}
\label{figmult3}
\end{figure}

\begin{figure}[htbp]
\centering        
\vspace*{-0.2cm}
\includegraphics[width=0.65\textwidth,height=3.0in]{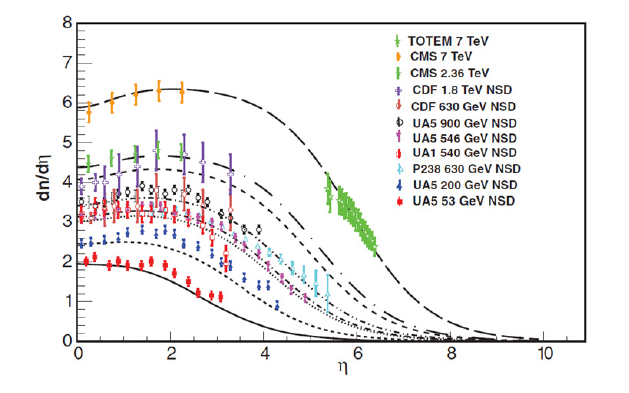}
\caption{Comparison of results from the evolution of $dn_{ch}/d\eta$ with dependence on pseudorapidity for pp collisions at different energies (lines) \cite{mult17}. }
\label{figmult4}
\end{figure}

\begin{figure}[htbp]
\centering        
%\vspace*{-2.0cm}
\includegraphics[width=0.9999\textwidth,height=3.5in]{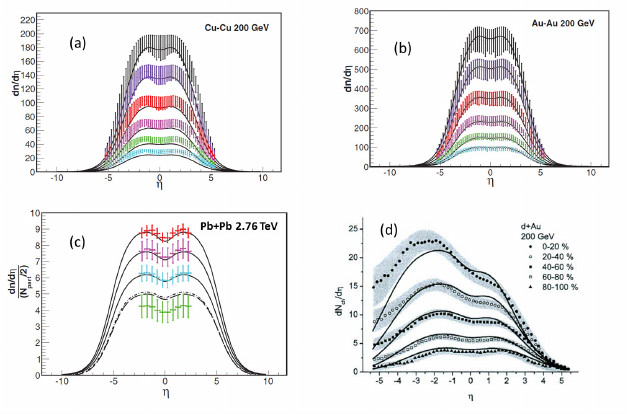}
\caption{Comparison of results from the evolution of $dn_{ch}/d\eta$ with dependence in pseudorapidity for (a).Cu-Cu at 200 GeV,  (b) Au-Au 200 GeV. Plot (c) show  $\frac {dn_{ch}}{d\eta}\frac{1}{(N_{part}/2)}$ for Pb-Pb collisions at 2.76 TeV (d) dn/d$\eta$ at different centralities for d+Au collisions at 200 GeV \cite{mult17}.}
\label{figmult5}
\end{figure}

Figure \ref{figmult2} shows a comparison of the results for the dependence of midrapidity multiplicity on the energy with data for pp \cite{mult5,mult6,levente,mult8,mult9,mult10,mult11} and central Cu-Cu ($N_{A}$ = 50, A = 63) \cite{mult12} and for Au-Au/Pb-Pb ($N_{A}$ = 175, A = 200) 
\cite{alice1}.
In Fig. \ref{figmult3} the result of the dependence of the multiplicity per participant nucleon on the number of participants is shown together with the experimental data for Cu-Cu ($\sqrt {s_{NN}}$ = 22.4, 62.4, 200 GeV) for Au-Au ($\sqrt {s_{NN}}$= 19.6, 62.4, 130 and 200 GeV) and for Pb-Pb ($\sqrt {s_{NN}}$ = 2.76, 3.2, 3.9, 5.5 TeV).
%n Table 1 we show the perdictions for pp and central Pb-Pb collisions at future LHC energies.
%
The evolution outside the central rapidity region has been studied extensively  extending  to all rapidities Eq. (\ref{multe7}) \cite{mult14,mult15,mult16,mult17}:
\beq
\frac {1}{N_{A}}\frac{dN^{N_{A}N_{A}}}{d\eta}= K J F(\xi_{p})N_{p}^{s}
\frac {\Big(1+\frac{F(\xi_{N_{A}})}{F(S_{p})}(N_{A}^{\alpha}-1)\Big)}{\exp\Big( \frac{(\eta-(1-\alpha)Y)}{\delta}\Big)+1}, 
\label{multe8}
\eeq
where $J$ is the Jacobean
\[
J = \frac{\cosh\eta}{\sqrt {k_{1}+\sinh^{2}\eta}}\] and \beq K= \frac {k}{J(\eta=0)}\Big(\exp(\frac{-(1-\alpha)Y}{\delta})+1\Big).
\label{multe9}
\eeq
Parameters $\alpha$ and $\delta$ are obtained from  fitting to the data ($\alpha=0.34, \delta= 0.84$). The pseudorapidity dependence is described by the same factor 
\beq
\frac {1}{\exp(\frac {\eta-(1-\alpha)Y}{\delta})+1}
\label{multe10}
\eeq
 in pp and AA collisions. This dependence gives rise to a smaller increase at central pseudorapidity  $\eta$ = 0 than at large pseudorapidity $\eta$ = Y \cite{mult15,mult16}. The limiting fragmentation property have been studied carefully, showing 
it is not exact and a violation of it should be more visible at the highest LHC energies \cite{mult14,mult15,mult16}.
In Fig. \ref{figmult4}  we show the comparison of Eq. (\ref{multe8}) with experimental data at different energies  (53 GeV $\leq \sqrt{s}\leq$ 7 TeV) \cite{mult18,mult19}. In Fig. \ref{figmult5}(a-c) we show the comparison between the results
for Cu-Cu \cite{mult19}, Au-Au \cite{mult20} and Pb-Pb \cite{mult21} collisions and the experimental data. 

The percolation of strings have been applied to the case of different projectile and target as well. In Fig. \ref{figmult5}(d)  the results for d+Au at different centralities compared to the experimental results \cite{phobos} are shown. Similar results were obtained in Ref. \cite{mult24}. 

The behavior obtained for $dN/dy$ in pp and AA collisions with  energy and  number of participants is very similar to the 
 Glasma picture in CGC. In fact, in CGC the multiplicity distribution is given by ~\cite{par39}
\beq
\frac {dN}{dy}\sim \frac{1}{\alpha_{s}(Q_{s})}Q_{s}^{2}R_{A}^{2}.
\label{multe12}
\eeq
Since  the saturation momentum squared $Q_{s}^{2}$ behaves like $N_{A}^{1/3}$, the multiplicity per participant is almost independent of $N_{A}$ and a weak dependence arises from the logarithmic dependence on $N_A$ of the running coupling constant $1/\alpha_s\sim  \log N_{A}$. In percolation the multiplicity per participant  is almost independent of $N_{A}$  as well and the only additional dependence arises from the factor $(1-\exp(-\xi))^{1/2}$ which grows weakly with $N_{A}$ above the percolation threshold. Concerning  the energy dependence,  $Q_{s}^{2}$  behaves like  $s^{\lambda}$, so that the same behavior is obtained in percolation. There is an extra energy dependence  due to the running coupling  constant $\alpha_{s}(Q_{s})$ in the CGC, which again corresponds to the energy dependence of the factor $(1-\exp(-\xi))^{1/2}$. It is not surprising that their exist a correspondence  between 1/$\alpha_{s}$, the occupation number or the number of gluons,  and the factor $1-\exp(-\xi)$ which represents the fraction of the collision area covered by strings. The larger the occupation number the larger the fraction is. The similarities between  the Glasma picture of CGC and the percolation of strings are visible not only  in multiplicities but in most of the other observables, as discussed later.

In order to explain  the faster rise of the multiplicity in central AA collisions than in pp collisions several possibilities have been proposed in CGC, such as enhanced parton showers in AA collisions due to the larger average transverse momentum of the initially produced minijets compared to pp collisions 
\cite{mult26} or non trivial Q effects intertwined with impact parameter 
dependence \cite{mult27}.

\subsection{Transverse momentum distributions}
In Section 2.4 we studied the effects of percolation of strings on the mean transverse momentum and the dispersion of the transverse momentum distribution and the experimental data related to it.

As explained in Sec. 2 the detailed study of both multiplicity and momentum distribution requires analysis of all overlaps of created strings which differ both in the number of the strings in the overlap and in the area of a particular overlap. With a large number of
strings created in AA collisions at high energy this is hardly feasible. So it is reasonable to search for some effective way to describe the observables in this situation.

Let us start with multiplicities. They come from a set of overlaps (``ministrings'') and depend on both the number of overlapped strings and the area of the overlap, which combine to give an average multiplicity $N$ from this overlap. We may characterize different
overlaps just by this average multiplicity. We call $N$ a ``size'' of the overlap, the quantity combining both the number of overlapped strings and the area. With a lot of overlaps $N$ will be changing practically continuously. We then can introduce a probability
$W(N)$ to have  overlaps with size $N$ in a collision and write the total distribution in multiplicity as
\beq
P(n)=\int dN W(N)P(N,n),
\label{p1}
\eeq
where $P(N,n)$ is the multiplicity distribution from the overlap of a given $N$, which we take to be Poissonian with the average 
multiplicity $N$
\beq
P(N,n)=\exp(-N)N^n/n!.
\label{p2}
\eeq
The normalization conditions $\sum_nP(n)=\sum_nP(N,n)=1$ and the condition $\sum_n nP(N,n)=N$ lead to relations
\beq
\int dN W(N)=1 ,\ \ <n>=<N>=\int dN NW(N).
\label{p3}
\eeq

\begin{figure}[htbp]
\centering        
\vspace*{-0.2cm}
\includegraphics[width=0.65\textwidth,height=3.0in]{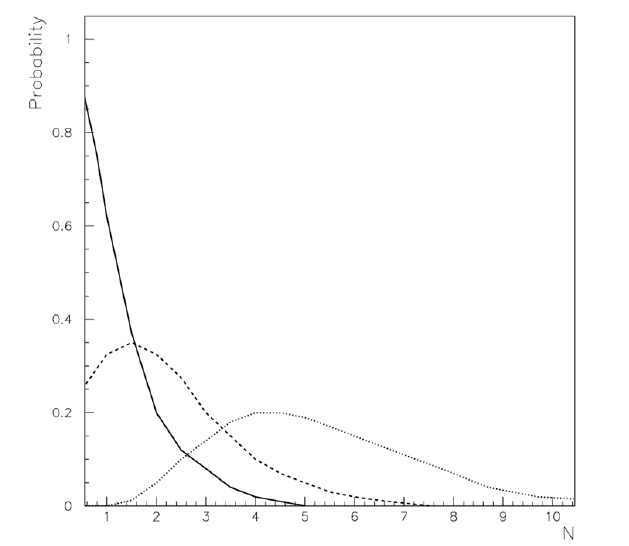}
\caption{Schematic representation of the number of clusters as a function of the number of strings of each cluster at three different centralities \cite{mult31}.}
\label{figmult11}
\end{figure}

For the weight function we assume the gamma distribution \cite{mult30,mult31,mult32}
\beq
W(N) =G(N,k_N,r_n)= \frac {r_N}{\Gamma(k_N)}(r_NN)^{k_N-1}\exp(-r_NN),  r_N=\frac {k_N}{<N>}.
\label{p4}
\eeq
There are several reason  for this choice. First, the  gamma  distribution  reproduces to a good  approximation  the cluster  size distributions  at different centralities. In fact, let us consider a peripheral  collision, where the density of strings is small and there are only  very few overlapping  strings. In this case, the cluster size is peaked at low values of the number of strings of the cluster. As the centrality increases, the density of strings increases as well ,and  there are more and more overlapping  strings. The cluster size distribution becomes strongly modified. Figure \ref{figmult11} shows a  plot  of three different cluster size distributions  corresponding to three centralities. Each  one can be described  by a gamma function corresponding  to different values of $k_N$.

There is another reason for choosing the gamma function, related to the re-normalization group. The growth of the centrality  can be seen as a transformation of the cluster size distribution. Start with a set of single strings with a few clusters formed of a few overlapping strings. As the centrality increases, there appear more strings and  more clusters composed of more strings. This change can be considered as substitution of strings in a cluster by newly formed clusters, defined by new $<n>$, corresponding to a higher color field
in the cluster. 
 This transformation, similar to the block transformations of Wilson type, can be seen as a transformation of the cluster size probability of the type:
\beq
P(x) \rightarrow \frac {xP(x)}{<x>}....\rightarrow \frac {x^{k}P(x)}{<x^{k}>}\rightarrow...
\label{pte6}
\eeq
 Transformations of this kind were studied long time ago by Jona-Lasinio in connection with the re-normalization group in probability theory \cite{mult33}, showing that the only
 probabilities $P(x)$ stable  under such transformations are the generalized gamma functions. Among the generalized  gamma functions the simplest one is the gamma function which in addition has one parameter less (This additional parameter could be used to refine the model for comparison with the data).
The transformations of type Eq. (\ref{pte6}) have been used previously to study the probability associated with some special events which are shadowed only by themselves and not for the total of events \cite{mult34,mult34x,mult34y,mult35,mult36}.

Notice that $W(N)$ satisfies KNO scaling, namely the product $ NW(N)$ is only a function of $ N/<N>$ and not of the energy. This property is a consequence of the invariance of the form of the gamma functions under transformations of type  Eq. (\ref{pte6}) \cite{mult37}. 
%%%%%%%%%%%%%%%%%%%%%%%%%%%%%%%%%%%%%%%%%%%%%%%%%%%%%%%%%%%%%%%%%%%
%%%%%%%%%%%%%%%%%%%%%%%%%%%%%%%%%%%%%%%%%%%%%%%%%%%%%%%%%%%%%%%

Now we pass to 
the transverse momentum distribution (TMD) $f(p_{t})$. As in the case of multiplicities it comes from overlaps of different number of strings having different areas and the TMD from an overlap depends on its both characteristics (we may again call this combination a ``size'' of the overlap, this time in respect to the TMD). Take the TMD from an overlap of a given size given by the Schwinger mechanism
\beq
f(p_t,x)=\exp (-xp_{t}^{2})
\label{p5}
\eeq
where $x$ is just this  ``size''. Assuming again that $x$ varies continuously one can write the total TMD, in all similarity to Eq. (\ref{p1}), as
\beq
f(p_{t}) = \int dx W_p(x)f(x,p_{t})
\label{pte1}
\eeq
with a certain positive weight $W_p(x)$.
To understand its property we use the normalization condition for TMD
\beq
\int dp_t^2f(p_t)=<n>
\eeq
This gives a relation
\beq
\int \frac{dx}{x}W_p(x)=<n>.
\label{p6}
\eeq
On the other hand, introducing integration variable $N=\alpha x$ we have from Eq. (\ref{p3})
\beq
<n>=\int dx x\alpha^2W(\alpha x).
\eeq
Comparing this with Eq. (\ref{p6}) we can make an identification
\beq
W_p(x)=(\alpha x)^2W(\alpha x)
\eeq
If we take the gamma distribution Eq. (\ref{p4}) for $W(N)$ then $W_p(x)$ turns out  up to a factor to be also the gamma distribution
bit with different $k$ and $r$
\beq
W_p(x)=\frac{r}{r_p}G(x,k+2,r_p)
\label{p7}          
\eeq
with $r_p=\alpha r$ (note that $\alpha$ is dimensionful).
 
So in the end both the distribution $P(n)$ and TMD $f(p_t^2)$ are given by a convolution of the cluster multiplicity and its TMD with the size probability $W$,
which in both cases can be taken as the gamma distribution although with different parameter $k_N,r_N$ and $k_p,r_p$ respectively. In the following having in mind that most applications will be devoted to the TMD, we denote $k_p$ and $r_p$ as simply $k$ and $r$ leaving notations $k'$ and $r'$ for $k_N$ and $r_N$. In many discussions the behavior of $k$ and $k'$ is similar and we do not specify  which of the two $k$'s we are discussing.

%%%%%%%%%%%%%%%%%%%%%%%%%%%%%%%%%%%%%%%%%%%%%%%%%%%%%%%%%%%%%%%
%%%%%%%%%%%%%%%%%%%%%%%%%%%%%%%%%%%%%%%%%%%%%%%%%%%%%%%%%%%%

%%%%%%%%%%%%%%%%%%%%%%%%%%%%%%%%%%%%%%%%%%%%%%%%%%%%%%%%%%%%%
%%%%%%%%%%%%%%%%%%%%%%%%%%%%%%%%%%%%%%%%%%%%%%%%%%%%%%%%%%%%%

 Introducing  Eq. (\ref{p4}) into  Eq. (\ref{pte1}) and  Eq. (\ref{p1}), we obtain
\beq
\frac{1}{(1+\frac{P_{t}^{2}}{r})^{k}}=\int_{0}^{\infty} dx{\exp(-p_{t}^{2}x)(\frac{r}{\Gamma(k)})(rx)^{k-1}\exp(-rx)}
\label{pte7}
\eeq
and
\[
\frac{\Gamma(n+k')}{\Gamma (n+1)\Gamma (k')}\frac{r^{`k}}{(1+r{'})(n+k')}\] \beq= \int_{0}^{\infty}dN \frac{e^{N}N^{n}}{n!}\frac{r^{'}}{\Gamma (k')}(r^{'}N)^{k'-1})\exp(-r^{'}N).
\label{pte8}
\eeq
The mean value and the dispersion of the distributions Eq. (\ref{pte7}) and  Eq. (\ref{pte8}) are
\beq
<x>=\frac{k}{r}, \frac {<x^{2}>-<x>^{2}}{<x>^{2}} = \frac{1}{k}
\label{pte9}
\eeq
\beq
<n>=<N>\frac{k'}{r^{'}},\ \ \frac {<N^{2}>-<N>^{2}}{<N>^{2}} = \frac{1}{k'},
\eeq
\beq 
 \frac {<n^{2}>-<n>^{2}>}{<n>^{2}} = \frac{1}{k'}+\frac{1}{<N>}
\label{pte10}
\eeq
The distribution given by Eq. (\ref{pte7}) is the negative binomial distribution.
 Equations (\ref{pte7}) and (\ref{pte8}) can be seen as a superposition of sources (clusters) where $1/k$ and $1/k'$ fix the transverse momentum  fluctuations and  the fluctuations on the number of strings in the clusters. At small string density  there are almost no strings overlapped, the strings are isolated and $k$ and $k'$ tend to infinity. When the density increases, there will be some overlapping strings forming clusters, and therefore $k$'s decrease. Their minimum is reached when fluctuations
in the number of strings per cluster reach their maximum. Above this point, with increasing string density, these fluctuations decrease and $k$'s increase.
We recall that $k$ and $k'$ are generally different. When comparing to the experimental data to fix them, one also has to take into account the experimental conditions, such as acceptance and range of variables studied.

\begin{figure}[htbp]
\centering        
\vspace*{-0.2cm}
\includegraphics[width=0.65\textwidth,height=3.0in]{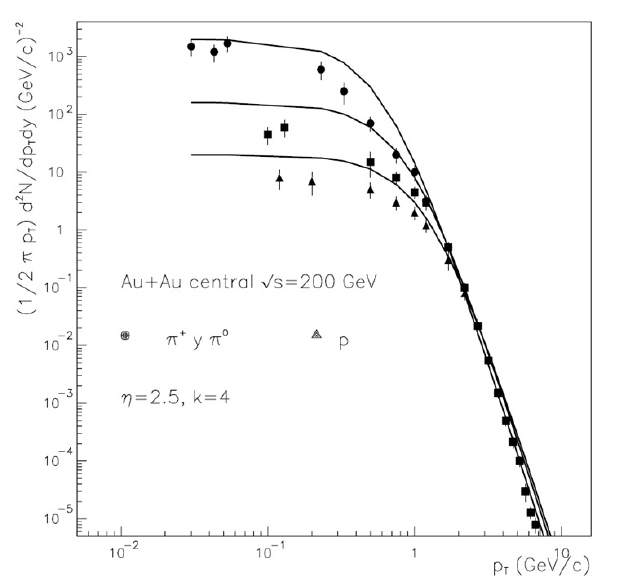}
\caption{Experimental PHOBOS data on low $p_{t}$ distributions for pions, kaons and protons along with our results for central Au-Au collisions at $\sqrt {s_{NN}}$ = 200 GeV \cite{mult31}.}
\label{figmult12}
\end{figure}

Now we have to  take into account that the mean transverse momentum and  mean multiplicity of a cluster of strings is given by  Eqs. (\ref{eq13}) and (\ref{eq2a}) and therefore we should incorporate this into  Eqs. (\ref{pte7}) and (\ref{pte8}).  In this way the respective distributions become
\beq
f(p_{t},y) = \frac {dN}{dp_{t}^{2}dy} = \frac{dN}{dy}\frac{k-1}{k}
\frac{F(\xi)}{<p_{t}^{2}>_{1}}\frac{1}
{\Big(1+\frac{F(\xi)p_t^2}{k<p_t^2>_1}\Big)^k}
\label{pte11}
\eeq
and
\beq
P(n)= \frac{\Gamma(n+k')}{\Gamma(n+1)\Gamma(k')}
\frac{\Big(\frac{k'}{<n>_{1}F(\xi)}\Big)^{k'}}{\Big(1+(\frac{k'}{<n>_{1}F(\xi)})\Big)^{n+k'}}
\label{pte12}
\eeq
We observe that 
\beq
<n>=F(\xi)N_{s}<n>_{1},\ \    <p_{t}^{2}>= \frac {k}{k-2} \frac {<p_{t}^{2}>_{1}}{F(\xi)}
\label{pte14}
\eeq
 Equations (\ref{pte11}) and (\ref{pte12})  give the distributions  for any projectile, target, energy and degree of centrality  and are universal functions which depend only on two parameters $<p_{t}^{2}>_{1}$ and $<n>_{1}$  the transverse momentum and multiplicity of particles produced  by one string. In the case of  identified secondary particles  
 the corresponding quantities for each identified particle,  $<p_{t}^{2}>_{1i}$ and $<n>_{1i}$  should be used. 
In addition to these parameters,  parameters $k$ and $k'$, actually are functions of  centrality 
and, as we said above, with the growth of string density first decrease up to a minimum close to the critical percolation value and afterwards increase .
 At $\xi \rightarrow \infty, k \rightarrow \infty$ and the distribution given by Eq. (\ref{pte11}) becomes  $\exp(-F(\xi)p_{t}^{2}/<p_{t}^{2}>_{1})$ very similar to the behavior at 
 $\xi \rightarrow 0$.
From Eq. (\ref{pte11}) we have.
\beq
\frac{dlnf}{dlnp_{t}} = \frac{-2F(\xi)}{\Big(1+\frac{F(\xi)p_{t}^{2}}{k<p_{t}^{2}>_{1i}}\Big)}\frac{p_{t}^{2}}{<p_{t}^{2}>_{1i}}.
 \label{pte15}
\eeq
At $p_{t}^{2}\rightarrow 0$ this reduces to $ -2F(\xi)p_{t}^{2}/<p_{t}^{2}>_{1i}$
and vanishes at $p_{t}$ = 0. On the other hand, as the mean $p_{t}$ of particles produced in a single string is larger for protons than for kaons and the latter larger than for pions, $ <p_{t}^{2}>_{1\pi}$ $<$  $<p_{t}^{2}>_{1k}$  $<$  $  <p_{t}^{2}>_{1p}$, the absolute value of the left-hand side of Eq. (\ref{pte15})
is larger for pions than for kaons and than for protons. This
is the well known hierarchy that is  very often advocated as evidence in favor of hydrodynamics models. In Fig. \ref{figmult12} the results of our distribution  Eq. (\ref{pte11})
are shown compared to the RHIC data of PHOBOS ~\cite{mult38} for central Au-Au collisions at $\sqrt {s_{NN}}$ = 200 GeV.  We use  the values $ <p_{t}^{2}>_{1\pi}=0.06, <p_{t}^{2}>_{1k} =0.14,\\  <p_{t}^{2}>_{1p} = 0.30 $ and $k=4$. The value of the string density is 2.5. We observe a good agreement taking into account that our result is   not a fit but the result of  Eq. (\ref{pte11}) fixing the normalization of the three curves at the same value at $p_{t}$ = 2 GeV/c.
Let us now discuss the interplay between low and high $p_{t}$. One defines the ratio  $R_{CP}(p_{t})$  between central and peripheral collisions as
\beq
R_{CP}(p_{t}) = \frac {f^{'}(p_{t},y=0)/N_{coll}^{'}}{f(p_{t},y=0)/N_{coll}},
\label{pte16}
\eeq
where the distributions in the numerator and denominator correspond to central and peripheral collisions respectively, $\xi^{'} > \xi$. The normalization  on
the number of collisions in Eq. (\ref{pte16}), essentially eliminates $N_{s}$, the number of strings, from dN/dy (this is true at midrapidity and not at forward or backward rapidities). From  Eqs. (\ref{pte1}) and (\ref{pte11}) we obtain
\beq
R_{CP}(p_{t}) = \frac {((k^{'}-1)/k^{'})}{((k-1)/k)}\Big(\frac {F(\xi^{'})}{F(\xi)}\Big)^{2}
\frac {\Big(1+\frac{F(\xi)p_{t}^{2}}{k<p_{t}^{2}>_{1i}}\Big)^{k}}{\Big(1+\frac{F(\xi^{'})p_{t}^{2}}{k^{'}<p_{t}^{2}>_{1i}}\Big)^{k^{'}}}.
\label{pte17}
\eeq
Here $k$ and $k'$ are values of the parameter $k$ for transverse momentum distribution for peripheral and central collisions.
In the limit $p_{t}^{2}\rightarrow 0$, as $F(\xi^{'}) < F(\xi)$ we obtain 
\beq
R_{CP}(0) \simeq \Big(\frac {F(\xi^{'})}{F(\xi)}\Big)^{2} < 1,
\label{pte18}
\eeq
which is approximately independent of $k$ and $k'$. As $\xi^{'}/\xi$ increases the ratio $R_{CP}$ decreases, in agreement with the experimental data. As $p_{t}$ increases, we have
\beq
R_{CP}(p_{t}) \sim \frac {1+\frac{F(\xi)p_{t}^{2}}{k<p_{t}^{2}>_{1i}}}
{1+\frac{F(\xi^{'})p_{t}^{2}}{k'<p_{t}^{2}>_{1i}}}
\label{pte19}
\eeq
and $R_{CP}$ increases with $p_{t}$ (again $F(\xi) > F(\xi^{'})$).
At large $p_{t}$
\beq
R_{CP}(p_{t}) \sim \frac {F(\xi)}{F(\xi^{'})} \frac{k^{'}}{k} p_{t}^{2(k-k')}.
\label{pte20}
\eeq
At low density in the region where $k$ decreases with the string
 density $k'< k$ and $R_{CP}(p_t) > 1 $. It is the Cronin effect. As $\xi^{'}/\xi$
 increases the ratio $R_{CP}$ increases.

With the growth of the energy of the collision, the energy density increases and therefore we arrive at the region where $k$ increases. Now as  $\xi^{'} > \xi$, $k^{'} > k$ and there will be a suppression of $p_{t}$.
\begin{figure}[htbp]
\centering        
\vspace*{-2.5cm}
\includegraphics[width=0.95\textwidth,height=3.5in]{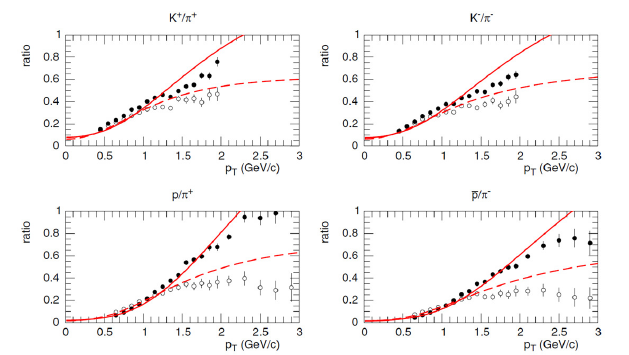}
\caption{Ratios for different distributions K/$\pi$, p/$\pi$ in Au-Au collisions at $\sqrt {s_{NN}}$ = 200 GeV at two different centralities :(0-5\%, solid circles, 60-70-\% open circles) in comparison with the data \cite{mult32}.}
\label{figmult14}
\end{figure}
In the forward rapidity region, the normalization of $N_{coll}$ in Eq. (\ref{pte16}) does not cancel $N_{s}$ from $ dN/dy$, since in this region $N_{s}$ is proportional to $N_{A}$ instead of $N_{coll}$. Therefore, an additional factor $\frac{N_{A}^{'}/N_{coll}^{'}}{N_{A}/N_{coll}}$ appears now in $R_{CP}(p_{t})$. As $N_{coll}^{'}-N_{A}^{'}$ for central collision is larger than the corresponding quantity for peripheral collisions, we have $R_{CP}(p_{t},y=3) < R_{CP}(p_{t},y=0)$, thus a further suppression occurs in agreement with the experimental data \cite{mult39}.

\begin{figure}[htbp]
\centering        
\vspace*{-1.0cm}
\includegraphics[width=0.65\textwidth,height=3.0in]{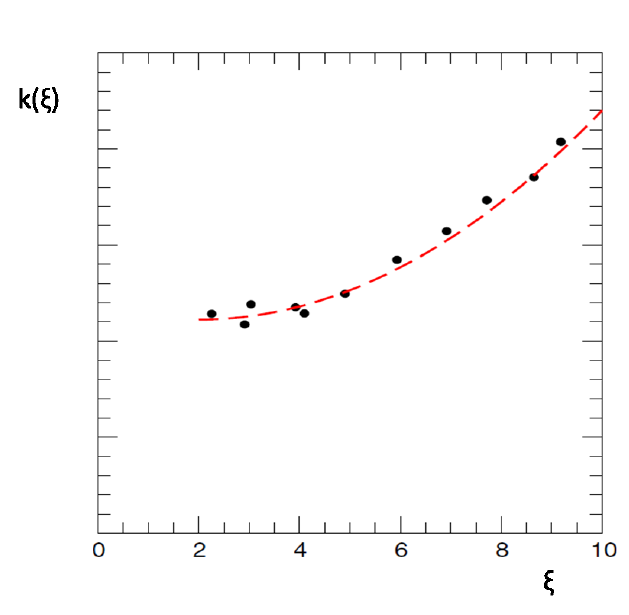}
\caption{$k(\xi)$ extracted from $\pi^{+}$ fits. The behavior expected in percolation is suggested by the data \cite{mult32}.}
\label{figmult15}
\end{figure}

 The results for the TMD for $\pi^{+}, K^{+}$ and p in Au-Au collisions at $\sqrt {s_{NN}}$ = 200 GeV are in good agreement with the PHENIX experimental data \cite{mult47}. In Fig. \ref{figmult14} we show the ratios kaons/pions and proton/pions as a function of $p_t$ at the two extreme centralities compared to the experimental data.
The obtained values of $k$ as a function of the string density evaluated for each centrality are shown in Fig. \ref{figmult15}. The expected increase of $k$ with the string density  is clearly seen.

The experimental data on pp in the range $\sqrt s$ =23, 200, 630, 1830, and 70000 GeV can  be described  by the distribution Eq. (\ref{pte11}) ~\cite{mult48} as well. In this case the values of $k$ decrease with energy as expected, a higher energy means a higher string density. Notice that for string densities above the critical percolation value, $k$ should increase, hence in pp a change in the behavior of $k$  is predicted. At 14 TeV the string density would be close to such critical value but probably would not reach it.

\begin{figure}[htbp]
\centering        
\vspace*{-0.2cm}
\includegraphics[width=0.99\textwidth,height=3.0in]{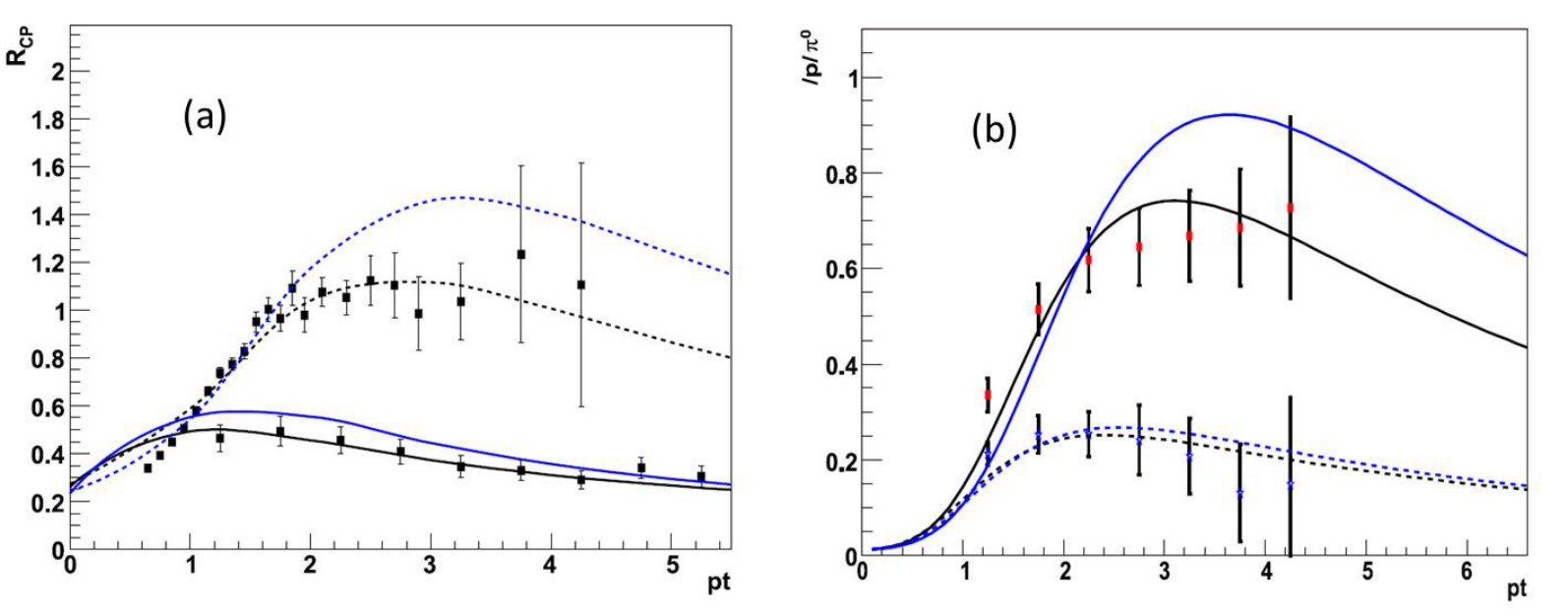}
\caption{ (a)$R_{CP}$ (0-10\% central/60-92\% peripheral) for pions(solid line) and (p+$\bar {p}$)/2 (dashed) compared to the PHENIX data. Blue line is LHC predictions. (b) $\bar {p} / \pi^{0}$ as a function of $p_{t}$ for 0-10\% (solid) and 60-92\% (dashed) centrality bins. The points are from PHENIX data. LHC predictions are in blue \cite{mult52}.}
\label{figmult16}
\end{figure}

Concerning the baryon $p_{t}$ distributions, the difference with the meson spectrum is not only due to the difference of masses which results in that
$<p_{t}>_{Mesons}$ $<$ $<p_{t}>_{Baryons}$. This effect only causes a shift of the maximum of the nuclear modified factor $R_{AA}$ but keeps the same height at the maximum contrary to the data. In the fragmentation of a cluster formed by several strings the flavor properties follow from the corresponding properties of the individual strings and hence the resulting flavor of the cluster $F$ is the flavor composition of the individual strings $f$ as well as the color composition. We obtain clusters with higher color and different flavored ends. In the fragmentation of a cluster we consider the creation of a pair of complexes $F\bar F$. After the decay the two new $F \bar F$ strings are treated in the same way and therefore decay into more $F \bar F$ strings until they come to objects with masses comparable to hadron masses, which are identified with observable hadrons by
combining  the produced flavor with statistical weights. In this way the production of baryons and antibaryons is enhanced. We observe that  additional quarks or antiquarks required to form a baryon or antibaryon are provided by the quarks or antiquarks of the overlapping strings that form the cluster. In some sense 
this coalescence is incorporated in a natural way ~\cite{mult49,mult50,mult51} in the CSPM. In order to take this into account, keeping a similar formula for the TMD, we modify the latter in an effective way \cite{mult52}. We take for the multiplicity (per unit rapidity) of baryons and antibaryons.
\beq
\frac {dN}{dy}= N_{s}^{1+\alpha}F(\xi)<n>_{1 B}
\label{pte21}
\eeq
instead  of Eq. (\ref{pte1}) fitting the parameter $\alpha$ to reproduce the experimental data ($\alpha$ =0.09). From Eq. (\ref{pte21}) we observe that when a (anti)baryon is triggered the effective number of strings is  $N_{s}^{1+\alpha}$ instead of $N_{s}$. This means that the string
density $\xi$ must be replaced by $N_{s}^{\alpha} \xi$. The (anti)baryons probe a higher density than the mesons for the same energy. The value of $\alpha$ is different for baryons  and antibaryons because the rise with centrality  is slightly different for them, depending on the specific kind of baryon or antibaryon  as well. In Fig. \ref{figmult16} we plot the results \cite{mult52} for $R_{CP}$ (0-10\% central)/(60-92\% peripheral) for pions and ($p+\bar p$)/2  compared to PHENIX data. At 5.5 TeV the LHC predictions  are shown. In  Fig. \ref{figmult16} we  plot the antiproton to neutral pion ratio for central and peripheral centrality bins compared to the PHENIX data together with the predictions at 5.5 TeV. More results can be seen in Ref. \cite{mult53}, where the the open charm production has been studied. Related to percolation, there are other extensive studies of the effects of the strong color field  on the string tension, including the effect of baryon junction. These studies extend to strangeness production \cite{mult54,mult55} and charm production \cite{mult56,mult57} in pp and AA collisions.

 Equation (\ref{pte11}) for the $p_{t}$ distribution does not include the high $p_{t}$ region as far as  we take a Gaussian shape for the fragmentation of a cluster. A more detailed  form for the cluster taking into account  perturbative QCD for the high $p_t$ region may extend the range of validity of our modified distribution.

\begin{figure}[htbp]
\centering        
\vspace*{-0.2cm}
\includegraphics[width=0.99\textwidth,height=3.0in]{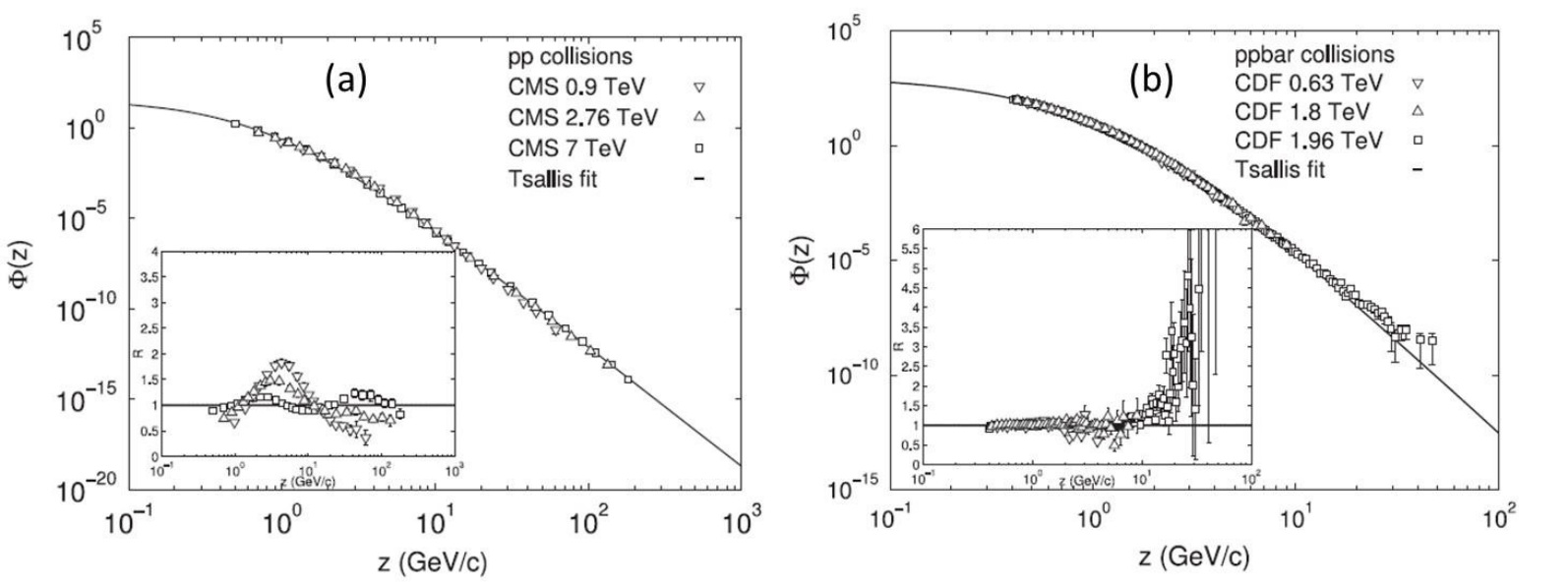}
\caption{ Scaling behavior of the charged hadron $p_{t}$ spectra presented in ${\it {z}}$ (a) in pp collisions and (b) $p\bar{p}$ collisions with different energy scales. The inset is the distribution of the ratio between the experimental data and the fitted results \cite{mult58}. }
\label{figmult18}
\end{figure}

There are several scaling properties found in the TMD related to string percolation. The experimental data for pp collisions exhibit a universal behavior
in a suitable variable $z=p_t/K$ 
\beq
\phi(z)=A \frac{d^{2}N}{2p_tdp_{t}d\eta}|_{p_{t}= Kz}.
\eeq
Here the  parameters $A$ and $K$ depend on the energy of the pp or p$\bar{p}$ collisions. This universal scaling behavior \cite{mult58}, as is shown in Fig. \ref{figmult18} for  pp at $\sqrt s$ =0.9, 2.36 and 7 TeV and  $p\bar p$ at 0.63, 1.8, 1.96 TeV. The value of ${\it K}$ increases with the energy. In string percolation this scaling is satisfied by the distribution Eq. (\ref{pte11}) with ${\it K}$ proportional to 1/F($\xi$). As F($\xi$) decreases with energy ${\it K}$ increases. A similar scaling has been found in Au-Au \cite{mult59,mult60} at different centralities for the distributions of kinetic energy of pions, kaons and protons. Again the percolation of strings distribution satisfies this scaling.
%
%skipping Tallis part and temp 
%
%

\begin{figure}[htbp]
\centering        
\vspace*{-0.2cm}
\includegraphics[width=0.75\textwidth,height=3.0in]{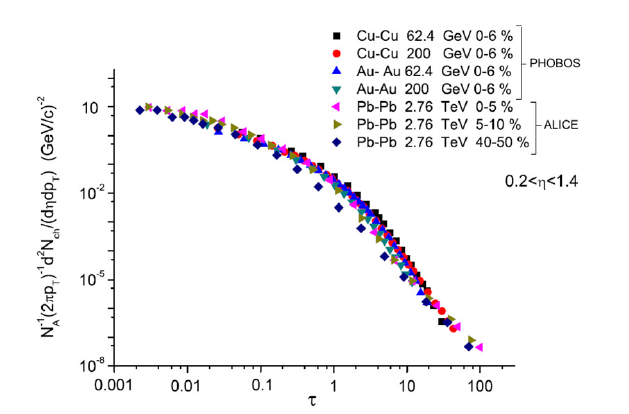}
\caption{ Charged particle multiplicity per participant at pseudorapidity $0.2 <\eta < 1.4 $ for Au-Au and Cu-Cu central collisions at two RHIC energies 62.4 and 200 GeV. and for Pb-Pb collisions at 2.76 TeV plotted as a function of $\tau$  for $\lambda = 0.27$ \cite{mult72}.}
\label{figmult20}
\end{figure}

Recently, it has been shown that in pp collisions the $p_{t}$ spectra of charged particles exhibit geometrical scaling, which is also satisfied in  nucleus-nucleus collisions \cite{mult70,mult71}. Using proper normalizations it was shown that in any type of collision for any centrality and energy the different distributions approximately are lying on the same curve for low and moderate $p_{t}$ \cite{mult72}, namely
\beq
\frac{1}{N_{A}} \frac{dN}{dp_{t}^{2}}= \phi(\tau),\ \  \tau = \frac{p_{t}^{2}}{Q_{s}^{2}}.
\label{pte23}
\eeq
This is illustrated in Fig. \ref{figmult20}, where the data for pp, pPb, Cu-Cu, Au-Au, and Pb-Pb at different energies and centralities  are plotted.
The percolation of strings, at low $p_{t}$ satisfies this scaling. In fact at low $p_{t}$, the distribution Eq. (\ref{pte11}) becomes  exp$(-p_{t} F(\xi)/<p_{t}>)$ which is a function only of $p_{t}/Q_{s}$ if $Q_{s}$ is proportional to 
$<p_{t}>/F^{1/2}(\xi)$.

The experimental data on the mean $p_{t}$ as a function of the multiplicity have shown an interesting systematic in pp, pPb, and Pb-Pb collisions at LHC. All of them grow with multiplicity, being larger in pp than in pPb and in Pb-Pb collisions \cite{mult73,mult74}. On the other hand, in Pb-Pb collisions the rise of mean $p_t$  with multiplicity is flattened above a certain low multiplicity. The same occurs for pPb collisions, even though in this case the change of behavior is at higher multiplicity. All this behavior is understood easily in string percolation as a consequence of Eq. (\ref{pte14}). In fact  factor $1/F^{2}(\xi)$ is responsible for the rise with multiplicity of mean  $p_{t}$ because $\xi$ grows with multiplicity. The flattening of the mean $p_{t}$ exhibited in Pb-Pb and pPb data is due to the dependence of the parameter $k$ in  Eq. (\ref{pte14}) on the string density $\xi$. In Pb-Pb for most of  multiplicities the corresponding string densities $\xi$ are above the critical percolation value. In this region, $k$ grows with $\xi$ and the rise of  $p_{t}$ is lowered. The same happens in pPb, but now we need higher multiplicities to have string densities above the critical percolation density. In the case of pp collisions most of the corresponding string densities lie below the critical percolation string density.
In this region $k$ is a decreasing function of $\xi$, hence there is no flattening. We expect that at higher energies even in pp collisions the critical density will be reached. However at 14 TeV probably the string density will be slightly below. This discussion  refers to charged particles but can be extended to identified particles. In the case of baryons, as noted before,  there are additional enhancements due to the higher flavor of the cluster. This can be taken into account in an effective way assuming that  baryons probe a higher string density. Due to that we expect a flattening of the mean $p_t$ of baryons
as a function of the multiplicity at future LHC energies.
\subsection{Collective flow and ridge }
\begin{figure}[htbp]
\centering        
\vspace*{-0.2cm}
\includegraphics[width=0.90\textwidth,height=2.0in]{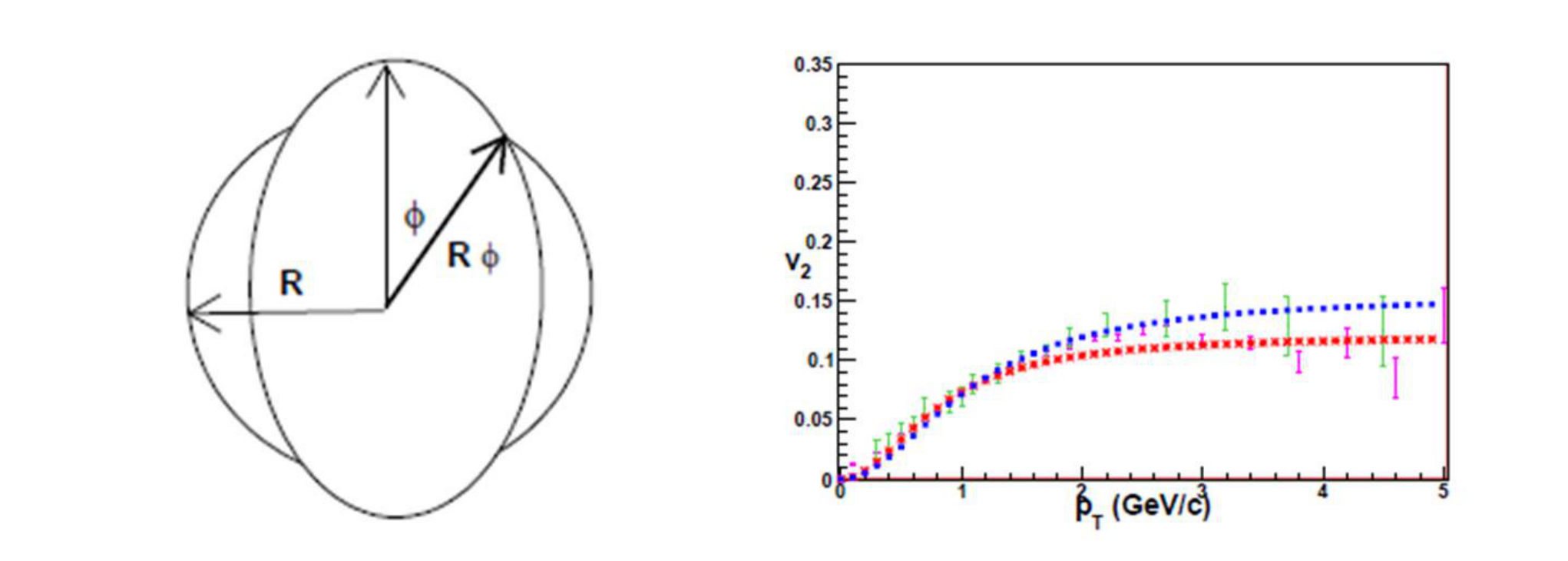}
\caption{Left plot:Azimuthal dependence of R. R being the radius of the projected circle \cite{bautista}. Right plot: Elliptic flow $v_{2}$ comparison with data and CSPM. The error bars in red and green are the results from Pb-Pb at $\sqrt {s_{NN}}$= 2.76 TeV and Au-Au  at $\sqrt {s_{NN}}$= 200 GeV. The CSPM results are shown as dotted blue and red lines \cite{flow2}.}
\label{figfl1}
\end{figure}

The cluster formed by the strings has generally an asymmetric form in the transverse plane and acquires dimensions comparable to the nuclear overlap. This azimuthal asymmetry is at the origin of the elliptic flow in CSPM. The partons emitted at some point inside the cluster have to pass through the strong color field before appearing on the surface. The energy loss by the parton is proportional to the length and therefore the $p_{t}$ of a particle will depend on the direction of the emission as shown in Fig. \ref{figfl1} (left plot) \cite{bautista}.
A Monte-Carlo simulation has been done taking into account
this energy loss. The results of this simulation for different
harmonics ~\cite{bautista,flow2} are in reasonable agreement with the experimental data on the $p_t$ and centrality dependencies. In order to get an analytical
expression, a simplification  was done encoding all the azimuthal
dependence in a new string density parameter $\xi_\phi$ in such a way
that the inclusive azimuthal
distribution is obtained by changing $\xi$ to $\xi_\phi$. In this way one obtains closed formula for $v_{2}$ \cite{flow2,bautista1,ridge1x}
\begin{equation}
v_{2}=\frac {2}{\pi} \int^{\pi}_{0}d\phi \cos(2\phi)
 \left (\frac{R_{\phi}}{R}\right)\left(\frac {e^{-\xi_\phi}-F(\xi_\phi)^{2}}{2F(\xi_\phi)^3}\right)\frac{R}{R-1}.
\label{fle}
\end{equation}
The transverse momentum dependence of $v_{2}$ computed using Eq. (\ref{fle}) for Pb-Pb at $\sqrt {s_{NN}}$ = 2.76 TeV and Au-Au  at $\sqrt {s_{NN}}$ = 200 GeV for 10-20\% centrality is shown in Fig. \ref{figfl1} (right plot). A good agreement is also obtained at all centralities and rapidities \cite{bautista,bautista2}.

%%%%%%%%%%%%%%%%%%%%%%%%%%%%%%%%%%%%%%%
The ridge structure was seen first at RHIC in Au-Au and Cu-Cu and later at LHC 
in Pb-Pb collisions. This structure has been also observed in proton-proton and proton-nucleus collisions at high multiplicity at LHC energies. This strong long range rapidity correlation collimated around  azimuthal angles 0 and $\pi$ is one of the most interesting features of the experimental data, especially the structures seen in pp an pPb are a challenge for most of the models, in particular the hydrodynamics description.

In simple string models azimuthal dependence was not generated, so that the arising correlations were flat in $\phi$. In string percolation correlations can arise from the superposition of many events with different number and type of strings. In this way there appears long range correlations in rapidity as discussed earlier. However passing to the azimuthal dependence, if the emission from strings is isotropic, the correlations due to their distribution in different events will also be isotropic. Also in the central rapidity region the inclusive cross section are practically independent of rapidity. This generates a plateau in the  $y-\phi$ distribution rather than a ridge, with only a narrow peak at small ${\it y}$ and $\phi$ due to short range correlations. This conclusion remains true if one averages the inclusive cross sections over all events with the resulting loss of azimuthal angle dependence. So the ridge can only be obtained on an event-by-event basis.

 In string percolation the ridge arises due to fluctuations in both string distributions and impact parameter \cite{ridge1x}. So it is important to study the averages relevant to the ridge formation. For a particular string configuration with a fixed azimuthal angle $\phi_{0}$ of the impact parameter, the single and double inclusive cross sections are
\[
I^{c}(y,\phi)= A^{c}(y)+2 \sum_{n=1}(B_{n}^{c}(y)\cos n(\phi-\phi_{0})+C_{m}^{c}\sin n(\phi-\phi_{0}))\] \begin{equation}
  = A^{c} (1+2 \sum_{n=1}(b_{n}^{c}(y)\cos n(\phi)+c_{m}^{c}\sin n(\phi))
\label{ridge1} 
\end{equation}
\[
I^{c}(y_{1},\phi_{1},y_{2},\phi_{2})= \Big[ A^{c}(y_{1})+2 \sum_{n=1}(B_{n}^{c}(y_{1})\cos n(\phi_{1}-\phi_{0})+C_{n}^{c}\sin n(\phi_{1}-\phi_{0})) \Big]X \] \begin{equation}
 \Big[A^{c}(y_{2})+2 \sum_{m=1}(B_{m}^{c}(y_{2})\cos m(\phi_{2}-\phi_{0})+C_{m}^{c}\sin m(\phi_{2}-\phi_{0}))\Big]. 
\label{ridge2}
\end{equation}
One must  average  this expression over string distributions and directions of the impact parameter. The latter reduces to integration over with weight 1/2$\pi$. Doing first this integration and then averaging over string distribution one obtains the measured double inclusive cross section 
\begin{equation}
I(y_{1},y_{2},\phi_{12}) =< A(y_{1}) A(y_{2})>+2\sum_{n}<W_{n}(y_{1},y_{2})> \cos n\phi_{12},
\label{ridge3}
\end{equation}
where $\phi_{12}=\phi_{1}-\phi_{2}$ and $W_{n}(y_{1},y_{2})= B_{n}(y_{1})B_{n}(y_{2})+C_{n}(y_{1})C_{n}(y_{2})$. Similar averaging of the single cross section eliminates all the azimuthal dependence $I(y_{1})$ = $<A(y_{1}>$. Thus the correlation function is
\[
C(y_{1},y_{2},\phi_{12})= \frac {< A(y_{1}) A(y_{2})>+2\sum_{n}<W_{n}(y_{1},y_{2})> \cos n\phi_{12}}{<A^{c}>^{2}}-1 \] \begin{equation}
 = \frac {DA(y_{1},y_{2})}{<A(y_{1})><A(y_{2})>}+2\sum_{n=1}<w_{n}(y_{1},y_{2})> \cos n\phi_{12},
\label{ridge4}
\end{equation}
where $DA(y_{1},y_{2})$ and $w_{n}(y_{1},y_{2})$ are given by
\beq
DA(y_{1},y_{2})= <A(y_{1})A(y_{2})>-<A(y_{1})><A(y_{2})>,
\eeq
\beq
w_{n}(y_{1},y_{2}) = \frac {<W_{n}(y_{1},y_{2})>}{<A>^{2}}.
\eeq
\begin{figure}[htbp]
\centering        
\vspace*{-2.0cm}
\includegraphics[width=0.99\textwidth,height=4.5in]{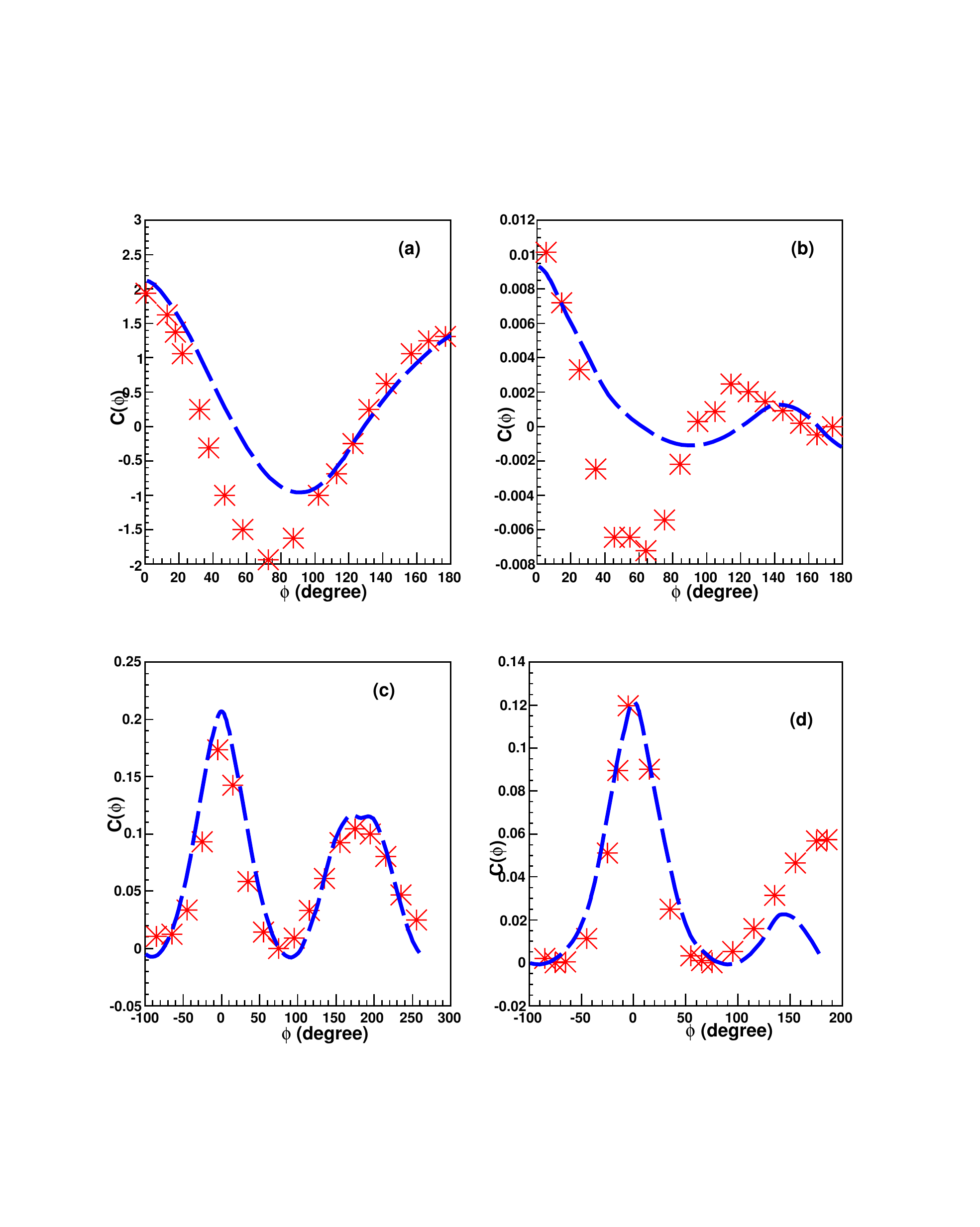}
\vspace*{-2.0cm}
\caption{ Correlation function $C(\phi)$ for (a). Au-Au at 200 GeV for 10\% of the most central events along with experimental data (b). Pb-Pb at 2.76 TeV. (c). p-Pb collisions at 5.02 TeV for central collisions (d) pp collisions at 7 TeV \cite{ridge1y}.}
\label{figridge1}
\end{figure}
The correlation contains two terms. The first is due to fluctuations in the total multiplicity and is independent of the angle. The ridge comes from the second term,which depends on the product of the initial coefficients $B_{n}$  and $C_{n}$ at different rapidities.

   In the central rapidity region the distributions are practically independent of rapidity. This creates a plateau in rapidity with the angular dependence of correlations given by
\begin{equation}
C(\phi_{12})=\frac {DA}{<A>^{2}} + 2\sum_{n=1}w_{n} \cos n\phi_{12}
\label{ridge5}
\end{equation}
with $DA$ = $<A^{2}>-<A>^{2}$, and $w_{n} = \frac {<B_{n}^{2}+C_{n}^{2}}{<A>^{2}}$.
 If one neglects fluctuations of the multiplicity at a given centrality and assumes 
 $<A^{2} >$ = $<A>^{2}$ then one finds $w_n$ = $<v_{n}^{2}>$, the average of the individual flow coefficients squared. This should be compared with the normally defined flow coefficients which are obtained by averaging of individual coefficients themselves. If fluctuations both in $A$ and in $v_{n}$  are negligible one can take $ <v_{n}^{2}> = <v_{n}>^{2} = v_{n}^{2}$ and thus find the ridge directly from the flow coefficients. 

 The general scheme of the calculations of $C(\phi_{12})$, Eq. (\ref{ridge5}), follows the Monte-Carlo code procedure used in the evaluations of the flow coefficients.
In this case the color and tension of the clusters formed from $n$ original strings are taken to be $\sqrt n$   greater than the original strings
without paying attention to the area of the formed cluster. This simplification is done to save computation time to do feasible the evaluations. In this way 
one computes the single and double inclusive cross-section  and averaging them finds  coefficients $v_{n},  w_{n}$  and the correlation $C(\phi_{12})$.

  Figures \ref{figridge1}(a) and (b) show the results \cite{ridge1y} of the correlation coefficient C($\phi$) for Au-Au at 200 GeV for  0-10\% centrality and Pb-Pb at 2.76 TeV for the most central events together the experimental data \cite{ridge2,par27}. A good agreement is observed in the regions around $\phi =0$ and $\pi$. It worsens at intermediate azimuthal angles.

  In pPb collisions we assume that the maximum number of strings attached to the nucleons of the target to be 18 at energies in the region 5-7 TeV, in accordance with the results on the multiplicity on pp collisions. The number of strings attached to the projectile proton will be $A^{1/3}$  times larger. The resulting $w_{n}$  and $C(\phi)$ are similar for energies 62.4, 200 GeV and 5.02 TeV. In Fig. \ref{figridge1}(c) the correlation coefficient at 5.02 TeV for central collisions compared to the data \cite{par28}  is shown. A very good agreement is observed.

  String percolation describes $ pp$ collisions in the same way as pA and AA collisions, therefore the same Monte-Carlo code procedure can be applied. From the multiplicity evaluations, it is known that the average number of formed strings at 62.4, 200 GeV and 7 TeV is 3, 4 and 9 respectively. For the distribution of hadronic matter in the proton a Gaussian of radius 0.8 fm is assumed. In Fig. \ref{figridge1}(d) we show the correlation coefficient $C(\phi)$ for pp collisions at 7 TeV with triple multiplicity compared to the experimental data \cite{par29}. We again observe a very good agreement.
\begin{figure}[htbp]
\centering        
\vspace*{-0.2cm}
\includegraphics[width=0.95\textwidth,height=2.5in]{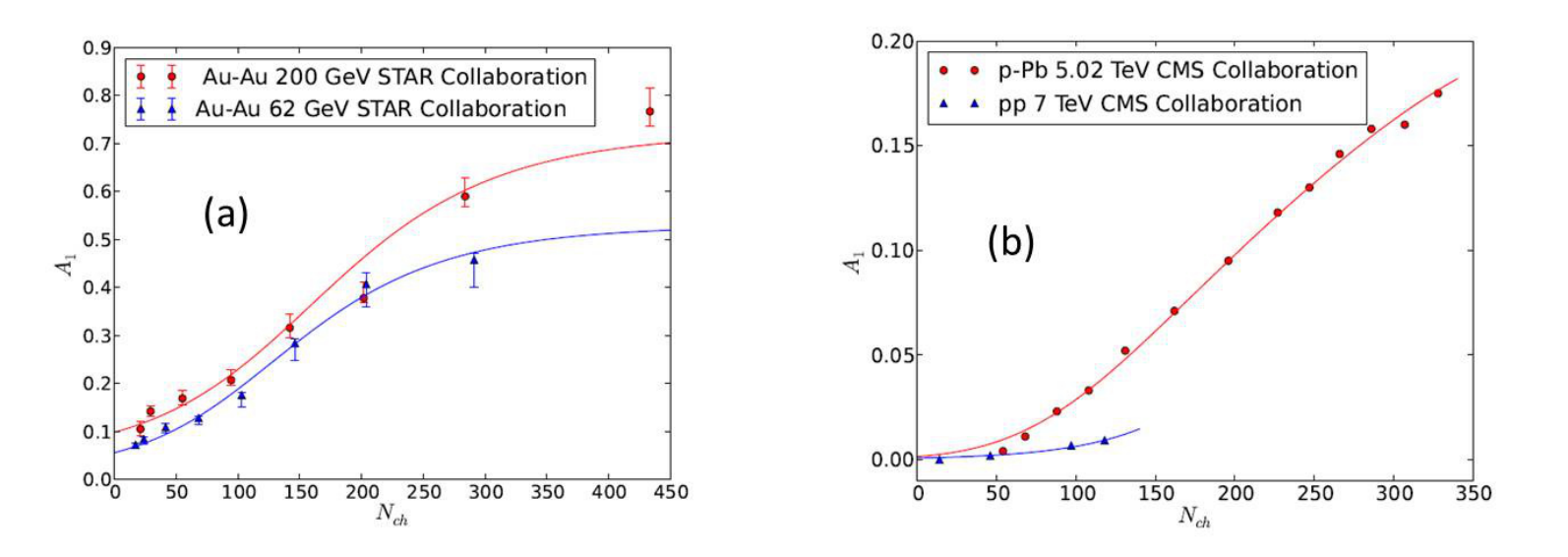}
\caption{CSPM results on the strength of the near-side ridge for (a). Au-Au collisions at two RHIC energies. (b) pp collisions at 7 TeV and p-Pb collisions at 5.02 TeV \cite{ridge7}.}
\label{figridge2}
\end{figure}

  Comparison with data at RHIC and LHC shows a good qualitative agreement in all cases. This agreement is also quantitative in cases of high multiplicity $pp$  and pPb collisions at 7 and 5.04 TeV respectively. The agreement with AA is reasonably good on the near side and on the away side but worse at intermediate angles. The reason for this probably has to do with simplifications in the simulations done to make calculations feasible. These simplifications  eliminate part of the fluctuations and  influence on the magnitude of  higher harmonics. For instance, the multiplicity of a cluster was taken $\sqrt n$  instead of $\sqrt {nS_{n} /S_{1}}$ times the multiplicity of the original strings.
  Concerning the pseudorapidity dependence and the strength of the near side ridge structure together with the onset of it, the experimental data show an interesting behavior. The pp ridge structure only appears for charged multiplicity larger than 110 in pp collisions \cite{par29} and around 50 in pPb collisions \cite{par28}. On the other hand, the RHIC data on Au-Au indicate a change in the strength of the near side ridge structure at a given multiplicity \cite{ridge6}. All these facts can be naturally explained in string percolation \cite{ridge7}. In fact, the ridge structure in pp at high multiplicity was anticipated by string percolation \cite{ridge8}.

It can be shown that the strength of the ridge structure $A_1$
defined as in ~\cite{ridge6} is proportional to
\begin{equation}
\frac{dN}{dy} \frac {<n(n-1)>-<n>^{2}}{<n>^{2}}, 
\end{equation}
which according to Eq. (\ref{pte10}) 
 of the previous section becomes $<N>/k$ where $N=N_sF(\xi)$ is the number of effective clusters and \\
 $k=<N>/(1-exp(-\xi))^{3/2}$ the negative binomial parameter. Hence the strength of the near-side structure, $A_{1}$, is proportional to
\begin{equation}
A_{1} = \frac {<N>}{k} = C ( 1 - \exp(-\xi))^{3/2}.
\end{equation}
The factor $ 1 - \exp(-\xi)$ stands  for an homogeneous profile. In the case of more realistic profiles as Gaussian or Wood-Saxon we should use Eq.(\ref{per4}),
 thus
\begin{equation}
A_{1} = C  \frac {1}{\Big(1 +a \exp(-(\xi-\xi_{c}))/b\Big)^{3/2}},
\end{equation}
where $C$ is a normalization constant which is different for each type of collision.

  Comparison of the experimental data for the dependence of Au-Au at 62.4 and 200 GeV on the multiplicity \cite{ridge6} and pPb and pp at 5.04 and 7 TeV respectively \cite{par28} is shown in Figs. \ref{figridge2}(a) and (b). In these figures the parameters $a$ and $\xi_{c}$   are kept fixed, with values 1.5 and 1.5 respectively (with these values one obtains the correct dependence of the fraction of the area covered by strings  on the string density). The values of $ b$ are 0.75 for Au-Au collisions and 0.35 for both pPb and pp collisions. We would expect $b\sim 1/R$, with $R$ the transverse dimension of the collision area, which is not far from the values found. Also a good agreement is found \cite{ridge7} for the behavior of the width in pseudorapidity of the ridge structure as a function of the multiplicity in Au-Au collisions at 62.4 and 200 GeV.

  We  conclude that string percolation is able to describe the ridge structures seen in pp high multiplicity, pPb and AA collisions. In this framework the ridge is obtained from the superposition of many events with different numbers and types of clusters of strings. In order to obtain anisotropy it is crucial to introduce  quenching of the emitted partons in the strong color field inside
the clusters of strings. In this way, it can be seen that the ridge structure in produced by an interplay of initial stage effects (formation of different clusters) and final state effects described by  quenching.

\subsection{Bose-Einstein correlations (BEC)}
Studies of correlations between two identical particles have been very fruitful to determine the extension of the source of multiparticle production in collisions of two hadrons. This has been used extensively in heavy ion experiments at RHIC at LHC ~\cite{be1,be2,be3,be4,be5,be6,be7}.
 These correlations provide information not only on the geometrical structure but also on the degree of coherence of the emitted particles.
% The correlation flmction is given by
%\beq
%C_{2}(p_{1},p_{2})= \frac{\rho(p_{1},p_{2})}{\rho(p_{1})\rho(p_{2})}
%\eeq
%where $\rho(p1,p2$ and $\rho(p)$ are the double and single inclusive
%distributions.
%Usually $C_2$ isparametrized by a Gaussian
%\beq
%C_{2}(q)= 1+\lambda exp(-(R_{in}^{2}q_{in}^{2}+R_{out}^{2}q_{out}^{2}+R_{long}^{2}q_{long}^{2}))
%\eeq
%with $\lambda$ describing the correlation  strength. 
The correlation strength is characterized by a parameter $\lambda$, which
can  also be interpreted as a measure of chaoticity or the degree of coherence of the collision ~\cite{be8,be9,be10}. In this interpretation $\lambda$ = 1 signals totally chaotic emission whereas $\lambda$ = 0 means radiation in a coherent way. However, this interpretation should be taken with caution. In fact, in $e^{+}-e^{-}$ annihilation, $\lambda$ is close to one at energies where there is no production of three or more jets and in the more complex pp collisions $\lambda$ decreases with increasing multiplicity. These observations would apparently suggests a systematic increase of the coherence from $e^{+}-e^{-}$ to pp collisions,\
which does not seem  reasonable.

There are many difficulties in extracting the value of $\lambda$ from measurements (disentanglement from the Coulomb repulsion, corrections from long lived resonances, extrapolation to low $p_{t}$, corrections due to experimental transverse resolution).
 However, the existing measurements at ISR, SPS, RHIC and LHC give valuable information which should be understood.
The experimental data on $\lambda$ have been obtained in different kinematic situations and also assuming  different  extrapolations, corrections and  parametrizations, which  makes rather  difficult  their  comparison  with  theoretical  models or  even  with different data. However some trends can be distinguished.
First, for a not  very large number of collisions, the data at SPS with p and Oxygen projectiles  show a decrease of $\lambda$ with  multiplicity \cite{be11,be12,be13}. As the number of  collisions increases the behavior  of $\lambda$  changes. It no longer decreases with  multiplicity  and even may increase. This is true at LHC where $\lambda$ reaches the values of 0.7-0.74. At SPS energies, the $\lambda$ values found  are larger at forward than at central rapidity (Notice that the particle multiplicity is larger at central  than  in forward  rapidity). This behavior 
can be	naturally understood in the framework of percolation of strings \cite{be14,be15} . 

\begin{figure}[htbp]
\centering        
\vspace*{-0.2cm}
\includegraphics[width=0.65\textwidth,height=3.0in]{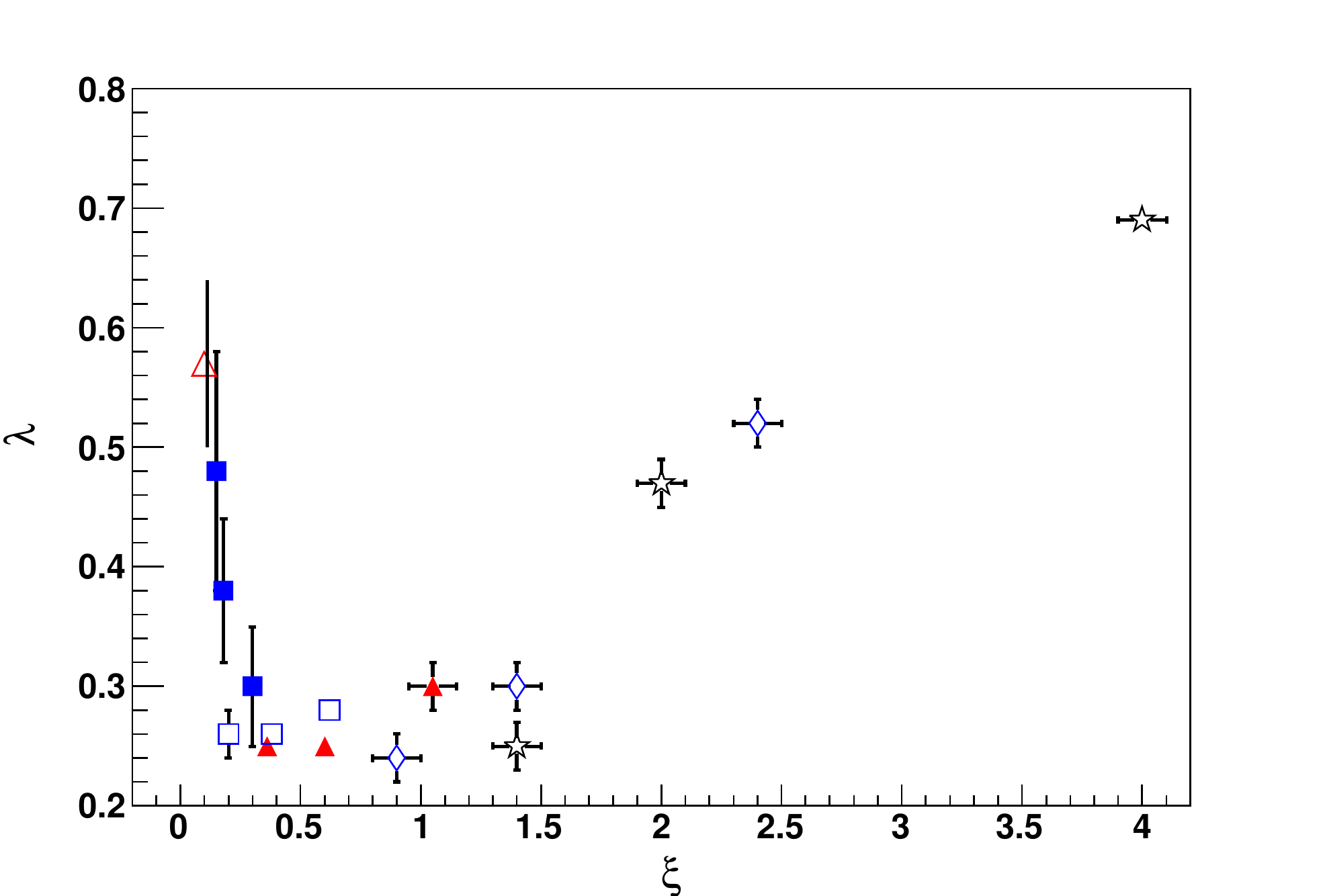}
\caption{Dependence of $\lambda$ on $\xi$ for different nucleus-nucleus collisions with percolating strings \cite{be14}.}
\label{figbe}
\end{figure}

The strings of the Lund type correspond to totally chaotic sources \cite{be17,be18}, $\lambda$ = 1, and usually it is assumed that there is no BEC from a pair of particles emitted from different strings (a discussion on this assumption can be seen in \cite{be19}). Under these assumption one finds
 \cite{be20}
\beq
\lambda = \frac{n_{s}}{n_{tot}}
\eeq
where $n_{s}$ is the mean number of identical particle pairs produced by fragmentation of the same string in a given rapidity and transverse momentum range and  $n_{tot}$ is the total number of identical pairs produced in the same kinematic range. The number of pairs of identical particles produced by each cluster is
\beq
n_{s}= \frac{1}{2}\mu_{1}^{2}<\sum_{n=1}^{N_{s}} \frac {a_{n}nS_{n}}{S_{1}}>
\eeq
and the total number of pair of identical particles produced is
\beq
n_{tot}= \frac{1}{2}\mu_{1}^{2}<\left (\sum_{n=1}^{N_{s}} \frac {a_{n} \sqrt nS_{n}}{S_{1}}\right)^{2}>
\eeq
where $a_{n}$ is the number of clusters with $ n$ strings, $\mu_{1}$ number of particles produced by a single string in the considered range.
A Monte-Carlo simulation taking into account energy conservation  was done at several energies and different projectiles and targets \cite{be14}. It was found that there is an approximate scaling in the string density, i.e. the chaoticity $\lambda$ only depends on it. In Fig. \ref{figbe} values  of $\lambda$ are shown as a function of the string density $\xi$. The details of the Monte-Carlo calculations together the SPS experimental results and the predictions for RHIC and LHC energies are shown in Ref. \cite{be14}. We observe a good agreement with the  data in particular with the change of the behavior with multiplicity. Notice that the value of $\lambda$ = 0.7 predicted at LHC 
(last point in Fig. (\ref{figbe}))  was confirmed by the LHC data \cite{be5,be6,be7}.
The three body BEC have been also studied in the framework of percolation \cite{be15}, predicting the strength of the genuine three particle Bose-Einstein correlations and obtaining  a general agreement with  experimental data ~\cite{be7,be22,be23}.

\subsection{J/$\Psi$ Suppression}
Melting of  the J/$\Psi$  due to the Debye screening in a deconfined medium was one of the first proposed signatures of the existence of quark gluon plasma \cite{par8,jp2}. Very early, at SPS energies, it was found  that J/$\Psi$ was suppressed in S-U and O-U collisions \cite{jp3,jp4}. However this suppression was naturally explained by absorption of the pre-resonant cc state in the nucleus  \cite{jp5,jp6}. From that time, many efforts have been devoted to disentangle the suppression due to nuclear effects from the actual melting in a deconfined medium. The RHIC and the LHC experimental data pointed out that, indeed, in addition to the nuclear effects there is a sequential suppression of J/$\Psi$ \cite{jp7} combined with an enhancement due to a recombination mechanism \cite{jp8} and to the existence of strong color fields above a certain matter density. On the other hand, ALICE collaboration \cite{jp9}  have shown that  J/$\Psi$ production in pp high multiplicity events,
 normalized to the minimum bias, as a function of the multiplicity in the central  rapidity region also normalized to the minimum bias increases linearly or even more rapidly at high multiplicity.
\begin{figure}[htbp]
\centering        
\vspace*{-0.2cm}
\includegraphics[width=0.65\textwidth,height=3.0in]{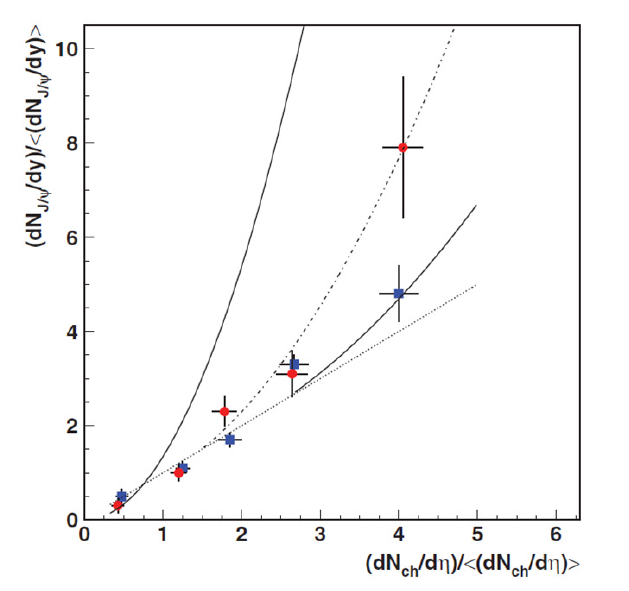}
\caption{Our results for pp collisions in the central ($|y| < 0.9$, dashed line) and forward (2.5 $< y < $4, dotted line) rapidity range, together with the experimental data for the central (circles) and forward (squares) rapidity regions from the ALICE collaboration \cite{jp9}. The linear behavior(solid line) and our prediction for p-Pb collisions(solid line) at 7 TeV are also plotted \cite{jp10}.}
\label{figjpsi}
\end{figure}

In string percolation it is expected that above the critical percolation threshold the J/$\Psi$ or at least the $\Psi'$s are suppressed. However, the string density reached at 7 TeV is much smaller than the percolation critical value even for high multiplicity events reached in the experiment. The behavior found by the ALICE collaboration can be naturally explained in the framework of string percolation \cite{jp10}. In fact, assuming that, as in  any hard process, the number of produced J/$\Psi$ is proportional to the number of collisions and thus is proportional to the number of strings $N_{s}$, we can write
\beq
\frac {n_{J/\Psi}}{<n_{J/\Psi}>} = \frac {N_{s}}{<N_{s}>}.
\label{jpsi1}
\eeq
On the other hand we have
\beq
\frac {dN/dy}{<dN/dy>}=\frac{N_{s}F(\xi)}{<N_{s}>F(<\xi>)}.
\label{jpsi2}
\eeq
From above equations we have
\beq
\frac {dN/dy}{<dN/dy>}=\left (\frac {n_{J/\Psi}}{<n_{J/\Psi}>}\right )^{1/2}
\left (\frac{1-\exp(-\frac{n_{J/\Psi}<\xi>}{<n_{J/\Psi}>}}{1-\exp(-<\xi>)}\right)^{1/2}.
\label{jpsi3}
\eeq
At low multiplicities where the number of strings $N_{s}$ is small, the above equation gives rise to the linear dependence
\beq
\frac {n_{J/\Psi}}{<n_{J/\Psi}>} = \frac {dN/dy}{<dN/dy>}.
\label{jpsi4}
\eeq
On the other hand, at high multiplicities, the bracket on the right hand side of Eq. (\ref{jpsi3}) can be approximated by $<\xi>^{-1/2}$. One obtains in this case
\beq
\frac {n_{J/\Psi}}{<n_{J/\Psi}>} =<\xi>\left (\frac{dN/dy}{<dN/dy>}\right )^{2}.
\label{jpsi5}
\eeq
Thus the linear dependence obtained previously changes to quadratic when the high multiplicity events are in play. At much higher multiplicity, about 7-8 times minimum bias we predict a rapid  suppression of this ratio. At 14 TeV the string density in pp increases  by about 20\% and thus at multiplicities of 5-7 
times greater than at minimum bias there will be a suppression of J/$\Psi$ free of nuclear effects.

In Fig. (\ref{figjpsi}) we show  the  results from Eq. (\ref{jpsi3}) (dashed line) in the central rapidity region together the experimental data \cite{jp9} (circles). Also  the results for the forward region (dotted line) together with the data (squares) are shown. A very good agreement is observed. Note that in the forward direction there are less number of strings, producing less multiplicity and thus the departure of the linear behavior would be at higher multiplicities.
%

%{\it END PAJARES+BRAUN+PAJARES.\\
%START OF BRIJESH}
%%%%%%%%%%%%%%%%%%%%%%%%%%%%%%%%%%%%%%%%%%%%%%
%%%%%%%%%%%%%%%%%%%%%%%%%%%%%%%%%%%%%%%%%%%%
% Brijesh write up Thermodynamics
%%%%%%%%%%%%%%%%%%%%%%%%%%%%%%%%%%%%%%%%
%%%%%%%%%%%%%%%%%%%%%%%%%%%%%%%%%%%%%%%
\section{Thermodynamic and transport properties}
A possible phase transition of strongly interacting matter from hadron to
quark-gluon plasma state has received considerable interest in the past. What conditions are necessary in the pre-equilibrium stage to achieve
deconfinement and perhaps subsequent Quark Gluon Plasma(QGP) formation ?

As discussed earlier in Section 2.2 the multiplicity
$\mu$ and the mean transverse momentum squared $\langle p_{t}^{2} \rangle$ of the particles produced by a cluster of ${\it n} $ strings \cite{pajares2}
\begin{equation}
\mu_{n} = \sqrt {\frac {n S_{n}}{S_{1}}}\mu_{1};\hspace{5mm}
%\end{equation}
%\begin{linenomath*}
 %\begin{equation}
\langle p_{t}^{2} \rangle = \sqrt {\frac {n S_{1}}{S_{n}}} {\langle p_{t}^{2} \rangle_{1}}
\end{equation}
%\end{linenomath*}
where $\mu_{1}$ and $\langle p_{t}^{2}\rangle_{1}$ are the mean
multiplicity and $\langle p_{t}^{2} \rangle$ of particles produced
from a single string with a transverse area $S_{1} = \pi r_{0}^2$.
In the thermodynamic limit, one obtains an analytic expression \cite{brapaj2000,pajares2}
\begin{equation}
\langle \frac {n S_{1}}{S_{n}} \rangle = \frac {\xi}{1-e^{-\xi}}\equiv \frac {1}{F(\xi)^2};\hspace{5mm}
F(\xi) = \sqrt {\frac {1-e^{-\xi}}{\xi}}
\label{ther1}
\end{equation}
Here
$\xi = \frac {N_{s} S_{1}}{S_{N}}$ is the percolation density parameter
with $N_S$ and $S_N$ being the total number of strings and interaction area respectively.
$F(\xi)$ is the color suppression factor and is related to $\mu_{n}$ and $\langle p_{t}^{2}\rangle_{n}$ through the relation  $\mu_{n}=F(\xi)\mu_{0}$,  $\langle p_{t}^{2}\rangle_{n} ={\langle p_{t}^{2} \rangle_{1}}/F(\xi)$.

 In this section thermodynamic and transport properties are discussed in the framework of clustering of
color sources at RHIC and LHC energies. 
The thermodynamics of clustering can be addressed by extracting
the temperature from the transverse momentum spectra of charged hadrons.
In the model a local temperature can be introduced. In fact the tension
of a cluster of  strings fluctuates around its mean value because the
chromoelectric field is not universal. Assuming a Gaussian form for these
fluctuations a thermal spectrum  is obtained with a temperature
inversely proportional to $\sqrt {2F(\xi)}$. It is remarkable that for
$\xi=1.2-1.5$  the corresponding temperature $T_{c}$ is not far from the
critical temperature  \cite{eos}. 
We extract the color suppression factor $F(\xi)$ using experimental data at RHIC energies.
 $F(\xi)$ and $\xi$ permit the determination of the initial temperature ${\it T}$ and the Bjorken initial energy density $\epsilon$.  Energy density $\epsilon$ was found to be a linear function of $\xi$. 
 As a result one can derive the equation of state and the transport coefficient  shear viscosity over entropy density ratio as a function of the temperature.
A good agreement between the model and lattice QCD results for energy density is obtained for $ 1 < T/T_{c} < 2$ \cite{lattice12}. 
% viscosity
Relativistic kinetic theory  expression for shear viscosity has been used to obtain shear viscosity over entropy ratio $\eta/s$ at RHIC and LHC energies \cite{eos2}.
$\eta/s$ has a minimum at the temperature close to the critical one,
rising slowly as the temperature increases.
The data based values are compared with theoretical values for both a weakly coupled QGP (wQGP) and a strongly coupled QGP (sQGP).
% trace anaomaly
The dimensionless quantities, the trace anomaly $\Delta = (\epsilon-3p)/T^{4}$ and the inverse of $\eta/s$, have identical  magnitudes  for $ 0.5 < T/T_{c} < 6$ which describes the transition from wQGP to sQGP and then back to wQGP. This 'Ansatz' provides an independent value for pressure at a given $\epsilon$ \cite{cpod13}. The sound velocity $C_{s}^{2}$ obtained using this pressure is in excellent agreement with the LQCD simulations \cite{lattice12}.

\subsection{Experimental determination of the color suppression factor $F(\xi)$}
 To evaluate the initial value of $\xi$ from data, a parameterization of p-p events at 200 GeV  is used to compute the $p_{t}$ distribution \cite{nucleo}
\begin{figure}[thbp]
\centering        
\vspace*{-0.2cm}
\includegraphics[width=0.70\textwidth]{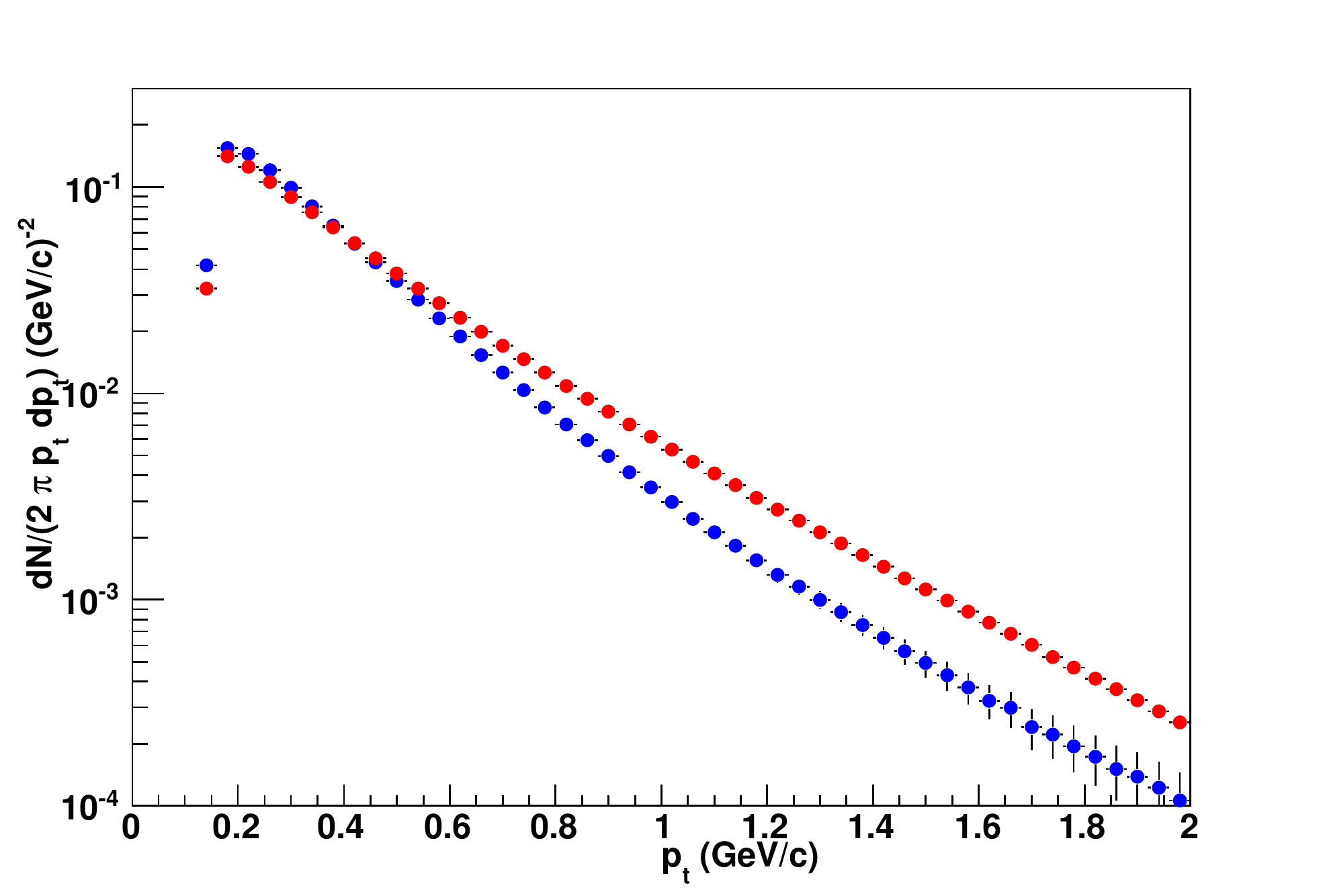}
%\vspace*{-0.5cm}
\caption{ Charged particle $p_{t}$ spectrum from pp collisions (blue color) and Au-Au collisions at 200 GeV for 0-10\% centrality \cite{levente}.} 
\label{ptspe}
\end{figure}
\begin{equation}
dN_{c}/dp_{t}^{2} = a/(p_{0}+p_{t})^{\alpha},
\end{equation}
where a, $p_{0}$, and $\alpha$ are parameters used to fit the data. 
Fig. \ref{ptspe} shows the normalized $p_{t}$ spectrum for pp collisions at $\sqrt s = $ 200 GeV. A fit to the data gives  $p_{0}$ = 1.982 and $\alpha$ = 12.877.
This parameterization is used for nucleus-nucleus collisions to take into account the interactions of the strings \cite{nucleo}
\begin{equation}
p_{0} \rightarrow p_{0} \left(\frac {\langle nS_{1}/S_{n} \rangle_{Au-Au}}{\langle nS_{1}/S_{n} \rangle_{pp}}\right)^{1/4},
\end{equation}
where $S_{n}$ corresponds to the area occupied by the $n$ overlapping strings.
The thermodynamic limit, i.e. letting $n$ and $S_{n}$ $\rightarrow \infty$  while keeping $\xi$ fixed, is  used to evaluate
\begin{equation}
\langle \frac {nS_{1}}{S_{n}} \rangle = 1/F^{2}(\xi)
\end{equation}
\begin{equation}
dN_{c}/dp_{t}^{2} = \frac {a}{{(p_{0}{\sqrt {F(\xi_{pp})/F(\xi)}}+p_{t})}^{\alpha}}.
\end{equation}
In pp collisions  $ \langle nS_{1} / S_{n} \rangle_{pp}$ $\sim$ 1  due to the low string overlap probability. The STAR analysis of charged hadrons for Au-Au collisions at $\sqrt{s_{NN}}$ = 200 and 62 GeV for  $\xi$  is shown in Fig. \ref{xinpart} as a function of number of participants \cite{nucleo}. It is observed that for all centralities, except for the most peripheral one, $\sqrt{s_{NN}}$ = 200 GeV data lie above the critical percolation threshold $\xi_{c}$ = 1.2. The value of $\xi$ increases with increasing collision centrality, an expected indication of additional string overlap in more central collisions. For  $\sqrt{s_{NN}}$ = 62.4 GeV, almost all centralities, except for the three most peripheral one lie above $\xi_{c}$. 

\begin{figure}[thbp]
\centering        
\vspace*{-0.2cm}
\includegraphics[width=0.65\textwidth,height=3.0in]{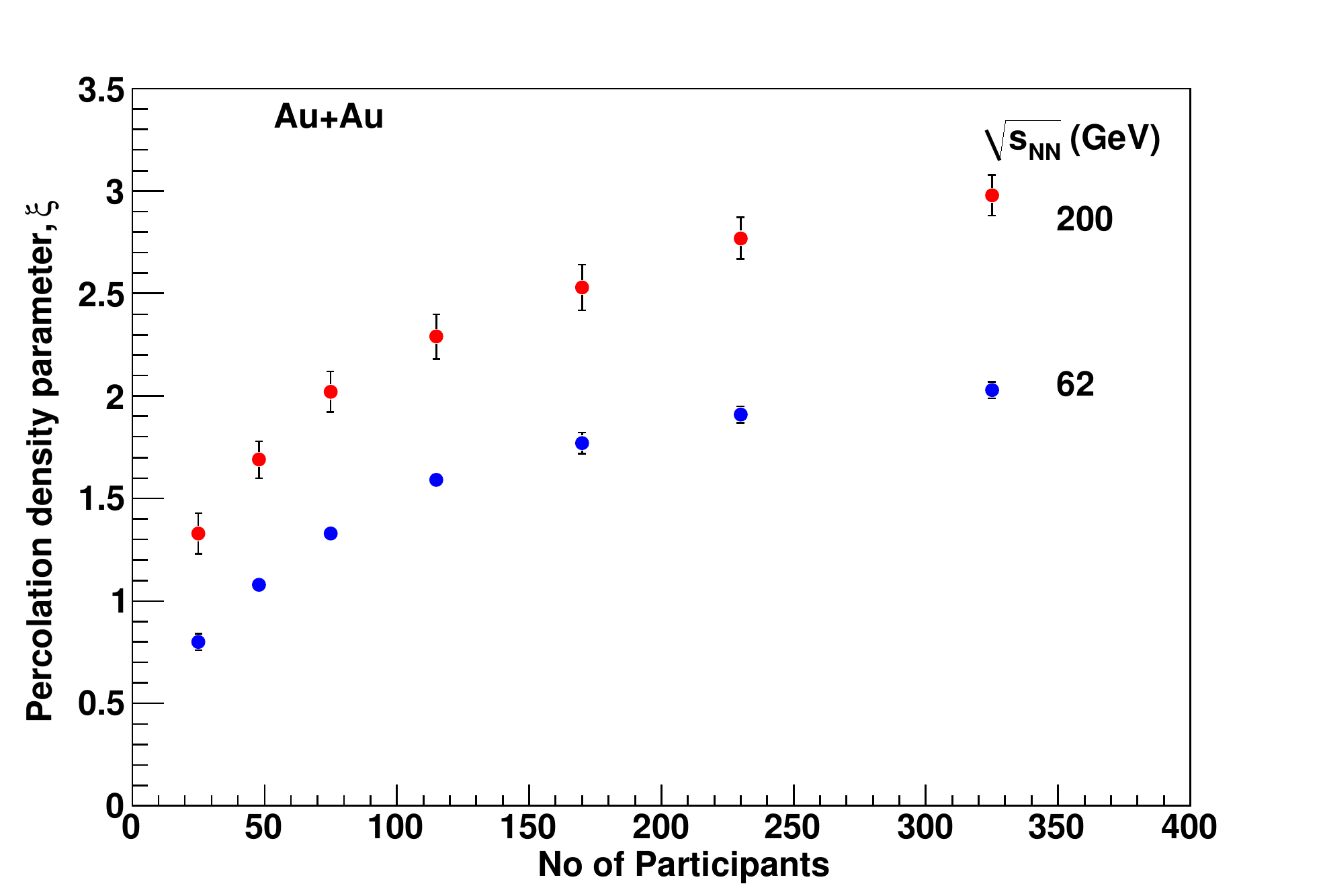}
\vspace*{-0.5cm}
\caption{Percolation density parameter $\xi$  as a function of number of participants for Au-Au collisions at  $\sqrt{s_{NN}}=$ 62 and 200 GeV \cite{nucleo}.}
\label{xinpart}
\end{figure}
\begin{figure}[thbp]
\centering        
\vspace*{-0.2cm}
\includegraphics[width=0.65\textwidth,height=3.0in]{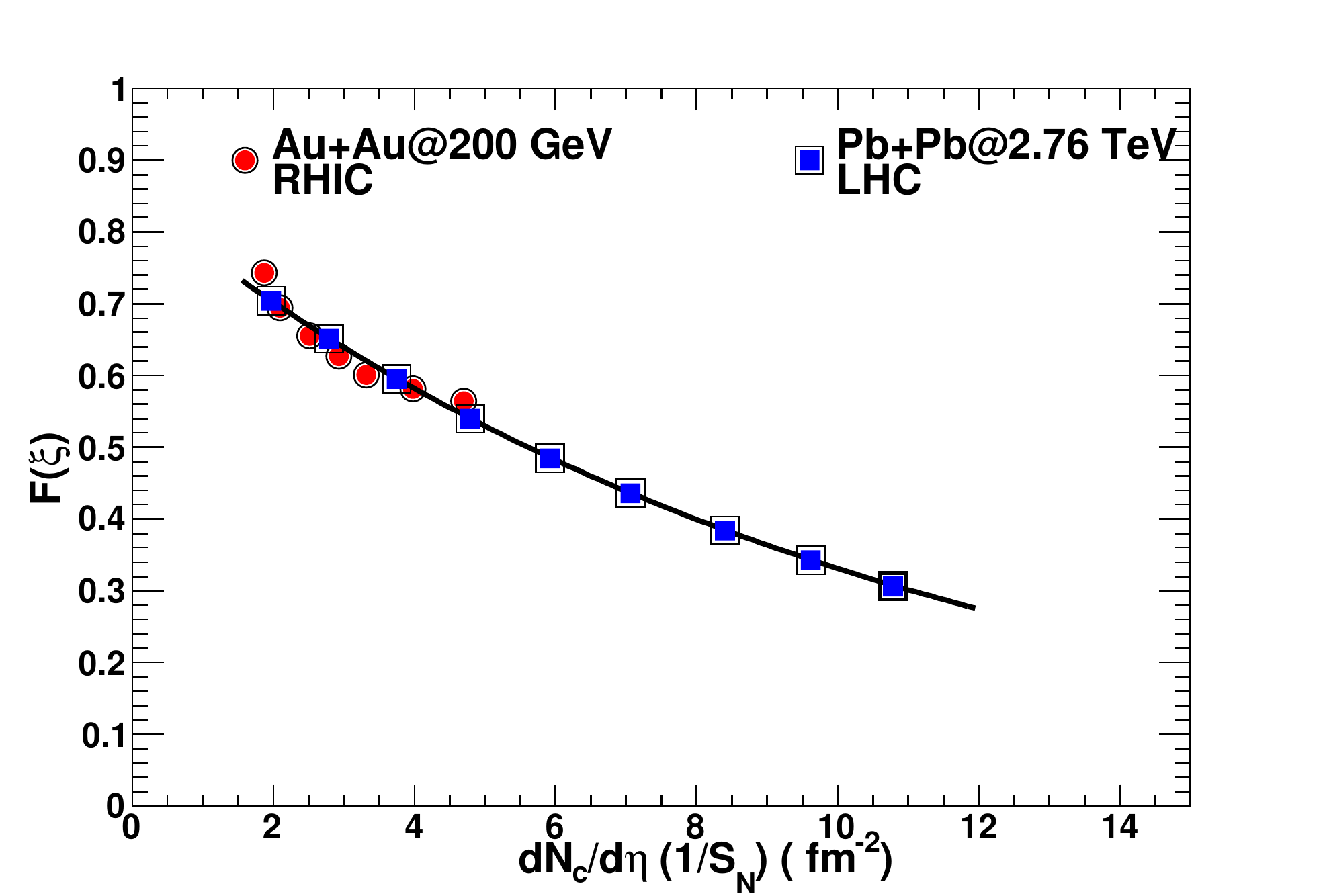}
\vspace*{-0.5cm}
\caption{Color suppression factor $F(\xi)$ as a function of $\frac {dN_{c}}{d\eta}/S_{N}(fm^{-2})$. 
The solid red circles are for Au-Au collisions at 200 GeV (STAR data) \cite{nucleo}. The error is smaller than the size of the symbol. The line is fit to the STAR data. The solid blue squares are for  Pb-Pb at 2.76 TeV \cite{eos2}.} 
\label{fetalhc}
\end{figure}

 
Figure \ref{fetalhc} shows a plot of $F(\xi)$ as a function of charged particle multiplicity per unit transverse area $\frac {dN_{c}}{d\eta}/S_{N}$ for Au-Au collisions at  $\sqrt{s_{NN}}$ = 200 GeV for various centralities for the STAR data \cite{levente,eos}. The error on  $F(\xi)$ is $\sim 3\%$. 
$F(\xi)$ decreases in going from peripheral to central collisions. The $\xi$ value is obtained using Eq. (\ref{ther1}), which increases with the increase in centrality. The fit to the Au-Au points has the functional form 
\begin{equation}
F(\xi) = \exp [-0.165 -0.094 \frac {dN_{c}}{d\eta}/S_{N}].
\label{eqfetalhc}
\end{equation}
Recently, the ALICE experiment at LHC published the charged-particle multiplicity density data as a function of centrality in Pb-Pb collisions at $\sqrt{s_{NN}}$ = 2.76 TeV \cite{alice1}. The STAR results for Au-Au collisions at $\sqrt{s_{N N}}$ = 200 GeV have been used to estimate F$(\xi)$ values for Pb-Pb collisions at different centralities using the fit function Eq. (\ref{eqfetalhc}) for Au-Au. 

\subsection{Energy density}
The QGP according to CSPM is born in local thermal equilibrium  because the temperature is determined at the string level. After the initial temperature $ T > T_{c}$ the  CSPM perfect fluid may expand according to Bjorken boost invariant 1D hydrodynamics \cite{bjorken}
\begin{equation}
\varepsilon = \frac {3}{2}\frac { {\frac {dN_{c}}{dy}}\langle m_{t}\rangle}{S_{n} \tau_{pro}},
\label{bjk}
\end{equation}
where $\varepsilon$ is the energy density, $S_{n}$ nuclear overlap area, and $\tau_{pro}$ the production time for a boson (gluon) \cite{wong}. 

The dynamics of massless particle production has been studied in QE2 quantum electrodynamics.
QE2 can be scaled from electrodynamics to quantum chromodynamics using the ratio of the coupling constants \cite{wong}. The $\tau_{pro}$ for a boson (gluon) is \cite{swinger}
 \begin{equation}
%\tau_{pro} = \frac {2.405\hbar}{mc^{2}}.
\tau_{pro} = \frac {2.405\hbar}{\langle m_{t}\rangle}.
\end{equation} 

Above the critical temperature only massless particles are present in CSPM. 
To evaluate $\varepsilon$ we use the charged pion multiplicity $dN_{c}/{dy}$ at midrapidity and $S_{n}$ values from STAR for 0-10\% central Au-Au collisions with $\sqrt{s_{NN}}=$200 GeV \cite{eos}. The factor 3/2 in Eq. (\ref{bjk}) accounts for the neutral pions. The average transverse mass $\langle m_{t}\rangle$ is given by $\langle m_{t}\rangle =\sqrt {\langle p_{t}\rangle^2 + m_{0}^2} $, where $\langle p_{t}\rangle$ is the transverse momentum of pion and $m_{0}$ being the mass of pion .

\begin{figure}[thbp]
\centering        
\vspace*{-0.2cm}
\includegraphics[width=0.65\textwidth,height=3.0in]{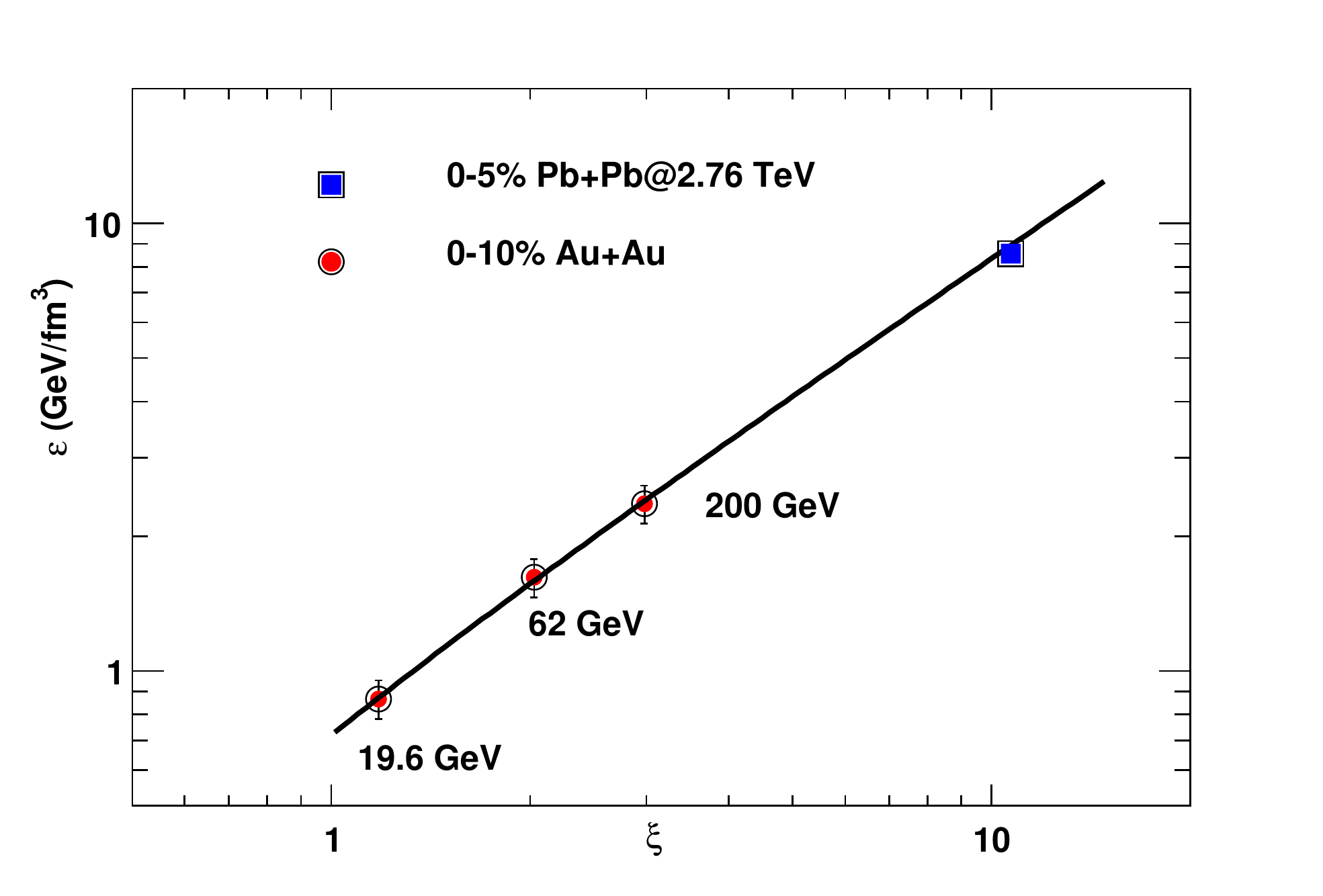}
\vspace*{-0.5cm}
\caption{Energy density $\epsilon$ as a function of the percolation density parameter $\xi$. The extrapolated value for LHC energy is shown as blue square \cite{eos2}.} 
\label{enereta}
\end{figure}

From the measured value of  $\xi$ and $\varepsilon$, as shown in Fig. \ref{enereta}, 
it is found  that $\varepsilon$ is proportional to $\xi$ for the range 
$1.2 < \xi < 2.88$, $\varepsilon = 0.788$ $\xi$ (GeV/$fm^{3}$) \cite{eos2}.  The extrapolated value of $\varepsilon$ for central Pb-Pb collision at 2.76 TeV is  8.32 $GeV/fm^3$ as  shown in Fig. \ref{enereta}.
\subsection{Temperature measurement and thermalization}
The connection between $\xi$ and the temperature $T(\xi)$ involves the Schwinger mechanism (SM) for particle production. 
The Schwinger distribution for massless particles is expressed in terms of $p_{t}^{2}$ \cite{swinger,wong}
\begin{equation}
dn/d{p_{t}^{2}} \sim \exp (-\pi p_{t}^{2}/x^{2})
\label{swing}
\end{equation}
where the average value of the string tension is  $\langle x^{2} \rangle$. The tension of the macroscopic cluster fluctuates around its mean value because the chromo-electric field is not constant.
The origin of the string fluctuation is related to the stochastic picture of 
the QCD vacuum. Since the average value of the color field strength must 
vanish, it can not be constant but changes randomly from point to point \cite{bialas,dosch}. Such fluctuations lead to a Gaussian distribution of the string tension
\beq
\frac{dn}{dp_{t}^{2}} \sim \sqrt \frac{2}{<x^{2}>}\int_{0}^{\infty}dx \exp (-\frac{x^{2}}{2<x^{2}>}) \exp (-\pi \frac{p_{t}^{2}}{x^{2}})
\eeq
 which gives rise to  thermal distribution \cite{bialas}
\begin{equation}
\frac{dn}{dp_{t}^{2}} \sim \exp (-p_{t} \sqrt {\frac {2\pi}{\langle x^{2} \rangle}} ),
\label{bia}
\end{equation}
with $\langle x^{2} \rangle$ = $\pi \langle p_{t}^{2} \rangle_{1}/F(\xi)$. 
The temperature is expressed as \cite{pajares3,eos}  
\begin{equation}
T(\xi) =  {\sqrt {\frac {\langle p_{t}^{2}\rangle_{1}}{ 2 F(\xi)}}}.
\label{temp}
\end{equation} 
%If we identified the percolation threshold with the critical phase transition temperature, one obtains
%\beq
%T_{c} =  {\sqrt {\frac {\langle p_{t}^{2}\rangle_{1}}{ 2 F(\xi_{c})}}}
%\label{temp}
%\eeq

We will adopt the point of view that the experimentally determined chemical freeze-out temperature is a good measure of the phase transition temperature, $T_{c}$ \cite{braunmun}. The single string average transverse momentum  ${\langle p_{t}^{2}\rangle_{1}}$ is calculated at $\xi_{c}$ = 1.2 with the  universal chemical freeze-out temperature of 167.7 $\pm$ 2.6 MeV \cite{bec1}. This gives $ \sqrt {\langle {p_{t}^{2}} \rangle _{1}}$  =  207.2 $\pm$ 3.3 MeV which is close to  $\simeq$ 200 MeV used previously in the calculation of percolation transition temperature \cite{pajares3}.

 Recently, it has been suggested that fast thermalization in heavy ion collisions can occur through the existence of an event horizon caused by a rapid deceleration of the colliding nuclei \cite{khar2}. The thermalization in this case is due the Hawking-Unruh effect \cite{hawk,unru}.
 
It is well known that the black holes evaporates by quantum pair production and behave as if they have an effective temperature of 
\beq
T_{H}= \frac {1}{8\pi GM},
\eeq
where 1/4GM is the acceleration of gravity at the surface of a black hole of mass M. The rate of pair production in the gravitational background of the black hole can be evaluated by considering the tunneling through the event horizon. Unruh showed that a similar effect arises in a uniformly accelerated frame, where an observer detects the thermal radiation with the  temperature T =a/2,
where $a$ is the acceleration. Similarly, in hadronic interactions the probability to produce states of masses M due to the chromoelectric field E and color charge is given by the Schwinger mechanism
\beq
W_{M} \sim \exp (\frac {-\pi M^{2}}{gE})\sim \exp(-M/T),
\eeq
which is similar to the Boltzmann weight in a heat bath with an effective temperature
\beq
T = \frac {a}{2\pi}, a= \frac {2gE}{M}.
\eeq

 In CSPM the strong color field inside the large cluster produces de-acceleration of the primary $q \bar q$ pair which can be seen as a thermal temperature by means of Hawking-Unruh effect. This implies that the radiation temperature is determined by the transverse extension of the color flux tube/cluster in terms of the string tension \cite{casto1,casto2}.
\begin{equation}
T =  {\sqrt \frac {\sigma}{2\pi}}
\label{haw}
\end{equation}
    
The string percolation density parameter $\xi$ which characterizes the percolation clusters measures the initial temperature of the system. Since this cluster 
covers most of the interaction area, this temperature becomes a global 
 temperature determined by the string density. In this way at $\xi_{c}$ = 1.2 the connectivity percolation transition at $T(\xi_{c})$ models the thermal deconfinement transition.
\begin{figure}[thbp]
\centering        
\vspace*{-0.2cm}
\includegraphics[width=0.65\textwidth,height=3.0in]{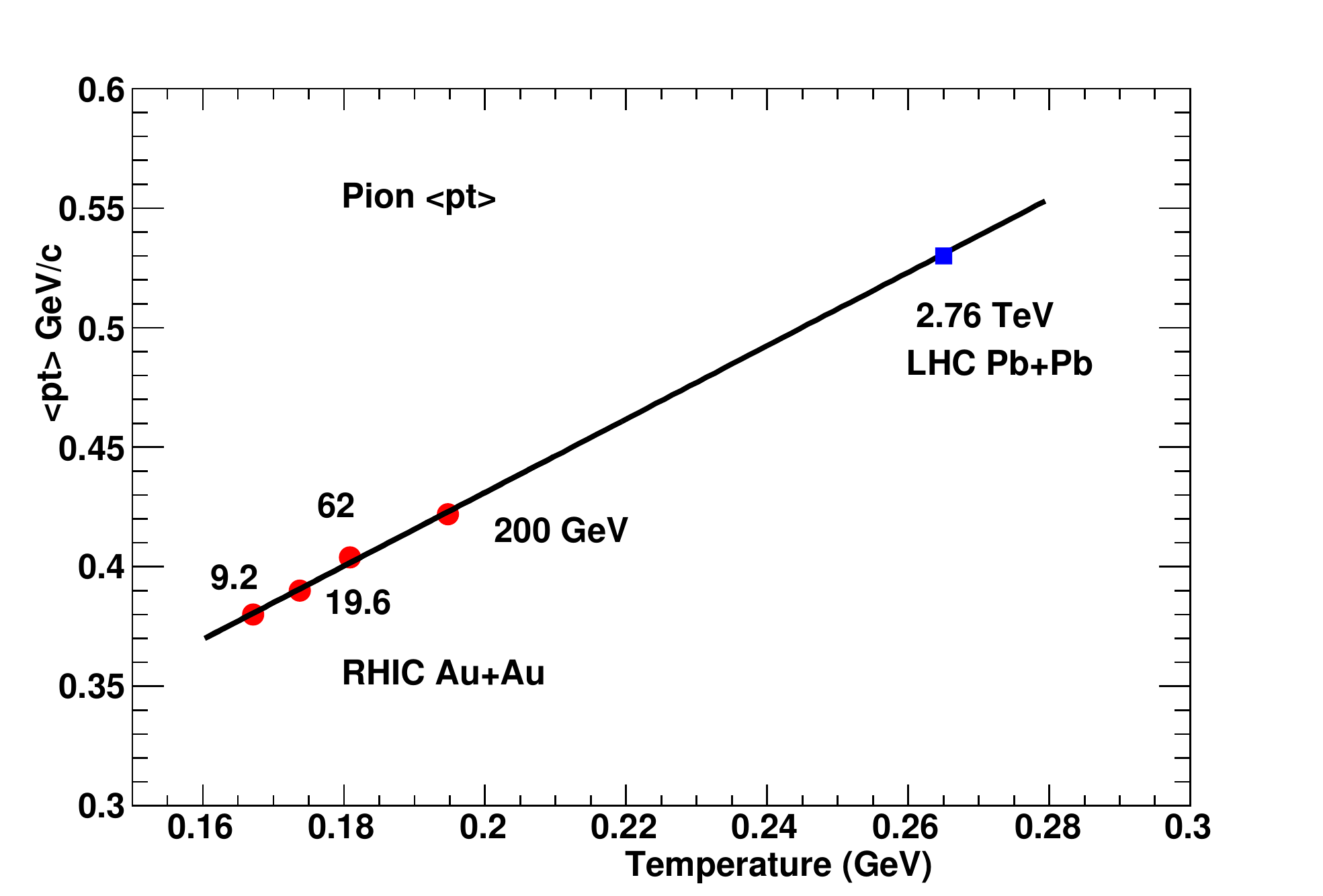}
\vspace*{-0.5cm}
\caption{Average $<p_t>$ of pions as a function of temperature obtained using Eq. (\ref{temp}).} 
\label{temppt}
\end{figure} 
The temperature obtained using Eq. (\ref{temp}) was $\sim$ 193.6 MeV for Au-Au collisions at $\sqrt{s_{NN}}$ = 200 GeV in reasonable agreement with $T_{i}$ = 221 $\pm 19^{stat} \pm 19^{sys}$ from the enhanced direct photon experiment measured by the 
PHENIX Collaboration \cite{phenix}.  
For Pb-Pb collisions at $\sqrt{s_{NN}}$ = 2.76 TeV the temperature is 
$\sim$ 262.2 MeV for 0-5$\%$ centrality, which is expected to be $\sim$ 35 $\%$ higher than the temperature from  Au-Au collisions \cite{eos}. A recent summary of the results from Pb-Pb collisions at the LHC has mentioned that the initial temperature
 increases at least by 30 $\%$ as compared to the top RHIC energy \cite{summ}. The direct photon measurements from ALICE gives the temperature of $T_{i}$  = 304 $\pm 51$ \cite{karel}. The agreement with the measured temperaure shows that the temperature obtained using Eq. (\ref{temp}) can be termed as the initial temperature of the percolation cluster. For an ultrarelativistic ideal gas $\langle p_{t} \rangle \propto T $. Figure \ref{temppt} shows a plot of $\langle p_{t} \rangle$ as a function of the temperature. When the transverse momentum is exponetially distributed with inverse slope T in a given event, $\langle p_{t} \rangle$ = 2 $T$ \cite{heiselberg}. 
Using this relation one can predict the $\langle p_{t} \rangle$ for  pions in Pb-Pb collisions at $\sqrt{s_{NN}}$ = 2.76 TeV. For 0-5 $\%$ the $\langle p_{t} \rangle$ = 0.53 GeV/c in agreement with the measured value of 0.517 $\pm$ 0.019 GeV/c \cite{alicept}.
Table I gives the CSPM  values  $\xi$, ${\it T}$, and $\varepsilon$  at $T/T_{c}$ = 0.88, 1, 1.16 and 1.57. 
% End of temperature

%
\begin{figure}[thbp]
\centering        
\vspace*{-0.2cm}
\includegraphics[width=0.65\textwidth,height=3.0in]{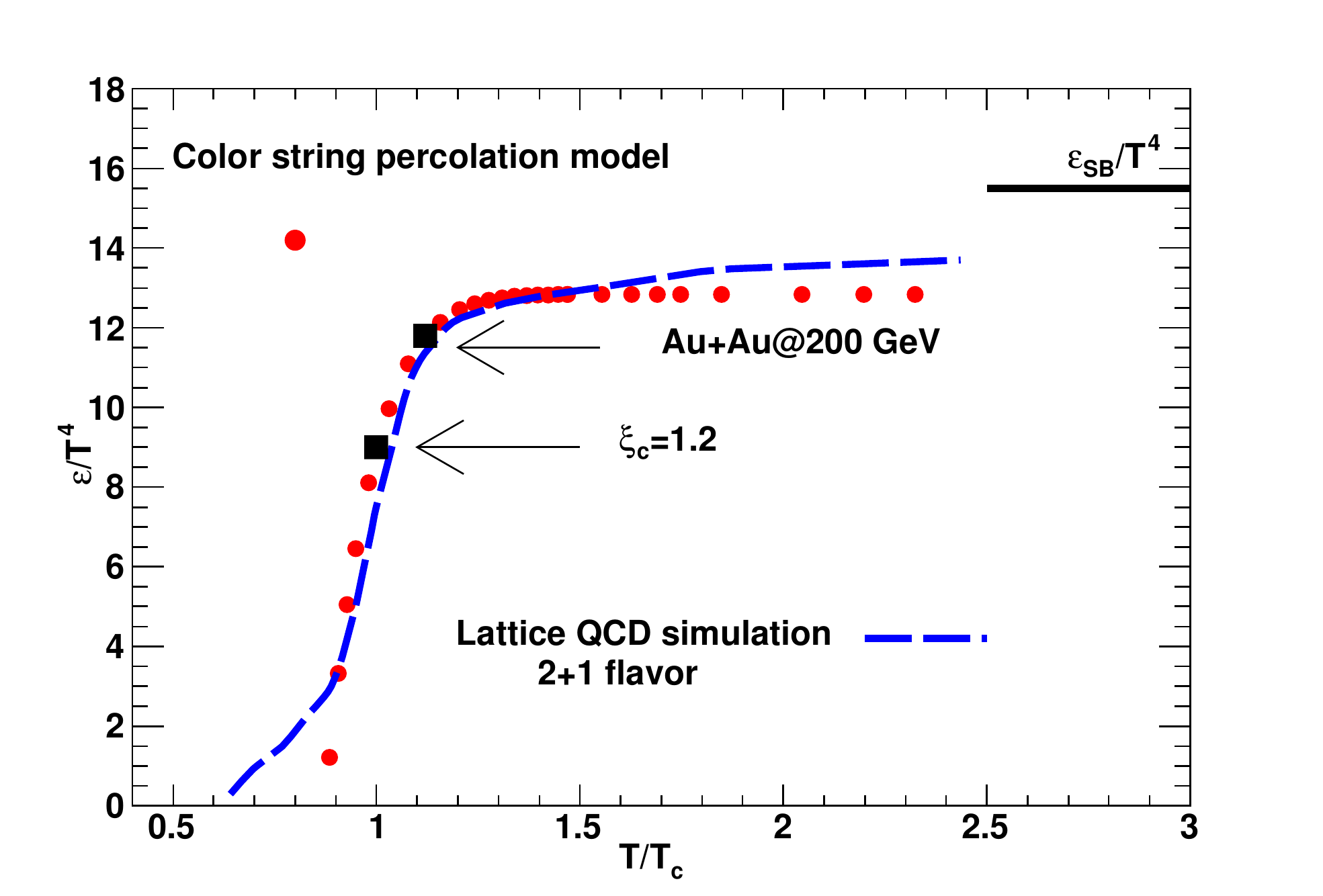}
\vspace*{-0.5cm}
\caption{$\epsilon/T^{4}$ versus $T/T_{c}$ from CSPM (red circles) and Lattice QCD (blue dash line) for 2+1 flavor and p4 action \cite{eos}.} 
\label{et4 }
\end{figure} 
\begin{table}
\caption{The percolation density parameter $\xi$, temperature ${\it T}$, $\tau{pro}$, and $\eta/s$ for meson gas \cite{prakash}, hadron to QGP transition in  Au-Au collisions at 200 GeV  and Pb-Pb at 2.76 TeV (estimated). Au-Au is for 0-10$\%$ and Pb-Pb is for 0-5$\%$ central events.}
\vspace*{0.5cm}
\begin{tabular}{|c|c|c|c|c|c|}\hline
System  & $\xi$ & T (MeV) & $\eta/s$ & $\tau_{pro}$ (fm) \\ \hline 
 Meson Gas    & 0.22 & 150.0 & 0.76 & \\ \hline
 Hadron to QGP & 1.2 & 167.7 $\pm 2.6$ & 0.240$\pm 0.012$ & \\ \hline
Au-Au    & 2.88$\pm0.09$ & 193.6 $\pm3.0$  & 0.204$\pm 0.020$ & 1.13 \\ \hline
Pb-Pb    & 10.56$\pm1.05$ & 262.2$\pm13.0$ & 0.260$\pm 0.026$ & 0.89 \\ \hline 
\end{tabular}
\label{table1}
\end{table}
%
%\section{Transport coefficient}
\subsection{Shear viscosity}
The observation of the large elliptic flow at RHIC in non-central heavy ion collisions suggest that the matter created is a nearly perfect fluid with very low shear viscosity \cite{star,phenix,phobos,brahms}. Recently, attention has been focused on the shear viscosity to entropy density ratio $\eta/s$  as a measure of the fluidity \cite{teaney,teaney1,lacey,rom}. The observed temperature averaged $\eta/s$ based on viscous hydrodynamics analyses of RHIC data , are suggestive of a strongly coupled plasma \cite{gul1,larry}. The effect of bulk viscosity is expected to be negligible. 
It has been conjectured, based on infinitely coupled super-symmetric Yang-Mills (SYM) gauge theory using the correspondence between Anti de-Sitter(AdS) space and conformal field theory (CFT), that the lower bound  for $\eta/s$ is $1/4\pi$ and is the universal minimal  $\eta/s$ even for QCD \cite{kss}. However, there are theories in which this lower bound can be violated \cite{buchel}. 

\begin{figure}[thbp]
\centering        
\vspace*{-0.2cm}
\includegraphics[width=0.65\textwidth,height=3.0in]{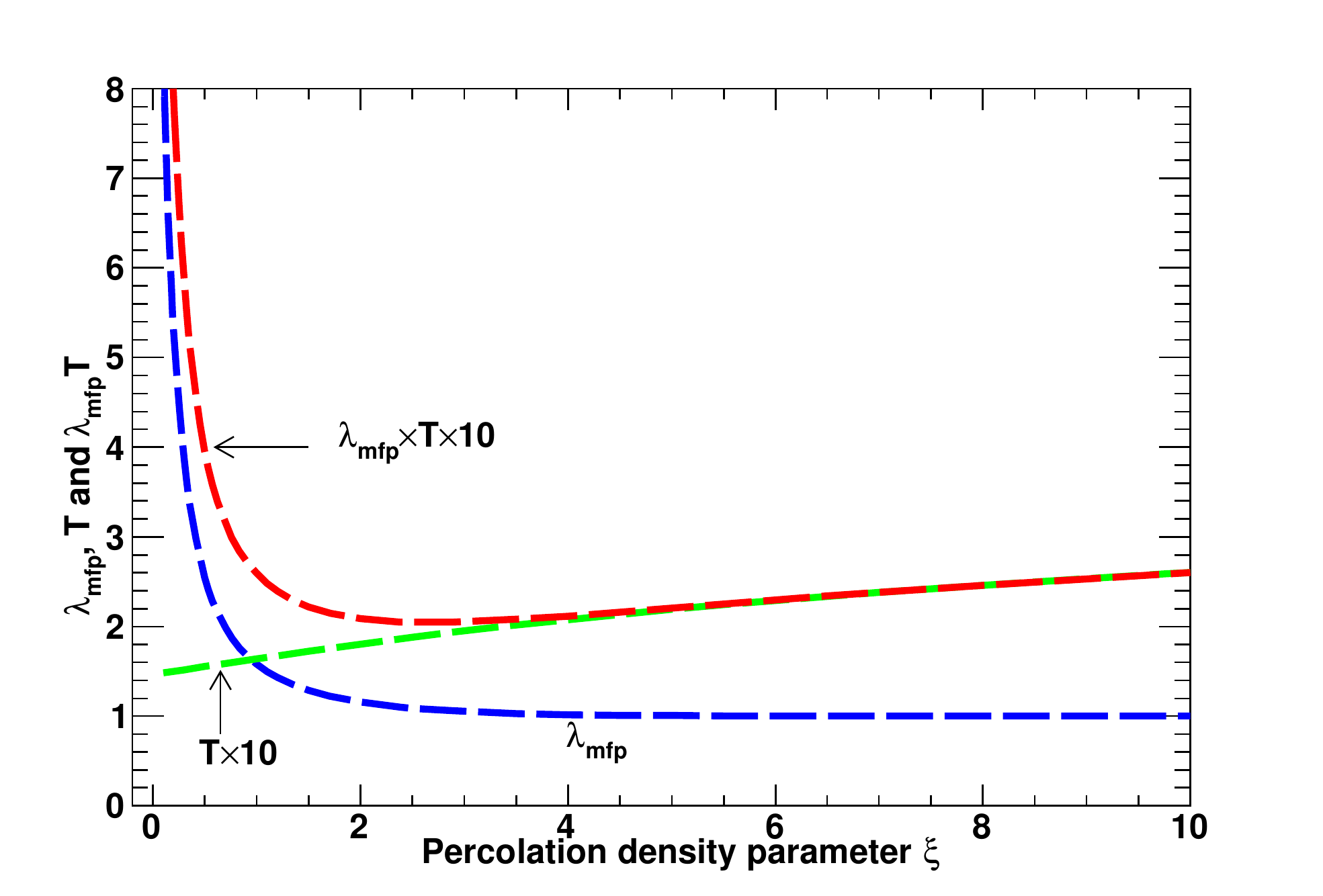}
\vspace*{-0.5cm}
\caption{$\lambda_{mfp}$ in fm ( blue line), temperature in GeV scaled by a factor of 10(green line) and $\lambda_{mfp} \times$ scaled T in GeV fm (red line) as a function of $\xi$. The minimum in $\eta/s$ is due to the combination of $\lambda_{mfp}$ and T. Meson gas \cite{prakash}, RHIC and LHC points are also shown as solid black square \cite{eos2}.} 
\label{meanfree}
\end{figure}
 The relativistic kinetic theory relation for the shear viscosity over entropy density ratio, $\eta/s$ is given by \cite{gul1,gul2}
\begin{equation}
\frac {\eta}{s} \simeq \frac {T \lambda_{mfp}}{5},     
\end{equation}
 %\begin{equation}
%\lambda_{mfp} \sim \frac {1}{(n\sigma_{tr})}
%\end{equation}
where T is the temperature and $\lambda_{mfp}$ is the mean free path. $\lambda_{mfp} \sim \frac {1}{(n\sigma_{tr})}$ where ${\it n }$ is the number density of an ideal gas of quarks and gluons and $\sigma_{tr}$ the transport cross section. 

In CSPM the number density is given by the effective number of sources per unit volume \cite{bautista,ICFP12}
\begin{equation}
n = \frac {N_{sources}}{S_{N}L}.
\label{vie1}
\end{equation}
 L is the longitudinal extension of the string $\sim$ 1 ${\it  fm} $ \cite{pajares3}. The area occupied by the strings is related to the percolation density parameter $\xi$ through the relation $(1-e^{-\xi})S_{N}$. Thus the effective no. of sources is given by the total area occupied by the strings divided by the area of an effective string $S_{1}F(\xi) $. 
\begin{equation}
N_{sources} = \frac {(1-e^{-\xi}) S_{N}}{S_{1} F(\xi)}. 
\label{vie2}
\end{equation}
 In general $N_{sources}$ is  smaller than the number of single strings. $N_{sources}$ equals the number of strings $N_{s}$ in the limit of $\xi $ = 0. 
The number density of sources from Eqs. (\ref{vie1}) and (\ref{vie2}) becomes
\begin{equation}
n = \frac {(1-e^{-\xi})}{S_{1}F(\xi) L}.
\end{equation}
The transport cross section $\sigma_{tr}$ is the transverse area of the effective string $S_{1}F(\xi)$. Thus $\sigma_{tr}$ is directly proportional to $\frac {1}{T^{2}}$, which is in agreement with the estimated dependence of $\sigma_{tr}$ on the temperature \cite{gul2,greco}. The mean free path is given by
\begin{equation}
\lambda_{mfp} = {\frac {L}{(1-e^{-\xi})}}. 
\end{equation}
 For large value of $\xi$ the $\lambda_{mfp}$ reaches a constant value.
$\eta/s$ is obtained from $\xi$ and the temperature
\begin{equation}
\frac {\eta}{s} ={\frac {TL}{5(1-e^{-\xi})}} 
\end{equation}
Below $\xi_{c}$, as the temperature increases, the string density increases and the area is filled rapidly  and $\lambda_{mfp}$ and $\eta/s$ decrease sharply. Above $\xi_{c}$, more than 2/3 of the area are already covered by strings, and therefore the area is not filling as fast and the relatively small decrease of $\lambda_{mfp}$ is compensated by the rising of temperature, resulting in a smooth increase of $\eta/s$. The behavior of $\eta/s$ is dominated by the fractional area covered by 
strings. This is not surprising because $\eta/s$ is the ability to transport momenta at large distances and that has to do with the density of voids in the matter.
Figure \ref{meanfree} shows a plot of $\lambda_{mfp}$, $ T$ and $\lambda_{mfp} \times T$ as a function of $\xi$. Thus the product  T($\xi$)$\times \lambda_{mfp}$ will have a minimum in $\eta/s$. It has been shown that $\eta/s$ has a minimum at the critical point for various substances for example helium, nitrogen and water \cite{larry}. Thus the measurement of $\eta/s$ as a function of temperature can locate the critical end point/crossover region in the QCD phase diagram $ T \sim $ 175-185 MeV.
\begin{figure}[thbp]
\centering        
\vspace*{-0.2cm}
\includegraphics[width=0.65\textwidth,height=3.0in]{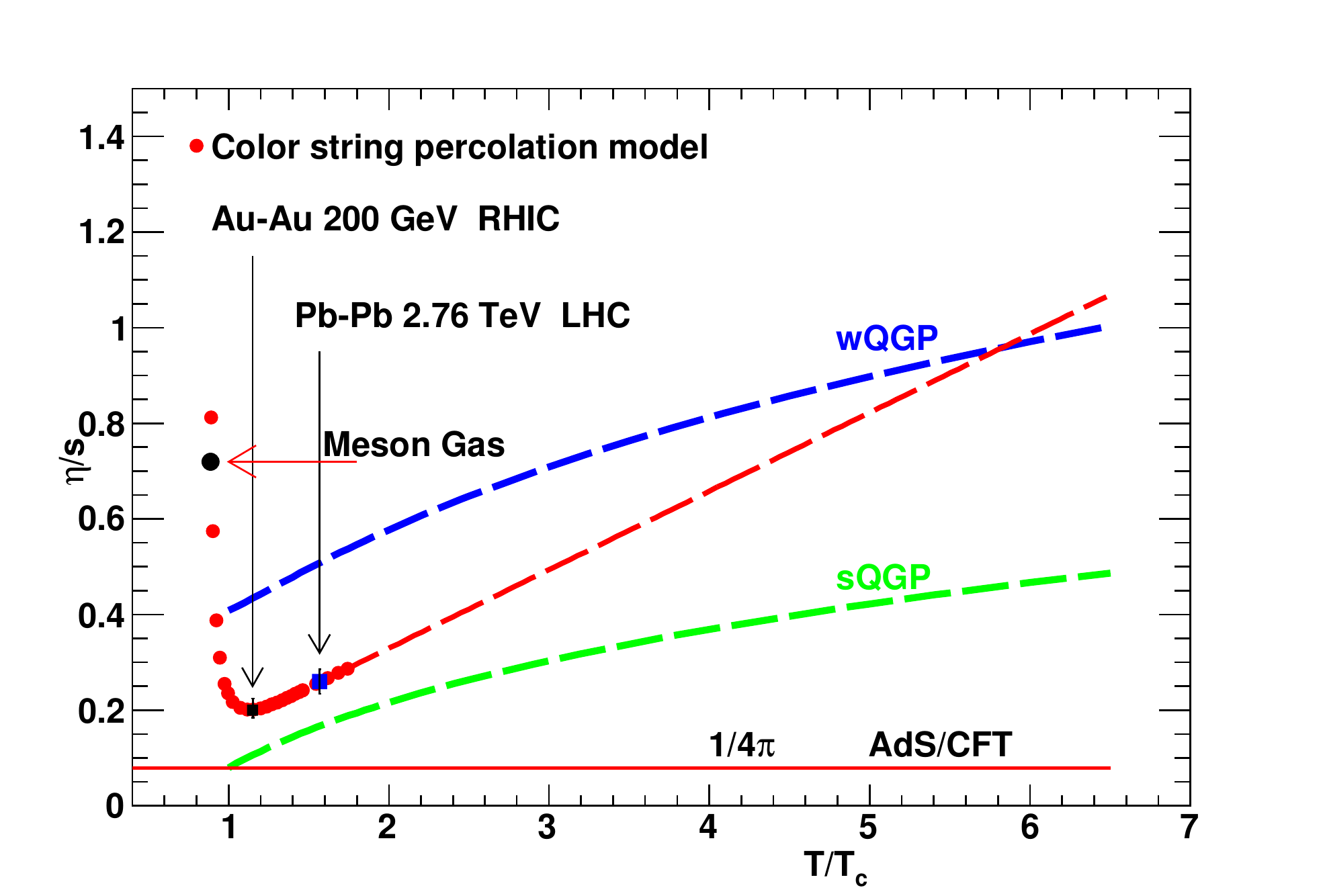}
\vspace*{-0.5cm}
\caption{$\eta/s$ as a function of T/$T_{c}$. Au-Au at 200 GeV for 0-10$\%$ centrality is shown as solid black square. wQGP and sQGP values are shown as dotted blue and green lines respectively \cite{gul1}. The red dotted line represents the extrapolation to higher temperatures from the CSPM \cite{eos2}. The hadron gas value for $\eta/s$ $\sim$ 0.7 is shown as solid black circle at T/$T_{c} \sim $0.88 \cite{prakash}. The red solid line is Ads/CFT limit \cite{kss}.} 
\label{etaovers}
\end{figure} 
Figure \ref{etaovers} shows a plot of $\eta/s$ as a function of T/$T_{c}$. The estimated value of $\eta/s$ for Pb-Pb is also shown in Fig. \ref{etaovers} at T/$T_{c}$ = 1.57.
The lower bound shown in Fig. \ref{etaovers} is given by AdS/CFT \cite{kss}. The results from STAR and ALICE show that the $\eta/s$ value is 2.5 and 3.3 times the KSS bound \cite{kss}. Extrapolating the $\eta/s$ CSPM values it is clear that it will approach the weak coupling limit near $T/T_{c}$ $\sim$ 5.8 as the temperature is increased. The CSPM  $\eta/s$ value for the hadron gas is in agreement with the calculated value using measured elastic cross sections for a gas of pions and kaons \cite{prakash}.  
\subsection{Jet quenching parameter}
One of the outstanding observation made in central collisions of heavy ions is that the final state hadrons are strongly suppressed at large transverse momenta \cite{jet1, jet2}. This phenomena is referred as jet quenching, is understood to be caused by gluon radiation induced by multiple collisions of the leading parton while traversing the medium \cite{jet3}. In QCD, the energy loss is expressed by the transport coefficient $\hat{q}$, which describes the rate at which medium transfers transverse momentum of the fast parton. 
 An important question is whether different transport coefficients are related with each other. It was shown that two transport coefficients shear viscosity $\eta$ and jet quenching parameter $\hat{q}$ are related \cite{jet3,jet4}. 
\begin{equation}
\frac {\eta}{s} \approx {\frac{3}{2} \frac {T^{3}}{\hat{q}}}.
\label{qhat1e}
\end{equation}
The relation associates a small ratio of $\eta/s$ to a large value of  $\hat{q}$. A large amount of theoretical work has been done to extract the jet transport parameter from jet quenching at RHIC and LHC energies \cite{jet1,jet3,jet4,jetqm}. The latest study by the JET Collaboration has extracted or calculated $\hat{q}$ from five different approaches to the parton energy loss in a dense medium. 
 The evolution of bulk medium in the study was given by 2+1D or 3+1D 
hydrodynamic models with the initial temperatures of $T_{RHIC}^{Hydro}$ = 346 - 373 MeV  and $T_{LHC}^{Hydro}$ = 447 - 486 MeV for most central Au-Au collisions at 
$\sqrt{s_{NN}}$ = 200 GeV and Pb-Pb collisions at $\sqrt{s_{NN}}$ = 2.76 TeV respectively. 
\begin{figure}[thbp]
\centering        
\vspace*{-0.2cm}
\includegraphics[width=0.65\textwidth,height=3.0in]{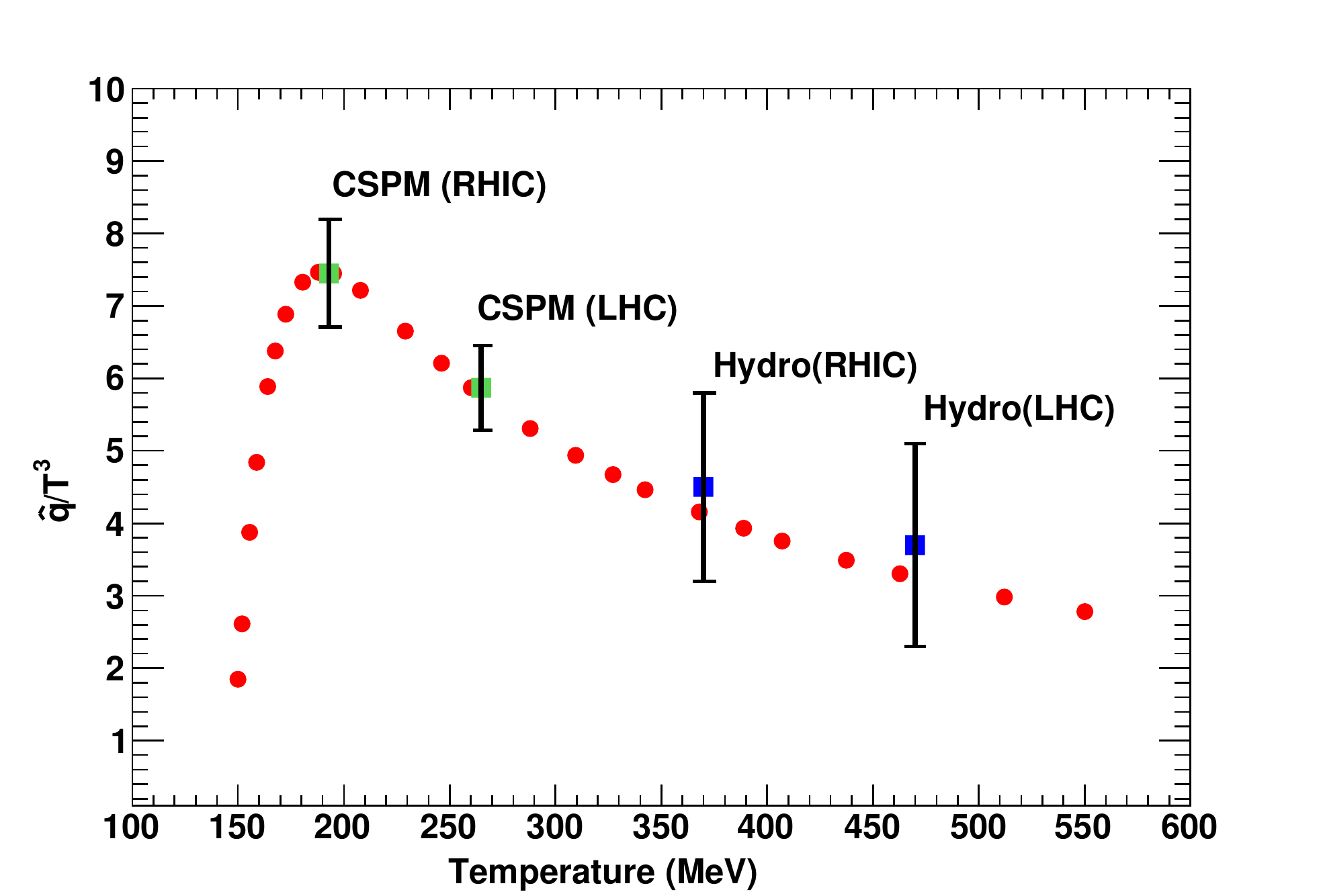}
%vspace*{-0.5cm}
\caption{Scaled jet quenching parameter $\frac {\hat {q}}{T^{3}}$ as a function of the temperature. The values shown in solid blue squares (Hydro(RHIC) and Hydro(LHC) are given by Eq. (\ref{qeq}). The CSPM values are shown in green solid squares CSPM(RHIC) and CSPM(LHC) for temperatures $\sim$193 and $\sim$262 MeV at RHIC and LHC energies respectively \cite{eos2}.}
\label{qhat}
\end{figure} 
The variation of $\hat{q}$ values between different models can be considered as theoretical uncertainties. One therefore can extract its range of values at RHIC and LHC \cite{jet1,jetqm}.
The scaled $\hat{q}$ at the highest temperatures reached in the most central Au-Au collisions at RHIC and Pb-Pb collisions at LHC are 
\begin{equation}
\frac{\hat{q}}{T^{3}} \approx \{^{4.5\pm 1.3 \   at \ RHIC}_ {3.7\pm1.4 \    at \ LHC}.
\label{qeq}
\end{equation}
 The corresponding absolute values for 
$\hat{q}$($GeV^{2}/fm$) for a 10 GeV quark jet are,  
\begin{equation}
\hat{q} \approx \{^{1.2\pm 0.3 \   }_ {1.9\pm0.7 } {^{T= 370 MeV}_{ T=470 MeV}},
\label{qeq2}
\end{equation}
at an initial time $\tau_{0} = 0.6 fm/c$. 
The temperature dependence of scaled jet transport parameter $\frac {\hat{q}}{T^{3}}$ is shown in Fig. \ref{qhat}. The CSPM values for $\frac {\hat {q}}{T^{3}}$, using the relation given by Eq. (\ref{qhat1e}), are shown in Fig. \ref{qhat} (solid red circles). The CSPM values for RHIC and LHC energies are shown as solid green squares while the theoretical values are shown as blue squares. It is observed that CSPM values are in agreement with the JET Collaboration results. 

The phase transition and $\hat{q}$ have  been studied in the framework of dynamical holographic QCD model \cite{liao}. It is found that both $\hat{q}/T^{3}$ and trace anomaly $\Delta = (\varepsilon - 3p)$ show a peak around the critical temperature $T_{c}$. This indicates that $\hat{q}$ can characterize the phase transition \cite{liao}. 

In the next section we show that in CSPM the inverse of $\eta/s$ is related to the trace anomaly.

%\section{Equation of State}
\subsection{Trace anomaly}
\begin{figure}[thbp]
\centering        
\vspace*{-0.2cm}
\includegraphics[width=0.65\textwidth,height=3.0in]{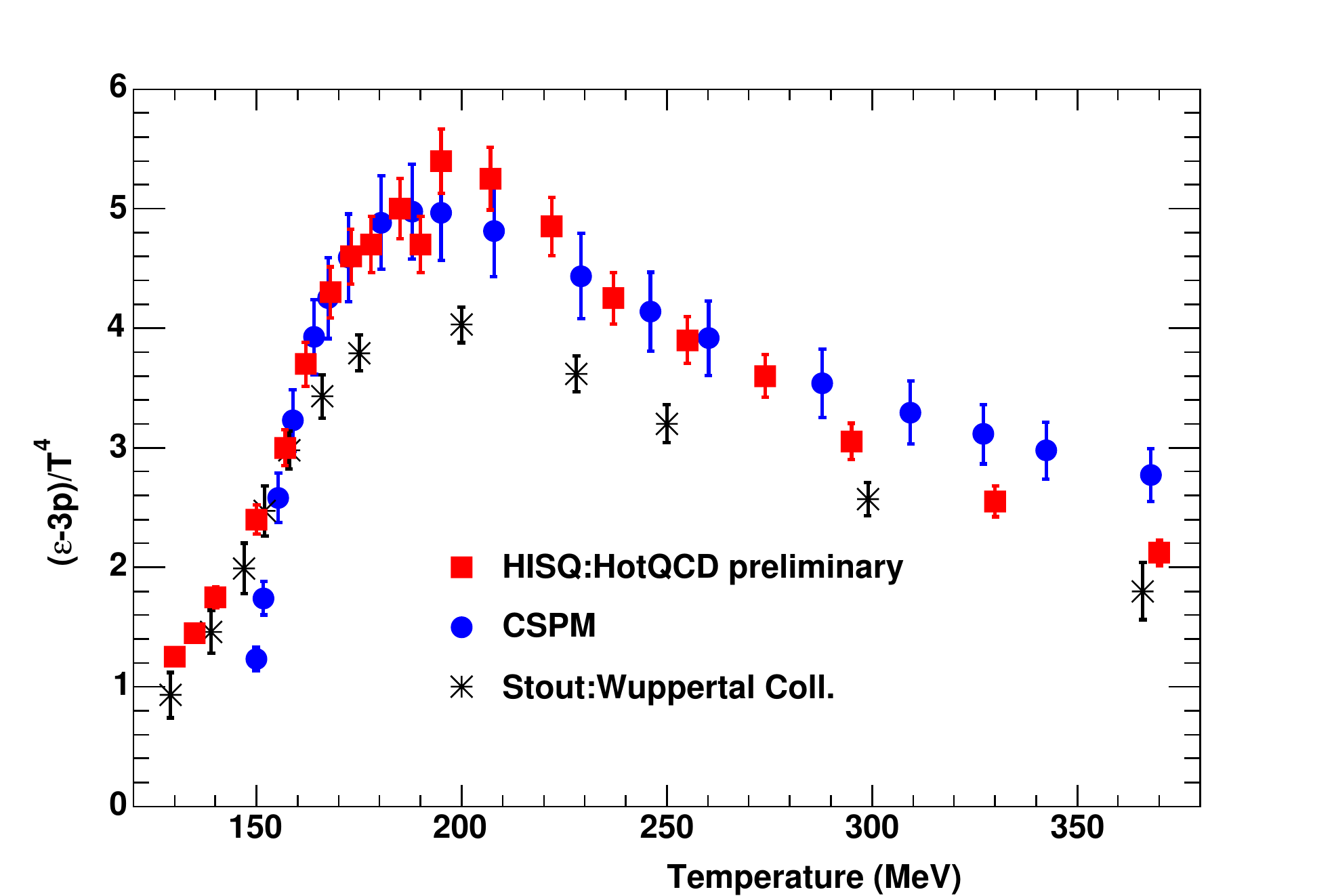}
%\vspace*{-0.5cm}
\caption{The trace anomaly $\Delta =(\varepsilon-3p)/T^{4}$ vs temperature \cite{lattice12}. Red solid squares are from HotQCD Collaboration \cite{lattice12}. Black stars are from Wuppertal Collaboration \cite{wuppe}.} 
\label{trace}
\end{figure}

The trace anomaly ($\Delta$) is the expectation value of the trace of the energy-momentum tensor, $\langle \Theta_{\mu}^{\mu}\rangle = (\varepsilon-3p)$, which measures the deviation from conformal behavior and thus identifies the interaction still present in the medium \cite{cheng}. We find that the reciprocal of $\eta/s$ is in quantitative agreement with $(\varepsilon-3p)/T^{4}$ over a wide range of temperatures \cite{cpod13,IS2013}. This result is shown in Fig.~\ref{trace}. The minimum in $\eta/s \sim 0.20$ determines the peak of the interaction measure $\sim$ 5 in agreement with the recent HotQCD values \cite{lattice12}. This happens  at the critical temperature of $T_{c} \sim 175$ MeV. Figure~\ref{trace} also shows the results from Wuppertal Collaboration \cite{wuppe}. 
The maximum in $\Delta$ corresponds to the minimum in $\eta/s$. Both $\Delta$ and $\eta/s$ describe the transition from a strongly coupled QGP to a weakly coupled QGP.
%
%\subsection{Sound Velocity $C_{S}^{2}$}
\subsection{Equation of State: Sound velocity }
We use CSPM coupled to a 1D Bjorken expansion. The input parameters are the initial temperature T, the initial energy density $\varepsilon$, and the trace anomaly $\Delta$ are determined by data. The Bjorken 1D expansion gives the sound velocity   
\begin{eqnarray}
\frac {1}{T} \frac {dT}{d\tau} = - C_{s}^{2}/\tau  \\
\frac {dT}{d\tau} = \frac {dT}{d\varepsilon} \frac {d\varepsilon}{d\tau} \\
\frac {d\varepsilon}{d\tau} = -T s/\tau, 
\end{eqnarray}
where $\varepsilon$ is the energy density, s the entropy density, $\tau$ the proper time, and $C_{s}$ the sound velocity. Since $s = (\varepsilon + P)/T$ and $P = (\varepsilon-\Delta T^{4})/3$ one gets
\begin{equation}
\frac {dT}{d\varepsilon} s = C_{s}^{2}. 
\end{equation}
From above equations $C_{s}^{2}$ can be expressed in terms of $\xi$
\begin{equation}
 C_{s}^{2} = (-0.33)\left(\frac {\xi e^{-\xi}}{1- e^{-\xi}}-1\right)+\\
             0.0191(\Delta/3)\left(\frac {\xi e^{-\xi}}{({1- e^{-\xi}})^2}-\frac {1}{1-e^{-\xi}} \right)
\end{equation}
\begin{figure}[thbp]
\centering        
\vspace*{-0.2cm}
\includegraphics[width=0.65\textwidth,height=3.0in]{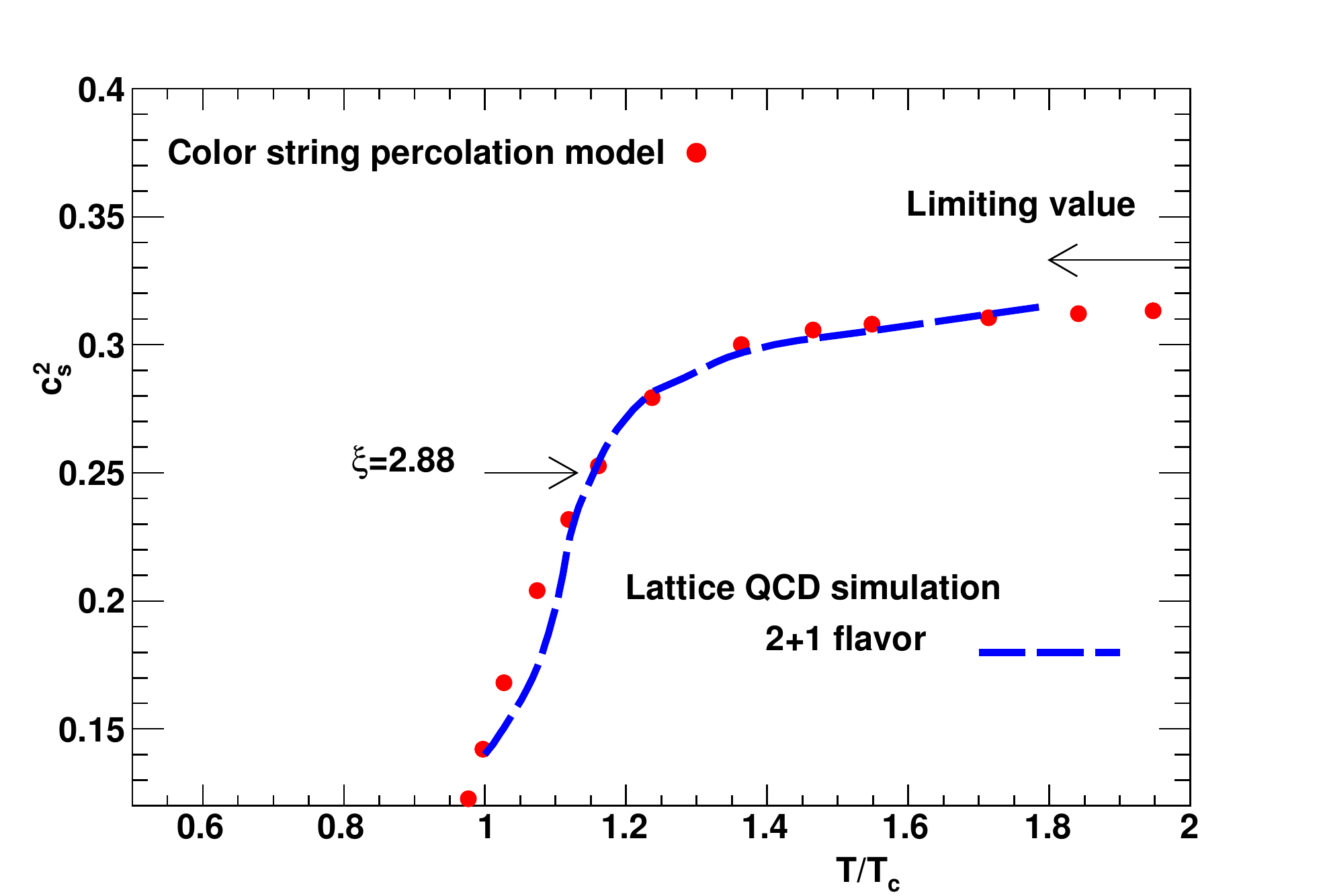}
\vspace*{-0.5cm}
\caption{The speed of sound from CSPM  versus $T/T_{c}^{CSPM}$(red circles) and Lattice QCD-p4  speed of sound versus  $T/T_{c}^{LQCD}$(blue dash line) \cite{hotqcd}.} 
\label{cs2}
\end{figure} 
Since there is no direct way to obtain pressure in CSPM, we have made the assumption that $\Delta = (\varepsilon- 3 P) \approx 1/(\eta/s)$ . Fig.~\ref{cs2} shows a plot of
 $C_{s}^{2}$ as a function of $T/T_{c}$. It is observed that the CSPM results are in very good agreement with the lattice calculations \cite{hotqcd}. This suggest that the $\Delta$ can be approximated to  $1/(\eta/s)$.
\section{Summary}
The string percolation gives a successfully description of most of the experimental data in the soft region, including rapidity distributions, probability distributions of multiplicities and transverse momentum, strength of  the B-E correlations, elliptic flow and ridge structures. The clustering of color sources has a clear physical basis, although is not derived directly from QCD, but has its most fundamental features. The non-Abelian character is reflected in the coherence sum of the color fields which as a consequence gives rise to an enhancement of the transverse momentum and a relative multiplicity suppression. The confinement is reflected in the small transverse size of the strings as well as in the length of the transverse correlations. The scaling observed in the transverse momentum distribution is a consequence of the invariance of the probability
distributions under the transformations of the size of the clusters of strings. 

The interpretation of data using string CSPM describes the equation of state close to the critical temperature and also obtains a reasonable dependence on the temperature of the shear viscosity over entropy density ratio. Thermodynamical results are also in agreement with the lattice QCD calculations. The percolation critical transition is known to represent a continuous phase transition. In central Au-Au collisions at $\sqrt{s_{NN}}=$ 200 GeV the QCD to hadron phase transition for baryon density $\mu_{B} \sim $ 0 is believed to be a cross-over transition which does not have a latent heat. The CSPM EOS correctly describes the QCD to hadron cross-over transition and provides an answer to the question of the origin of universal temperature observed in A-A, p-p and $e^{+} e^{-}$ collisions.
The percolation framework provides us with a microscopic picture which predicts the early thermalization required for hydrodynamical calculations. 
 
The minimum in $\eta/s$ can be studied as a function of the beam energy at RHIC that could locate the critical point/crossover in the QCD phase diagram seen in substances like helium, nitrogen and water. The accurate determination of $\eta/s$ is also important for the evaluation of another transport coefficient, the jet quenching parameter $\hat{q}$.
 
     There are many similarities between string percolation and the Glasma of the color glass condensate, in formal aspects and also in phenomenological applications. In this way string percolation can be regarded as a complementary view of CGC of the initial stage of a high hadronic collision. \\

%\subsection*{Acknowledgement}  

{\bf{ Acknowledgements}}\\

 This research was supported by the Office of Nuclear Physics within the U.S. Department of Energy  Office of Science under Grant No. DE-FG02-88ER40412. M.A.B. appreciates the support of S.Petersburg State University grant 11.38.197.2014 . J.D.D. thanks the support of the FCT/Portugal project PPCDT/FIS/575682004. C.P. was supported by the project FPA2011-022776 of MICINN the Spanish Consolider Ingenio 2010 program CPAN and Conselleria Education Xunta de Galicia.

%\subsection*{References}
%\maketitle
%\section*{References}
%\bibliographystyle{plainnat}
\bibliography{all}

\end{document}